\documentclass[a4paper,11pt]{article}
\usepackage{heppub}
\pdfoutput=1
\usepackage{xspace}
\usepackage{placeins}
\usepackage{xcolor}

\newcommand{\abs}[1]{\lvert#1\rvert}
\newcommand{\Abs}[1]{\bigl\lvert#1\bigr\rvert}
\newcommand{\ord}[1]{\mathcal{O}(#1)}
\newcommand{\Ord}[1]{\mathcal{O}\bigl(#1\bigr)}
\newcommand{\ORd}[1]{\mathcal{O}\Bigl(#1\Bigr)}
\newcommand{\ordsq}[1]{\mathcal{O}[#1]}

\newcommand{\df}{\mathrm{d}}
\newcommand{\img}{\mathrm{i}}

\newcommand{\eps}{\epsilon}

\newcommand{\w}{\omega}

\newcommand{\fm}{\,\mathrm{fm}}

\newcommand{\pb}{\,\mathrm{pb}}
\newcommand{\GeV}{\,\mathrm{GeV}}
\newcommand{\TeV}{\,\mathrm{TeV}}

\newcommand{\nn}{\nonumber}

\newcommand{\bt}{{\vec b}_T}
\newcommand{\qt}{{\vec q}_T}

\newcommand{\cF}{\mathcal{F}}
\newcommand{\cI}{\mathcal{I}}

\newcommand{\cO}{\mathcal{O}}

\newcommand{\Tau}{\mathcal{T}}

\newcommand{\tf}{\tilde{f}}

\newcommand{\tB}{\tilde{B}}

\newcommand{\tS}{\tilde{S}}
\newcommand{\tgamma}{\tilde{\gamma}}
\newcommand{\tsigma}{\tilde{\sigma}}

\newcommand{\as}{\alpha_s}
\newcommand{\aem}{\alpha_\mathrm{em}}

\newcommand{\lqcd}{\Lambda_\mathrm{QCD}}
\newcommand{\muZeromin}{\mu_0^\mathrm{min}}
\newcommand{\muBmin}{\mu_B^\mathrm{min}}
\newcommand{\mufmin}{\mu_f^\mathrm{min}}
\newcommand{\muinitmin}{\mu_\init^\mathrm{min}}
\newcommand{\muSmin}{\mu_S^\mathrm{min}}
\newcommand{\nuSmin}{\nu_S^\mathrm{min}}

\newcommand{\qTmax}{q_T^\mathrm{max}}
\newcommand{\zetainitmin}{\zeta_\init^\mathrm{min}}

\newcommand{\Ecm}{E_\mathrm{cm}}
\newcommand{\MSbar}{$\overline{\text{MS}}$\xspace}

\newcommand{\two}{{(2)}}
\newcommand{\three}{{(3)}}

\newcommand{\can}{\mathrm{can}}
\newcommand{\central}{\mathrm{central}}
\newcommand{\cusp}{\mathrm{cusp}}
\newcommand{\cut}{\mathrm{cut}}
\newcommand{\dglap}{\mathrm{DGLAP}}
\newcommand{\down}{\mathrm{down}}
\newcommand{\incl}{\mathrm{incl}}
\newcommand{\init}{\mathrm{init}}

\newcommand{\fin}{\mathrm{fin}}

\newcommand{\lp}{\mathrm{LP}}
\newcommand{\match}{\mathrm{match}}

\newcommand{\np}{\mathrm{np}}
\newcommand{\pert}{\mathrm{pert}}
\newcommand{\recoil}{\mathrm{recoil}}
\newcommand{\res}{\mathrm{res}}
\newcommand{\run}{\mathrm{run}}

\newcommand{\up}{\mathrm{up}}
\newcommand{\vary}{\mathrm{vary}}
\newcommand{\nons}{\mathrm{nons}}
\newcommand{\sing}{\mathrm{sing}}

\newcommand{\FO}{\mathrm{FO}}

\newcommand{\WidthTwoSubfigs}{0.48\textwidth}

\newcommand{\multrow}[1]{\begin{tabular}{@{}c@{}} #1 \end{tabular}}

\allowdisplaybreaks[3]

\title{\boldmath Drell-Yan Transverse-Momentum Spectra at N$^3$LL$'$
and Approximate N$^4$LL with SCETlib}

\author[a]{Georgios Billis,\hspace{-0.2ex}}
\emailAdd{georgios.billis@unimib.it}

\author[b,c,d]{Johannes K.~L.~Michel,\hspace{-0.2ex}}
\emailAdd{jklmich@mit.edu}

\author[e]{and Frank J.~Tackmann}
\emailAdd{frank.tackmann@desy.de}

\affiliation[a]{Universit\`{a} degli Studi di Milano-Bicocca \& INFN, Piazza della Scienza 3, Milano 20126, Italy\vspace{0.5ex}}
\affiliation[b]{Center for Theoretical Physics,\,Massachusetts Institute of Technology,\,Cambridge,\,MA\,02139,\,USA}
\affiliation[c]{Institute for Theoretical Physics Amsterdam and Delta Institute for Theoretical Physics, University of Amsterdam, Science Park 904, 1098 XH Amsterdam, The Netherlands}
\affiliation[d]{Nikhef, Theory Group, Science Park 105, 1098 XG, Amsterdam, The Netherlands}
\affiliation[e]{Deutsches Elektronen-Synchrotron DESY, Notkestr. 85, 22607 Hamburg, Germany}

\abstract{%
We provide state-of-the-art precision QCD predictions
for the fiducial $W$ and $Z$ boson transverse momentum spectra at the LHC
at N$^3$LL$'$ and approximate N$^4$LL in resummed perturbation theory,
matched to available $\mathcal{O}(\alpha_s^3)$ fixed-order results.
Our predictions consistently combine all information
from across the spectrum in a unified way,
ranging from the nonperturbative region of small transverse momenta
to the fixed-order tail,
with an emphasis on estimating the magnitude
of residual perturbative uncertainties,
and in particular of those related to the matching.
Parametric uncertainties related to the strong coupling,
the collinear PDFs, and the nonperturbative transverse momentum-dependent (TMD)
dynamics are studied in detail.
To assess the latter, we explicitly demonstrate how the full complexity
of flavor and Bjorken $x$-dependent TMD dynamics
can be captured by a single, effective nonperturbative function
for the resonant production of any given vector boson at a given collider.
We point out that the cumulative $p_T^Z$ cross section
at the level of precision enabled by our predictions
provides strong constraining power for PDF determinations
at full N$^3$LO.
}

\date{November 24, 2024}

\preprint{\vbox{%
\hbox{DESY 23-081}
\hbox{MIT-CTP/5572}
\hbox{Nikhef 2024-007}
}
}

\begin{document}

\maketitle

\section{Introduction}
\label{sec:intro}

The transverse momentum spectra of electroweak bosons produced in hadronic collisions
are flagship observables of the LHC precision program,
and have been measured to astonishing accuracy
by the ATLAS~\cite{Aad:2011fp, Aad:2014xaa, Aad:2015auj, Aaboud:2017ffb, Aad:2019wmn, ATLAS:2023lsr},
CMS~\cite{Chatrchyan:2011wt, Khachatryan:2015oaa, Khachatryan:2016nbe, Sirunyan:2017igm, Sirunyan:2019bzr},
and LHCb collaborations~\cite{LHCb:2015mad, LHCb:2016fbk}.
The $p_T$ spectrum of the $Z$ boson,
measured in the experimentally pristine Drell-Yan dilepton channel,
is of particular importance as an irreducible background
to many searches for physics beyond the Standard Model, see e.g.\ \refcite{Lindert:2017olm},
and likewise features great sensitivity
to key Standard Model parameters like the strong coupling~\cite{Camarda:2022qdg, ATLAS:2023lhg}.
On the other hand, the $p_T$ spectrum of the $W$ boson,
while much harder to access experimentally,
is a critical input to measurements of the $W$ boson mass
\cite{Aaboud:2017svj, LHCb:2021bjt, CDF:2022hxs, ATLAS:2024erm, CMS:2024lrd}
that rely on Jacobian peaks -- smeared out by the $p_T^W$ spectrum --
to achieve sensitivity.
Lastly, recent progress in understanding the complex interplay
of fiducial experimental selection cuts and QCD perturbation theory
has revealed that even seemingly inclusive quantities
like (fiducial) rapidity spectra can feature sensitivity
to small-$p_T$ physics, requiring a precise understanding
of the underlying $p_T$ spectrum as well~\cite{Ebert:2020dfc, Billis:2021ecs, Salam:2021tbm, Amoroso:2022lxw, Alekhin:2024mrq}.
In turn, a thorough understanding of these percent-level effects
on fiducial rapidity spectra is required
e.g.\ to deliver constraining power on collinear parton distribution functions (PDFs),
whose determination has advanced to approximate N$^3$LO accuracy~\cite{McGowan:2022nag, NNPDF:2024nan, Cridge:2024exf, Cooper-Sarkar:2024crx, MSHT:2024tdn}.

The phenomenological importance of the $p_T^Z$ and $p_T^W$ spectra
has motivated a large-scale effort by the theory community
to provide precision predictions for them,
in turn making them crucial tests of our understanding
of QCD and the electroweak sector of the Standard Model.
Fixed-order predictions in QCD perturbation theory at finite $p_T^{Z,W}$ have
reached an impressive $\ord{\as^3}$ accuracy~\cite{Ridder:2015dxa, Ridder:2016nkl, Boughezal:2015dva, Boughezal:2015ded, Boughezal:2016isb, Boughezal:2016dtm, Gehrmann-DeRidder:2017mvr, Campbell:2019gmd, Neumann:2022lft},
i.e., NNLO$_1$ relative to the tree-level LO$_1$
process where one hard parton is radiated into the final state.
On the other hand, at small transverse momenta $p_T^V \ll m_{V}$,
where the vector boson $V$ recoils against soft and collinear radiation,
the infrared singularities of gauge theory
enhance the contribution to the cross section at each perturbative order
by large double logarithms $\as^n \ln^{2n} (p_T^V/m_V)$,
upsetting the convergence and validity of the fixed-order series.
The all-order resummation of these dominant singular terms,
which is based on factorization theorems and the renormalization group,
has in the meantime been achieved at N$^3$LL$'$~\cite{Billis:2021ecs, Ju:2021lah, Re:2021con, Chen:2022cgv, Camarda:2022qdg}
and approximate N$^4$LL order~\cite{Neumann:2022lft, Camarda:2023dqn, Moos:2023yfa, Piloneta:2024aac}
by several groups.
The analytic resummation at this order also provides an important
ingredient to extend the combination of high-order calculations with parton showers
to N$^3$LO$+$PS using the methods of \refscite{Alioli:2015toa, Alioli:2021qbf}.
(Another ingredient is the NNLL$'$ or N$^3$LL resummation
for the $V+\text{1 jet}$ process~\cite{Alioli:2023rxx}.)
The region of even smaller transverse momenta $p_T^V \lesssim \lqcd$
close to the QCD confinement scale is of particular interest
because it provides access to the nonperturbative
transverse momentum-dependent (TMD) dynamics
of partons within the proton, see e.g.\ \refscite{Bacchetta:2022awv, Moos:2023yfa, Bacchetta:2024qre}.
Beyond (non)perturbative QCD, mixed strong and weak or electromagnetic corrections
have been studied both at fixed order~\cite{Dittmaier:2014qza, Dittmaier:2015rxo, deFlorian:2018wcj, Delto:2019ewv, Bonciani:2019nuy, Cieri:2020ikq, Buccioni:2020cfi}
and to all orders in the limit of small transverse momentum~\cite{Cieri:2018sfk, Bacchetta:2018dcq, Billis:2019evv, Buonocore:2024xmy}.

In this paper, we present predictions for the resummed and matched $q_T \equiv p_T^{W,Z}$ spectrum
as implemented in \texttt{SCETlib}~\cite{scetlib}, a C$++$ library for numerical calculations in QCD
and Soft-Collinear effective theory (SCET).
Our emphasis lies on consistently combining all information
from across the spectrum, ranging from nonperturbative to fixed-order scales,
where we in particular go beyond the status of the literature
by including a thorough assessment of the associated matching uncertainties.
This is part of an overall, careful estimate of the magnitude of residual perturbative uncertainties,
which we supplement with detailed studies of the parametric strong coupling,
PDF, and nonperturbative TMD uncertainties.
We further present a novel way of rigorously defining
effective nonperturbative TMD functions for resonant $p_T^V$
and multi-differential spectra,
which has served as an important ingredient of the theoretical model
underlying the recent CMS measurement of the $W$ boson mass~\cite{CMS:2024lrd}.
As another application of our predictions, we consider
the possibility of constraining collinear PDFs at complete three-loop accuracy
using the cumulative $p_T^Z$ distribution.
\enlargethispage{-\baselineskip}

The paper is structured as follows:
In \sec{fact_and_pert_ingred}, we review the factorization theorem underlying the resummation,
give an overview of the perturbative ingredients entering our predictions,
and introduce our matching formalism that unifies the nonperturbative limit,
the perturbative resummation, and the fixed-order tail.
In \sec{np_model}, we review the nonperturbative structure of the factorization theorem,
and develop our framework of effective nonperturbative functions.
In \sec{uncerts} we present our predictions for the $p_T^Z$ spectrum,
discussing the perturbative uncertainty, the impact of nonperturbative physics
the parametric $\as$, and the parametric PDF uncertainties in turn,
before showcasing the impact of approximate N$^3$LO PDFs and N$^4$LL
Sudakov ingredients on the predicted $p_T^Z$ spectrum.
In \sec{results_Z_cumulants}, we discuss the cumulative fiducial cross section
and its sensitivity to PDFs as an immediate application.
In \sec{results_W}, we present our results for the $p_T^W$ spectrum.
We conclude and summarize our results in \sec{conclusions}.

\section{Factorization, perturbative ingredients, and matching}
\label{sec:fact_and_pert_ingred}

\subsection{Review of factorization}
\label{sec:factorization}

We consider the production of a dilepton pair through $Z/\gamma^*$
or of a lepton-neutrino final state through a $W$ boson at the LHC.
At leading order in the electromagnetic interaction,
the fiducial cross section differential in the total transverse momentum $q_T$ of the respective vector boson,
$q_T = p_T^Z = p_T^{\ell\ell}$ or $q_T = p_T^W = p_T^{\ell\nu}$,
is given by~\cite{Ebert:2020dfc}
\begin{align}
\frac{1}{\pi q_T} \frac{\df\sigma(\Theta)}{\df q_T}
&=  \frac{1}{4\Ecm^2} \int \! \df Q^2 \, \df Y
\sum_{i = -1}^7 \sum_{V, V'} L_{i\,VV'}(q, \Theta)\,W_{i\,VV'}(q, P_a, P_b)
\,,\end{align}
where $q$ is the momentum carried by the (generally off-shell) vector boson
with invariant mass $Q^2 \equiv q^2$ and rapidity $Y$.
Here we have decomposed the hadronic tensor
into nine scalar hadronic structure functions $W_{i\,VV'}(q, P_a, P_b)$, with $i = -1, 0, \dots, 7$,
and summed over the possible vector bosons $V, V'$ interfering with each other.
The $W_{i\,VV'}$ encode the hadronic production dynamics.
They depend on the vector boson momentum $q^\mu$ and the incoming proton momenta $P_{a,b}$,
and can be interpreted as entries in the $3 \times 3$ spin density matrix
for the polarization of the vector boson.
By contrast, the leptonic tensor projections $L_{i\,VV'}(q, \Theta)$
that describe the propagation and decay of the vector boson
only depend on the intermediate momentum $q$ and the set of fiducial acceptance cuts applied on the dilepton or lepton-neutrino final state,
which we collectively denote by $\Theta$.
Explicitly, the leptonic tensor projections are given by
\begin{align}
L_{i\,VV'}(q, \Theta)
&=  \frac{3}{16\pi}
\int_{-1}^1 \! \df \cos \theta \int_0^{2\pi} \df \varphi \, \hat{\Theta}(q, \theta, \varphi) \,
L_{\pm(i)\,VV'}(q, \theta, \varphi) \, g_i(\theta, \varphi)
\,,\end{align}
where $\cos \theta$ and $\varphi$ are spherical coordinates parametrizing
the momentum of the matter particle
(i.e., the negatively charged lepton, or the neutrino)
in a suitable vector boson rest frame,
which we take to be the Collins-Soper frame~\cite{Collins:1977iv}.
The prefactors $L_{\pm(i)\,VV'}$ contain the vector-boson propagator and leptonic couplings
and depend on whether the hadronic structure function is parity even, $\pm(i) = +$ for $i = -1, 0, 1, 2, 5, 6$,
or parity-odd, $\pm(i) = -$ for $i = 3, 4, 7$.
It is convenient to also define the so-called helicity cross sections
\begin{align}
\frac{\df \sigma_i}{\df^4 q} \equiv \frac{1}{2\Ecm^2} \sum_{V, V'} L_{\pm(i)\,VV'}(q^2) \, W_{i\,VV'}(q, P_a, P_b)
\end{align}
multiplying the spherical harmonics $g_i$
at the level of the six-fold differential cross section
\begin{align} \label{eq:helicity_xsecs}
\frac{\df \sigma}{\df^4 q \, \df \cos \theta \, \df \phi}
= \frac{3}{16\pi} \sum_i \frac{\df \sigma_i}{\df^4 q} \, g_i(\theta, \varphi)
\,.\end{align}
Only the parity-even structure functions contribute to the $p_T^Z$ spectrum
as long as fiducial cuts are applied independent of lepton charges,
while all structure functions in general contribute to the $p_T^W$ spectrum
due to the cuts generically being asymmetric under $\ell \leftrightarrow \nu$.
The $g_i(\theta, \varphi)$ are real spherical harmonics
that describe the distribution of decay products for a given intermediate polarization state.
Finally, $\hat{\Theta}(q, \theta, \varphi)$
encodes the action of the fiducial cuts on a dilepton phase-space point
parametrized by the two rest-frame angles $\theta$ and $\varphi$,
and an additional total boost by $q$ from the lab frame.%
\footnote{Note that this specifies the vector boson rest frame only up to Wigner rotations.
The Collins-Soper frame specifically is defined
by first performing a longitudinal boost along the beam axis into the frame where the vector boson has vanishing rapidity,
followed by a transverse boost by $q_T$.
In the limit of massless protons (and only in this limit~\cite{Ebert:2020dfc}),
this definition is equivalent to demanding that the incoming protons lie in the plane spanned by the $x$ and $z$ axis of the rest frame,
and that they have equal and opposite angles to the $z$ axis.}
One notable special case is $\hat{\Theta} = 1$,
which corresponds to the inclusive $q_T$ spectrum proportional to the linear combination $W_\incl \equiv W_{-1} + W_0/2$,
where $g_{-1,0} = 1 \pm \cos^2 \theta$.
Another important special case is $\hat{\Theta} = \operatorname{sgn}(Y) \bigl[ \Theta(\cos \theta ) - \Theta(- \cos \theta) \bigr]$,
which projects out $g_4 = \cos \theta$, i.e., the forward-backward asymmetry $A_\mathrm{FB}(q) \propto \operatorname{sgn}(Y) \, W_4/W_\incl$
differential in the vector boson kinematics (but applying no additional fiducial cuts).
We stress that while it is convenient to parametrize the decay kinematics in a definite frame,
the corresponding projectors acting on the hadronic tensor can in fact be defined in a fully covariant way~\cite{Ebert:2020dfc},
such that the $W_{i\,VV'}$ are genuine Lorentz scalars.

We are interested in the region of small transverse momentum, $q_T \ll Q$.
In this limit, the hadronic structure functions corresponding
to the inclusive $q_T$ spectrum and the $q_T$-dependent forward-backward asymmetry
satisfy the following factorization theorem
\cite{Collins:1981uk, Collins:1981va, Collins:1984kg, Collins:1350496, Bauer:2000ew, Bauer:2000yr, Bauer:2001yt, Bauer:2002nz, Becher:2010tm, GarciaEchevarria:2011rb, Chiu:2012ir, Li:2016axz},
which is valid for $i = -1$ and $i = 4$,
\begin{align} \label{eq:tmd_factorization_m1_4}
W_{i\,VV'}
&= W^\lp_{i \, VV^\prime} (q, P_a, P_b)
\, \Bigl[ 1 + \ORd{\frac{q_T^2}{Q^2}, \frac{\lqcd^2}{Q^2}} \Bigr]
\nn \\
&= \sum_{a,b} H_{i\,VV'\,ab}(Q^2, \mu) \,
[B_a B_b S](Q^2, x_a, x_b, \qt, \mu) \,
\Bigl[ 1 + \ORd{\frac{q_T^2}{Q^2}, \frac{\lqcd^2}{Q^2}} \Bigr]
\,,\end{align}
where $x_{a,b} \equiv Q/\Ecm \, e^{\pm Y}$.
As indicated, the factorization receives power corrections in $(q_T/Q)^2$ and $(\lqcd/Q)^2$,
but remains valid in the nonperturbative regime $q_T \sim \lqcd$.
We will exploit this in \sec{np_model} by explicitly parametrizing
the leading nonperturbative corrections $\ord{\lqcd^2/q_T^2}$ at small $q_T$
in a way consistent with the field-theoretic structure of the factorization theorem,
and by capturing yet higher corrections through model functions.

In \refcite{Ebert:2020dfc}, it was shown that all linear power corrections $\ord{q_T/Q}$ to the fiducial $q_T$ spectrum
can be predicted from factorization and resummed to all orders
by retaining the exact dependence of the leptonic tensor projections $L_i(q, \Theta)$
on $q_T$, which naively would be power suppressed,
\begin{align} \label{eq:tmd_factorization_fid_spectrum}
\frac{1}{\pi q_T} \frac{\df\sigma(\Theta)}{\df q_T}
&=  \frac{1}{2\Ecm^2} \int \! \df Q^2 \, \df Y
\sum_{i = -1,4} \sum_{V, V'} L_{i\,VV'}(q, \Theta)\,W_{i\,VV'}^\lp(q, P_a, P_b)
\nn \\ & \quad \times
\Bigl[ 1 + \ORd{\frac{q_T^2}{Q^2}, \frac{\lqcd^2}{Q^2}} + \ORd{\frac{\lqcd^2}{q_T Q}}_\text{DBM} \, \Bigr]
\,,\end{align}
where the $W_{i\,VV'}^\lp$ for $i = -1,4$ are given in \eq{tmd_factorization_m1_4}.
As discussed in \refcite{Ebert:2020dfc}, this in fact holds for a broader class of observables that are azimuthally symmetric at leading power,
and also generalizes to so-called leptonic power corrections
that arise for leptonic observables sensitive to the edge of Born phase space.

The first set of power corrections in \eq{tmd_factorization_fid_spectrum}
arises from the quadratic power corrections to the $i = -1,4$ structure functions,
whose leptonic coefficient functions scale as $L_{-1, 4\,VV'} \sim (q_T/Q)^0$,
as well as from the hadronic structure functions for $i = 0, 1, 3, 5, 6, 7$
that start at most at linear order in $q_T/Q$,
but whose leptonic coefficients are all in addition suppressed by $(q_T/Q)$.
The second set of power corrections, indicated by the subscript ``DBM''
for Double Boer-Mulders effect,
arises from the two structure functions $W_{2,5\,VV'}$.
These have a linearly suppressed leptonic coefficient $L_{2,5\,VV'} \sim q_T/Q$,
but only receive a contribution at leading power in $q_T/Q
$ from the product of two Boer-Mulders functions,
which in turn are suppressed by one power of $\lqcd/q_T$ each
in the case of massless $n_f = 5$ QCD~\cite{Bacchetta:2008xw},
and by at least two powers of $\as$ each for massive quarks~\cite{vonKuk:2023jfd}.
The region where both expansion parameters $\lqcd/q_T$ and $q_T/Q$ are not small
is negligible for resonant $Z$ or $W$ production at the LHC,
and we therefore ignore this contribution.%
\footnote{This is further supported by the observation in \sec{uncerts_np}
that the dominant estimated
nonperturbative contribution arises from the Collins-Soper (or rapidity) evolution,
which is common to both $i = -1,4$ and $i = 2,5$.
See also \ftn{dbm} for the case of leptonic observables directly sensitive to low scales,
such as $\phi^*_\eta$ or the $p_T^\ell$ spectrum near the Jacobian peak.}

We now return to the individual ingredients in \eq{tmd_factorization_m1_4}.
Here $H_{i\,VV'\,ab}$ denotes the hard function,
which encodes virtual corrections to the production amplitudes $ab \to V, V'$ in the underlying hard interaction.
The \MSbar result for $H_{i\,VV'\,ab}$ can, for instance, be obtained as the IR-finite part of the
corresponding quark form factors squared, using dimensional regularization to regulate IR divergences.
Explicit expressions for the leptonic prefactors $L_{\pm\,VV'}$ and the hard functions
in terms of IR-finite parts of quark form factors (i.e., SCET Wilson coefficients)
can be found in the appendices of \refcite{Ebert:2020dfc}.
At our working order in this paper, we require the complete three-loop results for the hard function.
This involves the three-loop results for the quark nonsinglet form factor~\cite{Gehrmann:2010ue, Baikov:2009bg}.
In addition, starting at two loops, there are contributions
to the quark singlet axial and vector form factors from closed fermion loops~\cite{Dicus:1985wx, Kniehl:1989qu, Bernreuther:2005rw, Gehrmann:2021ahy, Chen:2021rft},
which we discuss in more detail in \sec{singlet_coeffs}.

The second term in \eq{tmd_factorization_m1_4} encodes physics at the low scale $\mu \sim q_T$,
and can be written in two prototypical forms,
\begin{subequations} \label{eq:tmd_factorization}
\begin{align} \label{eq:tmd_factorization_bbs}
&[B_a B_b S](Q^2, x_a, x_b, \qt, \mu)
\nn \\[0.4em]
&\equiv \frac{1}{2\pi} \int_0^\infty \df b_T \, b_T J_0(b_T q_T) \,
   \tB_a(x_a, b_T, \mu, \nu/\w_a) \, \tB_b(x_b, b_T, \mu, \nu/\w_b) \,
   \tS(b_T, \mu, \nu)
\\ \label{eq:tmd_factorization_ff}
&= \frac{1}{2\pi} \int_0^\infty \df b_T \, b_T J_0(b_T q_T) \,
   \tf_a(x_a, b_T, \mu, Q^2) \, \tf_b(x_b, b_T, \mu, Q^2)
\,.\end{align}
\end{subequations}
In \eq{tmd_factorization_bbs}, the beam functions $B_i(x, \vec{k}_T, \mu, \nu/\w)$ describe
the extraction of an unpolarized parton $i$ with longitudinal momentum fraction $x$
and transverse momentum $\vec{k}_T$ from an unpolarized proton,
where we have taken a Fourier transform from momentum ($\vec{k}_T$) to position ($\vec{b}_T$) space.
Similarly, the soft function $S(\vec{k}_T, \mu, \nu)$
encodes wide-angle soft radiation with total transverse momentum $\vec{k}_T$,
which is again Fourier conjugate to $\vec{b}_T$.
Evaluating the two beam functions and the soft function at a common argument $\vec{b}_T$
and taking the inverse Fourier transform as above implements momentum conservation in position space.
We have also used the fact that the quark beam and soft functions (TMD PDFs) only depend on the magnitude of $\vec{b}_T$
to freely integrate the Fourier phase $e^{\img \bt \cdot \qt}$ over the azimuth of $\vec{b}_T$ in \eq{tmd_factorization},
yielding a zeroth-order Bessel function $J_0(b_T q_T)$ of the first kind.

Equivalently, one can write this as shown in \eq{tmd_factorization_ff},
where the transverse-momentum dependent beam and soft functions have been combined
into transverse-momentum dependent PDFs (TMD PDFs)
\begin{align} \label{eq:TMDPDF}
\tf_i(x, b_T, \mu, \zeta) = \tB_i\Bigl(x,b_T,\mu,\frac{\nu}{\sqrt{\zeta}}\Bigr) \sqrt{\tS(b_T,\mu,\nu)}
\,.\end{align}
A key feature of transverse-momentum dependent factorization
is the explicit dependence of the low-energy matrix elements
on the energy of the colliding parton,
encoded either in its lightcone momentum $\w$ or in the Collins-Soper scale $\zeta$,
where
\begin{align} \label{eq:def_x_i_w_i}
x_{a,b} = \frac{\w_{a,b}}{\Ecm} e^{\pm Y}
\,, \qquad
\w_{a,b} = Q
\,, \qquad
\zeta_{a,b} \propto \w_{a,b}^2
\,, \qquad
(\w_a \w_b)^2 = \zeta_a \zeta_b = Q^4
\,,\end{align}
and $\Ecm$ is the total hadronic center-of-mass energy.
This explicit energy dependence sets the TMD PDF apart from the  usual collinear PDFs,
which only depend on the momentum fraction $x$,
and is a remnant of so-called rapidity divergences~\cite{Collins:1981uk,Collins:1992tv,Collins:2008ht,Becher:2010tm,GarciaEchevarria:2011rb,Chiu:2011qc,Chiu:2012ir}.
Regulating and renormalizing them separately
in the individual beam and soft functions in \eq{tmd_factorization_bbs}
introduces an additional scale $\nu$,
which is analogous to the \MSbar scale $\mu$ from renormalizing UV divergences.
The dependence on $\nu$ (or equivalently, the rapidity divergences in the bare objects)
cancels between the beam and soft function in the TMD PDF,
leaving behind the explicit dependence on $\omega$ (or $\zeta$).
Since both ways of writing \eq{tmd_factorization} are frequently encountered in the literature,
we will continue to present e.g.\ the construction of our central profile scale functions
in both notations for the benefit of the reader.

The beam and soft functions (and thus the TMD PDFs) are defined as proton and vacuum matrix elements
of renormalized operators without making reference to perturbation theory,
and thus allow for a rigorous field-theoretic treatment of the $\qt$ spectrum
in the nonperturbative regime $q_T \sim \lqcd$.
For perturbative $|\vec k_T| \sim 1/b_T \gg \lqcd$,
they can be matched onto collinear PDFs
and soft vacuum condensates by performing an operator product expansion~\cite{Collins:1981uw,Collins:1984kg},
\begin{align} \label{eq:ope_beam_soft_tmd}
\tilde B_i\Bigl(x, b_T, \mu, \frac{\nu}{\w}\Bigr)
&= \sum_j \int\! \frac{\df z}{z} \, \tilde \cI_{ij}\Bigl(z, b_T, \mu, \frac{\nu}{\w}\Bigr) \, f_{j}\Bigl( \frac{x}{z}, \mu \Bigr) + \ord{\lqcd^2 b_T^2}
\,, \nn \\
\tilde{S}(b_T, \mu, \nu) &= \tilde{S}_\pert(b_T, \mu, \nu) + \Ord{\lqcd^2 b_T^2}
\,, \nn \\[0.4em]
\tf_i(x, b_T, \mu, \zeta)
&= \sum_j \int\! \frac{\df z}{z} \, \tilde C_{ij}(z, b_T, \mu, \zeta) \, f_{j}\Bigl( \frac{x}{z}, \mu \Bigr) + \ord{\lqcd^2 b_T^2}
\,,\end{align}
where the perturbatively calculable matching coefficients are related by
\begin{align}
\tilde C_{ij}(z, b_T, \mu, \zeta) = \tilde \cI_{ij}\Bigl(z, b_T, \mu, \frac{\nu}{\sqrt{\zeta}}\Bigr) \sqrt{\tilde{S}_\pert(b_T, \mu, \nu)}
\,.\end{align}
The perturbative soft, beam functions and TMD PDF matching coefficients are all known
to three loops~\cite{Catani:2011kr,Catani:2012qa,Gehrmann:2014yya,Luebbert:2016itl,Echevarria:2015byo,Echevarria:2016scs,Li:2016ctv,Luo:2019hmp,Luo:2019szz, Ebert:2020yqt, Li:2016ctv, Lubbert:2016rku},
as required at our perturbative working order in this paper.
Our treatment of the nonperturbative corrections to \eq{ope_beam_soft_tmd} is described in \sec{np_model}.

In practice, we find it important to be able
to evaluate the PDFs in \eq{ope_beam_soft_tmd}
at a scale that is potentially different from
(but still parametrically similar to)
the overall scale of the beam function or TMD PDF.
In this case we have
\begin{align} \label{eq:ope_beam_muf_neq_mu}
\tilde B_i(x, b_T, \mu, \nu/\w)
&= \sum_j \int\! \frac{\df z}{z} \, \tilde \cI_{ij}\Bigl(z, b_T, \mu, \frac{\mu_f}{\mu}, \frac{\nu}{\w}\Bigr) \, f_{j}\Bigl( \frac{x}{z}, \mu_f \Bigr) + \ord{\lqcd^2 b_T^2}
\,,\end{align}
where we analytically evaluate the Mellin convolution
\begin{align}
\tilde \cI_{ij}\Bigl(z, b_T, \mu, \frac{\mu_f}{\mu}, \frac{\nu}{\w}\Bigr)
\equiv \sum_k \int \! \frac{\df z'}{z'} \, \tilde \cI_{ik}\Bigl(\frac{z}{z'}, b_T, \mu, \frac{\nu}{\w}\Bigr) \,
U_{kj}\Bigl(z', \mu, \frac{\mu_f}{\mu} \Bigr)
\,,\end{align}
consistently reexpanding and truncating in $\as(\mu)$,
and $U_{kj}(z', \mu, \mu_f/\mu)$ is the DGLAP evolution operator
that evolves the PDF from $\mu_f$ to $\mu$,
likewise expanded in terms of powers of $\as(\mu)$.

\subsection{Renormalization group evolution}
\label{sec:rges}

For observables like $q_T$ that are sensitive to soft and collinear radiation,
higher-order perturbative corrections induce double-logarithmic terms $\as^n\ln^{m} q_T/Q,\,\, m \!\leq\! 2n$
in the hadronic structure functions $W_{i\,VV^\prime}$.
In the singular limit, $q_T\to0$, these logarithms
grow arbitrarily large and invalidate its perturbative convergence.
The factorization theorem in \eq{tmd_factorization_bbs},
effectively splits these logarithms according to their
dominant contribution from the respective physical modes,
resulting in single-scale functions.
By solving the renormalization group equations (RGEs),
each function is evaluated at its intrinsic (canonical) $\mu$ or $\nu$ scale
and evolved to a common overall $\mu$ and $\nu$,
resumming the logarithically enhanced terms to all orders in perturbation theory.
A standard way to obtain the solutions of the RGEs
is to make use of the $b_T$-space form of the cross section
in \eq{tmd_factorization_bbs}, where the RGEs assume a simple multiplicative form
and the boundary scales that eliminate all large logarithms
have the canonical scaling
\begin{align} \label{eq:canonical_scales_bbs}
\mu_B^\can = \mu_S^\can = \nu_S^\can = \mu_0^\can = \mu_f^\can = \frac{b_0}{b_T}
\,, \qquad
\nu_B^\can = \mu_H^\can = Q
\,,\end{align}
where $\mu_0$ is the scale at which the boundary condition
of the resummed rapidity anomalous dimension or Collins-Soper kernel is evaluated (see below),
and the factor of $b_0 = 2 e^{-\gamma_E} \approx 1.12292$ is conventional.
In practice, this amounts to setting the scales in position space, performing
the evolution, and finally transforming back to (momentum) $q_T$-space
by numerically integrating over $b_T$.
An exactly equivalent way of obtaining the same resummed result
is by setting the boundary scales for the TMD PDFs in \eq{tmd_factorization_ff} as
\begin{align} \label{eq:canonical_scales_ff}
\mu_0^\can = \mu_\init^\can = \mu_f^\can = \sqrt{\zeta_\init^\can} = \frac{b_0}{b_T}
\,,\end{align}
and evolve them to the overall $\mu_\mathrm{final} = \sqrt{\zeta_\mathrm{final}} = Q$.
This approach bypasses unphysical singularities~\cite{Frixione:1998dw}
that arise when naively setting scales or counting logarithms in momentum space,
although approaches to momentum-space resummation exist~\cite{Monni:2016ktx, Ebert:2016gcn};
the formal momentum-space RG solutions of \refcite{Ebert:2016gcn}
are equivalent to the $b_T$-space approach up to differences
in the fixed-order boundary conditions.
Recently, similar unphysical singularities were pointed out for the first time in the resummation
of an observable not involving transverse momentum~\cite{Bhattacharya:2022dtm}
(the heavy-jet mass distribution near the Sudakov shoulder),
and a successful resolution using position-space scale setting
was demonstrated in \refcite{Bhattacharya:2023qet}.

To implement the resummation of rapidity logarithms
in the individual beam and soft functions in \eq{tmd_factorization_bbs},
we employ the exponential regulator~\cite{Li:2016axz}
together with the rapidity renormalization group framework~\cite{Chiu:2012ir}.
In $b_T$ space, the quark beam and soft functions satisfy the virtuality RGEs,
\begin{align}\label{eq:mu_RGEs}
 \mu\frac{\df}{\df\mu} \ln\tB_q(x,b_T,\mu,\nu/\omega)
 &= \tilde\gamma_B^q(\mu,\nu/\omega)
 = 2 \Gamma_\cusp^q[\as(\mu)] \ln\frac{\nu}{\omega} + \tilde\gamma_B^q[\as(\mu)]
\,,\nn\\
 \mu\frac{\df}{\df\mu} \ln\tS_q(b_T,\mu,\nu)
 &= \tilde\gamma_S^q(\mu,\nu)
 = 4 \Gamma_\cusp^q[\as(\mu)] \ln\frac{\mu}{\nu} + \tilde\gamma_S^q[\as(\mu)]
\,.\end{align}
Here the cusp anomalous dimension $\Gamma_\cusp^q[\as(\mu)]$ is in the fundamental
representation. Its coefficients are known analytically to four loops~\cite{Korchemsky:1987wg, Moch:2004pa, Vogt:2004mw, Lee:2016ixa,Moch:2017uml,Lee:2019zop,Henn:2019rmi,Bruser:2019auj, Henn:2019swt, vonManteuffel:2020vjv}
(see \refcite{Bruser:2019auj} for a complete list of earlier references), which are necessary for the N$^3$LL and N$^3$LL$^\prime$ Sudakov evolution together with the four-loop QCD beta function~\cite{Tarasov:1980au, Larin:1993tp, vanRitbergen:1997va, Czakon:2004bu}.
At N$^4$LL, we in addition require the five-loop beta~\cite{Herzog:2017ohr} and
cusp coefficient, where the latter is currently known only approximately~\cite{Herzog:2018kwj}.
The soft noncusp anomalous dimension $\tilde\gamma_S^q[\as(\mu)]$
in the exponential regulator scheme is directly related to the soft threshold noncusp anomalous dimension,
which is known to four loops~\cite{Davies:2016jie, Moch:2017uml, Moch:2018wjh, Das:2019btv, Das:2020adl}.
The beam noncusp anomalous dimension $\tilde\gamma_B^q[\as(\mu)]$ then follows
from consistency, since the \MSbar hard anomalous dimension
is likewise known to four loops~\cite{vonManteuffel:2020vjv, Agarwal:2021zft}.
In turn, these noncusp anomalous dimensions are also related through consistency relations~\cite{Billis:2019vxg}
to those of the inclusive jet function (or the $0$-jettiness beam function) and the thrust and $0$-jettiness soft functions,
which have been independently obtained by direct calculations up to three
loops~\cite{Bruser:2018rad, Ebert:2020unb}.

The rapidity RGEs that the soft and beam functions satisfy are given by
\begin{align} \label{eq:nu_RGEs}
 -2\nu\frac{\df}{\df\nu} \ln\tB_q(x,b_T,\mu,\nu/\omega)
 =\nu\frac{\df}{\df\nu} \ln\tS_q(b_T,\mu,\nu)
 = \tilde\gamma_\nu^q(b_T,\mu)
\,,\end{align}
where the $\mu$ and $\nu$ anomalous dimensions satisfy an all-order integrability condition,
\begin{align} \label{eq:gamma_nu_RGE}
\mu \frac{\df}{\df \mu} \tilde\gamma^q_\nu(b_T, \mu) = - 4\Gamma^q_\cusp[\as(\mu)] = \nu \frac{\df}{\df \nu} \tilde\gamma_S^q(\mu, \nu)
\,.\end{align}
\Eq{gamma_nu_RGE} predicts the all-order logarithmic ($b_T$) structure of the rapidity
anomalous dimension in perturbation theory,
\begin{align} \label{eq:resummed_rapidity_anom_dim}
\tgamma^q_{\nu}(b_T, \mu)
&= -4\eta^q_\Gamma(\mu_0, \mu) + \tgamma^q_{\nu}(b_T, \mu_0) + \ord{b_T^2}
\,, \nn \\
\eta_\Gamma^{q}(\mu_0,\mu)
&\equiv \int_{\mu_0}^\mu \! \frac{\df \mu'}{\mu'} \, \Gamma^{q}_{\rm cusp}[\as(\mu')]
\,,\end{align}
up to the fixed-order boundary term $\tgamma^q_{\nu}(b_T, \mu_0)$ that necessitates
an explicit calculation and is currently known to four
loops~\cite{Luebbert:2016itl, Li:2016ctv, Vladimirov:2016dll, Duhr:2022yyp, Moult:2022xzt}.
The four-loop result
has been obtained in \refscite{Duhr:2022yyp, Moult:2022xzt}
using an expansion around the critical number of dimensions where QCD is conformal~\cite{Vladimirov:2016dll},
such that the four-loop rapidity anomalous dimension can
be evaluated in terms of the four-loop soft threshold noncusp anomalous dimension~\cite{Davies:2016jie, Moch:2017uml, Moch:2018wjh, Das:2019btv, Das:2020adl}
and a correction term that involves explicit calculations of the three-loop threshold soft function
to subleading order in the dimensional regulator~\cite{Duhr:2022yyp, Moult:2022xzt, Duhr:2022cob}.
In addition, the above expression, which holds for $\mu \gtrsim \mu_0 \sim 1/b_T$,
receives nonperturbative power corrections of $\ord{\lqcd^2 b_T^2}$
to which we return in \sec{np_model}.

Solving the coupled system in \eqs{mu_RGEs}{nu_RGEs} yields the resummed beam and soft functions,
\begin{align} \label{eq:RGevolution}
 \tB_q\Bigl(x,b_T,\mu,\frac{\nu}{\omega}\Bigr) &
 = \tB_q\Bigl(x,b_T,\mu_B,\frac{\nu_B}{\omega}\Bigr)
   \exp\biggl[-\frac12 \ln\frac{\nu}{\nu_B} \tgamma^q_\nu(b_T,\mu_B) \biggr]
   \exp\biggl[\int_{\mu_B}^\mu \frac{\df\mu'}{\mu'} \tgamma_B^q(\mu',\nu/\omega) \biggr]
\,,\nn\\
 \tS_q(b_T,\mu,\nu) &
 = \tS_q(b_T,\mu_S,\nu_S)
   \exp\biggl[\ln\frac{\nu}{\nu_S} \tgamma^q_\nu(b_T,\mu_S) \biggr]
   \exp\biggl[\int_{\mu_S}^\mu \frac{\df\mu'}{\mu'} \tgamma_S^q(\mu',\nu) \biggr]
\,,\end{align}
where the exponentials correspond to the Sudakov evolution kernels via which
resummation is achieved. Crucially, choosing appropriately the boundary scales $\mu_{H,B,S}$
and $\nu_{B,S}$, the hard, beam, and soft functions are free of large (double) logarithms
which results in their well-behaved perturbative convergence, and allows for their evaluation
at fixed-order. The beam and soft functions up to N$^3$LO have been obtained
in \refscite{Li:2016ctv, Lubbert:2016rku, Ebert:2017uel, Ebert:2020yqt, Luo:2020epw}
and are collected in our notation in \refcite{Billis:2019vxg}.
The flavor nonsinglet contribution to the massless quark (vector or axial) form factor,
which determines the hard function in \eq{tmd_factorization_m1_4},
is likewise known to N$^3$LO~\cite{Baikov:2009bg, Lee:2010cga, Gehrmann:2010ue}.
Explicit expressions for the hard functions and nonsinglet matching coefficients
in our notation can be found in \refcite{Ebert:2017uel, Ebert:2020dfc}.

In \eq{RGevolution} we choose to perform the evolution first in $\nu$ and then in $\mu$.
We stress that any other resummation path in the two-dimensional $(\mu, \nu)$ plane is equivalent
as a result of the RG consistency relations that the anomalous dimensions satisfy.

It has been shown that the evaluation of the Sudakov evolution kernels in
\eqs{resummed_rapidity_anom_dim}{RGevolution} based on approximate analytical methods
can lead to a numerical discrepancy that can no longer be considered
a higher-order effect~\cite{Billis:2019evv},
see also \refscite{Bertone:2022sso, Bertone:2024snr},
and in addition upsets the so-called closure condition,
i.e., the group property of the renormalization group evolution,
and the exact path independence in the $(\mu, \nu)$ plane.
Recently, in \refcite{Ebert:2021aoo} an exact solution for the evolution kernels and
for the running of the strong coupling was derived
by recasting the original integrands in a form
amenable to partial fractioning and the residue theorem.
In our N$^3$LL$'$ predictions we use these exact analytic solutions
for the $\beta$ function and Sudakov kernels,
which implies that all RGE running
is exact and free of any of the usually employed approximations.%
\footnote{An exception are the predictions
in \figss{Z_an3lopdf_ratio_msht20}{Z_an3lopdf_ratio_nnpdf40}{Z_an3lopdf_np_unc}.
There, we consistently employ
the approximate unexpanded analytic evolution kernels
(or ``iterative'' solutions)~\cite{Billis:2019evv}
extended to this order for simplicity,
see \app{rge_solutions_n4ll},
although in principle, the method for obtaining their exact solution~\cite{Ebert:2021aoo}
should also be applicable at N$^4$LL.
In this case we also consistently use the iterative solutions
at lower orders to ensure that comparisons to lower orders are one to one.}

\subsection{Flavor-singlet hard matching coefficients}
\label{sec:singlet_coeffs}

SCET Wilson coefficients are the building blocks of the hard function $H_{i\,VV^\prime\, ab}$.
They are obtained by matching the full-QCD quark form factors
for the underlying hard process $ab \to V,V^\prime$ onto SCET.
Specifically, for the neutral current Drell-Yan process ($V, V^\prime = Z/\gamma^\ast$),
the QCD corrections to the form factors are categorized differently
depending on whether $a, b$ directly couple to $V, V^\prime$
or if a quark loop of possibly different flavor couples to $V, V^\prime$.
The former contribution is referred to as the (flavor) nonsinglet matching coefficient,
and the latter as the singlet matching coefficient.
While the nonsinglet coefficient is the same for axial and vector terms
due to chirality and the fact that the external quarks are massless at the hard scale,
the axial and vector singlet coefficients differ from each other.

Using the notation of \refcite{Ebert:2020dfc}, the axial singlet matching coefficient
summed over all quark flavors in the Standard Model running in the loop is given by
\begin{align}
\sum_f a_f \, C_{a\, f}(q^2, m_f^2, \mu)
= a_t \, \Delta C_{a\,t,b}(q^2, m_t^2, \mu) + \ORd{\frac{m_b^2}{q^2}}
\,,\end{align}
where the contributions from the other (approximately massless) generations cancel exactly.
The two-loop coefficient $\Delta C_{a\,t,b}^\two(q^2, m_t^2)$
is well known~\cite{Dicus:1985wx, Kniehl:1989qu},
including its exact dependence on $m_t$~\cite{Bernreuther:2005rw, Chen:2021rft}.
We extract the three-loop expression for $\Delta C_{a\,t,b}^\three(q^2, m_t^2, \mu)$
from the recent calculation of all three-loop singlet contributions
to the quark form factors in \refscite{Gehrmann:2021ahy, Chen:2021rft}.
Our notation relates to that of \refcite{Chen:2021rft} as $\Delta C_{a\,t,b} = \cF^{\prime\,A}_{s,t} - \cF^{\prime\,A}_{s,b}$,
where $\cF^{\prime\,A}_{s,q}$ is the axial singlet contribution
to the quark form factor computed in pure dimensional regularization and with infrared poles subtracted in \MSbar.
We account for the one-loop decoupling relation between $\as^{(6)}(\mu)$ and $\as^{(5)}(\mu)$
to extract the individual coefficients $\Delta C_{a\,t,b}^{(2,3)}$
in a truncated expansion in the five-flavor running coupling $\as^{(5)}(\mu)$.
We have checked that the result satisfies the renormalization group running
expected from consistency with an effective quark-antiquark operator in SCET with $n_f = 5$ massless quark flavors.
We stress the necessity of using the three-loop massless axial vector coefficient~\cite{Gehrmann:2021ahy}
together with the corresponding massive piece~\cite{Chen:2021rft}
to ensure that their $\mu$ dependence properly cancels.
The vector singlet coefficient starts at $\ord{\as^3}$ and receives contributions
from all massless quark flavors, which can readily be extracted
from the term multiplying the $N_{F,V}$ coefficient in \refcite{Gehrmann:2010ue}.
In addition, there are contributions from closed top-quark loops
at $\ord{\as^3 Q^2/(4m_t^2)}$, which were likewise calculated in \refcite{Chen:2021rft}.
Here the $m_t$ dependence is purely power suppressed
and not logarithmically enhanced
since it results from the matching of the conserved vector current.
However, based on the observed, negligible numerical effect
of the corresponding $\ord{Q^2/(4m_t^2)}$ corrections
in the axial coefficient (discussed below),
we choose to ignore the latter contribution for simplicity
and work to leading power in $Q/(2m_t)$.

\begin{figure*}
\includegraphics[width=\WidthTwoSubfigs]{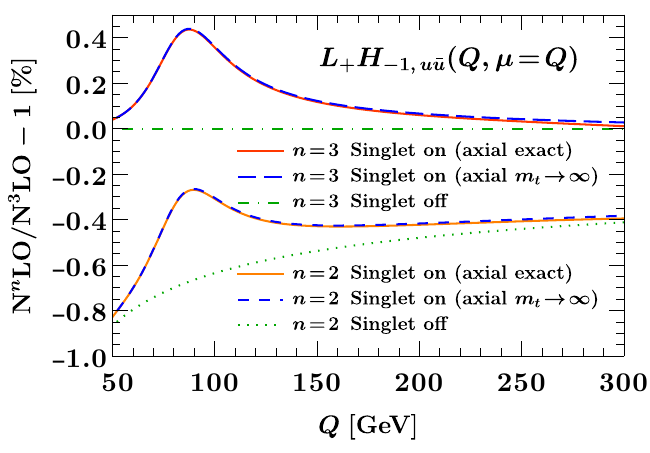}%
\hfill%
\includegraphics[width=\WidthTwoSubfigs]{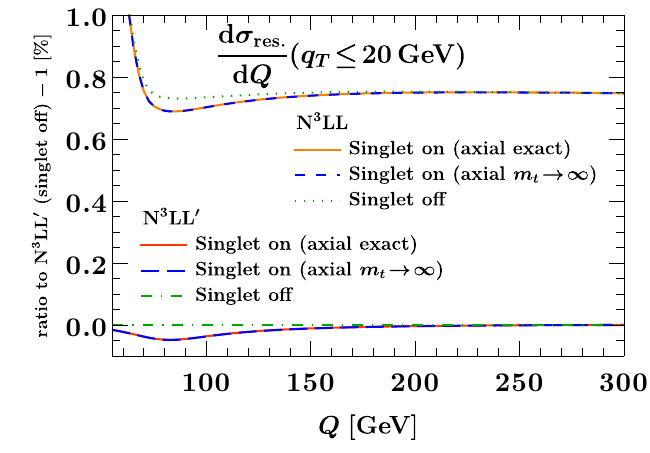}%
\caption{Relative impact of the singlet contributions to the combined
inclusive hard function and leptonic coefficient
$L_+ H_{-1,u \bar{u}} \equiv \sum_{VV'} L_{+\,VV'} H_{i\,VV'\,u\bar{u}}$
(left) and the resummed cross section (right) for neutral-current Drell-Yan
as a function of the dilepton invariant mass $Q$.
}
\label{fig:axial_terms_impact}
\end{figure*}

In the left panel of \fig{axial_terms_impact}
we illustrate the numerical impact of the singlet contributions
to the hard function for the $i = -1$ (unpolarized) helicity channel
in the $u\bar{u}$ flavor channel at NNLO and at N$^3$LO.
Excluding the vector and axial vector singlet terms at both orders results
in a $\sim 0.5\%$ deviation from the exact result,
especially in the region $Q \sim m_Z$
where the relative impact of the latter terms
is enhanced due to the $Z$ resonance.
On the other hand,
it is evident that the power corrections to the axial coefficient
between the exact and the LP $Q \ll m_t$ result are negligible
in the range of $Q$ of interest.
This justifies using the LP expressions for both the axial and vector singlet contributions.
In the right panel of \fig{axial_terms_impact}
we perform the same comparison at the level of the resummed cross section
at NNLL$'$ and N$^3$LL$'$.
The contributions of the axial singlet coefficient
to the hard function for up-type and down-type quarks
have opposite sign because the axial singlet coefficient
is interfered with the nonsinglet one weighted with the respective opposite weak charges.
This leads to a large degree of cancellation at the level of the cross section
that leaves behind a relative effect of $0.1 \%$.
Despite the small size we nevertheless include the
full set of singlet coefficients (to leading power in $Q \ll 2m_t$)
in our nominal predictions,
since (a)~they are a part of the nominal NNLL$'$ and N$^3$LL$'$ accuracy,
respectively, and (b)~the degree of cancellation between flavor channels
is contingent on the underlying collinear PDF set,
and also in general different for the $i = 4$ helicity cross section,
i.e., the forward-backward asymmetry.

We stress that keeping these virtual terms
to improve the resummed singular cross section is not inconsistent
with dropping the corresponding real contributions
in fixed-order predictions used to extract
the nonsingular cross section
as long as one restricts to a range in $q_T$ where the latter
is still power-suppressed and numerically small (see the next section).
In general, these contributions arise
from heavy-quark pair production or from hard gluon radiation out of heavy-quark loops.
To leading power in $Q, q_T \ll 2m_t$ the only contribution of this kind,
starting at $\ord{\as^3}$,
comes from the three-parton real-emission amplitudes involving
the effective operator from integrating out the axial contribution of the top quark.

\subsection{Nonsingular cross section}
\label{sec:nons_cross_section}

\begin{figure*}
\centering%
\includegraphics[width=\WidthTwoSubfigs]{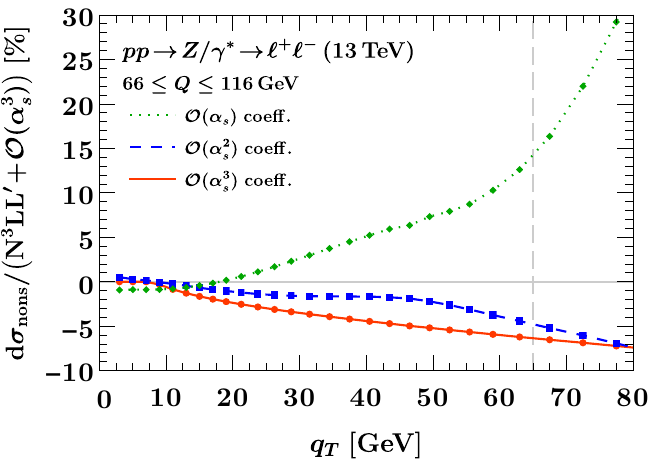}%
\hfill%
\includegraphics[width=\WidthTwoSubfigs]{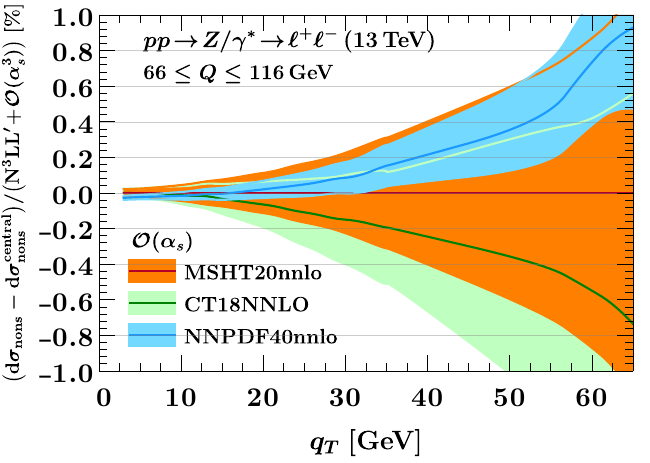}%
\caption{
Left:
Relative contributions
from the nonsingular cross section coefficients
at different fixed orders
to the total fiducial $q_T$ spectrum
at N$^3$LL$^\prime+\ord{\as^3}$.
We use the ATLAS $13 \TeV$ fiducial cuts for definiteness, see \eq{fiducial_cuts}.
Right:
Impact of parametric PDF variations and alternate PDF sets
on the $\ord{\as}$ nonsingular cross section
with the central \texttt{MSHT20nnlo} result as baseline,
likewise normalized
to the total fiducial $q_T$ spectrum
at N$^3$LL$^\prime+\ord{\as^3}$.
The PDF variations around the solid red line on the right,
which give an upper estimate of the corresponding effect
on the nonsingular cross section coefficient at $\ord{\as^2}$ and $\ord{\as^3}$,
can be thought of as dressing the lines in the plot on the left
in the same units (but note the difference in vertical scale).
}
\label{fig:nnlojet_nons_impact}
\end{figure*}

While the singular cross section in \eq{tmd_factorization_fid_spectrum}
dominates the spectrum as $q_T\to0$ and captures all singularities
$\delta(q_T)$ and $\as^n \ln^m(q_T/Q)/q_T$,
it corresponds only to the leading term in the
expansion of the full cross section in this limit.
To include the remaining tower of power-suppressed contributions
of $\ord{q_T^2/Q^2}$ and higher,
usually referred to as the nonsingular cross section,
and extend the prediction into the fixed-order (FO) region,
we perform an additive matching.
In practice, the nonsingular cross section (or ``the nonsingular'', for short)
is obtained by a differential $q_T$ subtraction~\cite{Catani:2007vq, Gaunt:2015pea},
\begin{align} \label{eq:nons_def}
\frac{\df\sigma_\nons}{\df q_T}
= \Bigl[ \frac{\df \sigma_{\FO}}{\df q_T} - \frac{\df\sigma_\sing}{\df q_T} \Bigr]_{q_T>0}
\,,\end{align}
where $\df\sigma_\FO/\df q_T$ is the FO cross section for the $Z/W^\pm + 1$
parton configuration, and $\df\sigma_\sing/\df q_T$ is
the singular cross section at the appropriate perturbative order.
Both are, in practice, calculated using strict twist-2 collinear factorization,
i.e., in terms of standard leading-twist collinear PDFs.
Since the outcome of the subtraction in \eq{nons_def} is suppressed
by $q_T^2/Q^2$,
i.e. numerically significantly smaller in the resummation region $q_T \ll Q$,
it typically is sufficient to evaluate it at a fixed scale $\mu \sim \mu_\FO \sim Q$
and ignore the shape effect due to its (unknown) all-order resummation.
We will further restrict most of our predictions
in this paper to a range of $q_T \leq 60-65 \GeV$,
on which we empirically find that the nonsingular cross section
-- extracted as described in the next paragraphs --
only amounts to a few-percent contribution to the total $q_T$ spectrum,
see the left panel of \fig{nnlojet_nons_impact}.
This has the substantial benefit that the nonsingular cross section,
e.g.\ when evaluating parametric PDF uncertainties,
may simply be kept at the central member of the PDF set in question
(or may even use another, fixed PDF set),
since propagating these changes into the nonsingular amounts
only to a few-percent change on top of a few-percent contribution.
This substantially reduces the numerical effort involved
in obtaining the nonsingular cross section.
To make this approximation quantitative,
the right panel of \fig{nnlojet_nons_impact}
shows the impact of PDF variations and alternate PDF sets
on the $\ord{\as}$ nonsingular cross section,
where we normalize the difference to the central
$\ord{\as}$ \texttt{MSHT20nnlo} nonsingular cross section
to our best prediction for the total $q_T$ spectrum.
The impact of these variations,
which are computed as described in \sec{uncerts_as_pdf},
amounts to less than two permille for $q_T \leq 30 \GeV$,
and less than a percent for $q_T \leq 65 \GeV$.
This effect on the leading $\ord{\as}$ nonsingular cross section
(where PDF variations are computationally cheap to perform,
and fully retained in all our predictions later on)
can serve as an upper estimate of the error one makes
when neglecting PDF variations in the smaller contributions
from the $\ord{\as^2}$ and $\ord{\as^3}$ nonsingular coefficients,
and thus fully justifies this approximation,
e.g.\ when comparing to our final estimate of the total
perturbative uncertainty in \fig{Z_qT_fid_predictions},
which reaches $1 \%$ in the
range $5 \leq q_T \leq 30 \GeV$.
Several additional approximations to the nonsingular cross section,
which we describe in the following,
can be performed by a similar argument
without affecting the overall accuracy.
We note that these manipulations are not justified
when extending the prediction into the far tail
where the complete FO prediction must be recovered,
and in particular parametric variations
must consistently be propagated everywhere.
(In \app{results_Z_nnlojet_13tev_full_range} we present such predictions
valid for all $q_T$ for a fixed reference PDF set.)

At $\ord{\as}$ we obtain the full cross section for $q_T > 0$,
including arbitrary cuts or measurements on the leptonic decay products,
from an in-house implementation of the analytic LO$_1$ structure
functions~\cite{Cleymans:1978ip, Chaichian:1981va, Mirkes:1992hu}.
The $\cO(\as^2)$ coefficient is obtained from dedicated, high-precision runs
of the publicly available Monte-Carlo program \texttt{MCFM}~\cite{Campbell:2015qma, Boughezal:2016wmq}.
For the evaluation of both coefficients we employ the \texttt{MSHT20nnlo} PDF set~\cite{Bailey:2020ooq},
which we will also use as our default PDF set in final predictions.

For $Z+\mathrm{jet}$ production with the ATLAS $13\TeV$ cuts at $\ord{\as^3}$,
we use publicly available, high-precision
FO data from the \texttt{NNLOjet} collaboration~\cite{Chen:2022cgv}.
In these, we observe that the NNLO$_1$ $K$-factor,
$K = \df\sigma_{\text{NNLO$_1$}}/ \df\sigma_{\text{NLO$_1$}}$,
is quite small for $q_T \gtrsim 20\GeV$
and follows an almost constant trend with $K \approx 0.08$.
We furthermore find that the model function
\begin{align} \label{eq:nons_model_func_def}
f_\nons(q_T) = \Theta(q_T - q_T^*) \biggl(
   c_1 \ln\frac{q_T}{q_T^\ast} + c_2 \ln^2\frac{q_T}{q_T^\ast}
\biggr)
\,,\end{align}
captures reasonably well the $q_T$ dependence of the ratio
\begin{align} \label{eq:ratio_nons_to_nnlo}
\frac{\df\sigma_\nons^{(3)}/\df q_T}{\df\sigma_{\text{NLO$_1$}}/\df q_T}
\,,\end{align}
which we use to perform a fit of the coefficients $c_{1, 2}$.
By multiplying the fitted $f_\nons$
with $\df\sigma_{\text{NLO$_1$}}/\df q_T$,
we obtain an approximate $\ord{\as^3}$ nonsingular coefficient
that in particular is free of the bin-to-bin statistical fluctuations
visible in the raw nonsingular result.
Note that we use \texttt{NNLOjet} data~\cite{Chen:2022cgv}
that employ the \texttt{NNPDF40nnlo} set~\cite{NNPDF:2021njg}
for the estimation of the model function,
which we then multiply with $\df\sigma_{\mathrm{NNLO}}/\df q_T$
evaluated using the \texttt{MSHT20nnlo} PDF set
to obtain an approximate nonsingular for the latter (default) set.

The specific choice for the parametrization of $f_\nons(q_T)$
is motivated by the expected
functional form of the nonsingular cross section.
Specifically, the $q_T$ spectrum receives Sudakov double logarithms
at each order in $\as$ at any order in the power expansion.
At N$^3$LO, there are at most two extra powers of the logarithm
when dividing out the NNLO full cross section,
which we expect to dominate the shape of the ratio.
Another notable feature of the model function in \eq{nons_model_func_def}
is that we set the $\ord{\as^3}$ nonsingular coefficient to zero at
$q_T \leq q_T^\ast$ to avoid biasing the shape of the spectrum
through the unresummed logarithms of $q_T/Q$ in the nonsingular
at very small values of $q_T$.
A priori, such a modification is not compatible
with maintaining (even approximate) N$^3$LO accuracy
on the normalization of the spectrum,
since it effectively amounts to performing a $q_T$ subtraction
with an unreasonably high technical cutoff,
which in turn is known to substantially bias the total integral.
In the present case, however, we rely
on a remarkable feature of the available fixed-order data~\cite{Chen:2022cgv},
which indicate that the $\ord{\as^3}$ nonsingular coefficient
undergoes a sign change around $q_T = 2 \GeV$
such that its integral up to $q_T^\ast = 8 \GeV$ is in fact compatible with zero
within the quoted numerical uncertainty.
Specifically, we find that the integral of the $\ord{\as^3}$
nonsingular coefficient between the actual technical cutoff of \refcite{Chen:2022cgv}
at $q_T^\mathrm{cut} = 0.447 \GeV$ and $q_T^\ast = 8 \GeV$
amounts to only
\begin{align}
\int_{0}^{q_T^\ast} \! \df q_T \, \frac{\df \sigma^{(3)}_\nons}{\df q_T}
\approx \int_{q_T^\cut}^{q_T^\ast} \! \df q_T \, \frac{\df \sigma^{(3)}_\nons}{\df q_T}
= (-0.34 \pm 0.76) \pb
\,,\end{align}
which should be compared to the integral of the prediction
at N$^3$LL$'$$+$$\ord{\as^3}$ over the reference range $q_T \in [0, 65] \GeV$,
for which we find $673.4 \pb$, see \tab{norm_factors_fid_pTZ}.
Conveniently, this choice of $q_T^\ast$ therefore lets us preserve
both an unbiased shape of the matched spectrum at small $q_T < q_T^\ast$
as well as the approximate N$^3$LO accuracy of its normalization
up to $q_T \approx 70 \GeV$,
i.e., within the approximations in the choice of model function,
and accounting for the fact
that we use $f_\nons$ to port the nonsingular from one PDF set to another.

\begin{figure*}
\centering%
\includegraphics[width=\WidthTwoSubfigs]{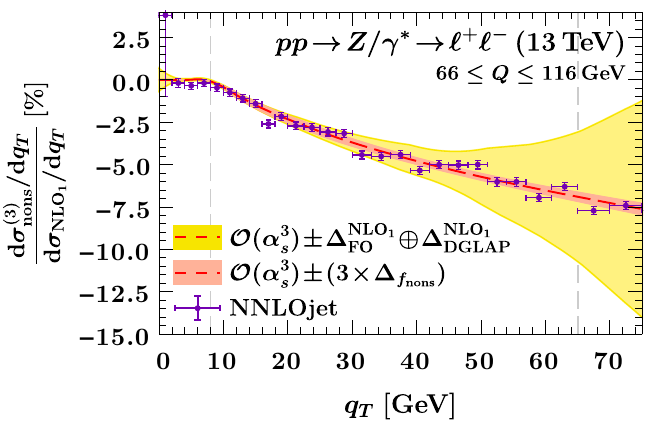}%
\hfill%
\includegraphics[width=\WidthTwoSubfigs]{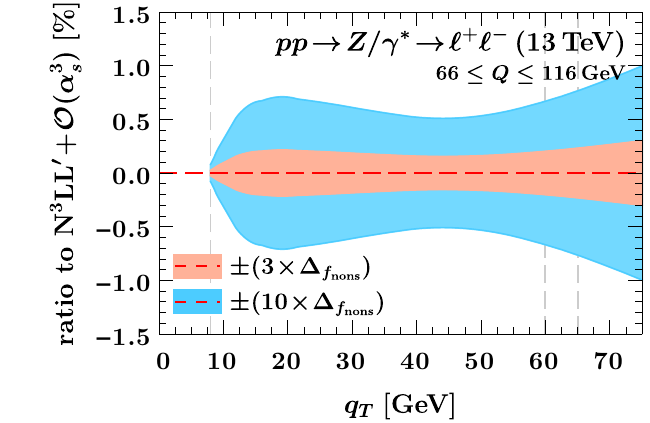}%
\\%
\includegraphics[width=\WidthTwoSubfigs]{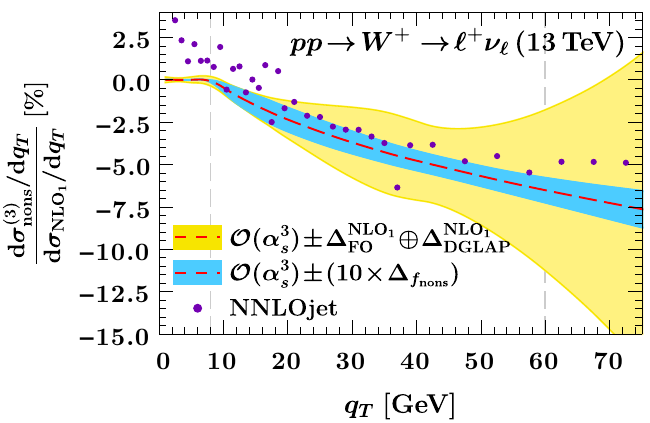}%
\hfill%
\includegraphics[width=\WidthTwoSubfigs]{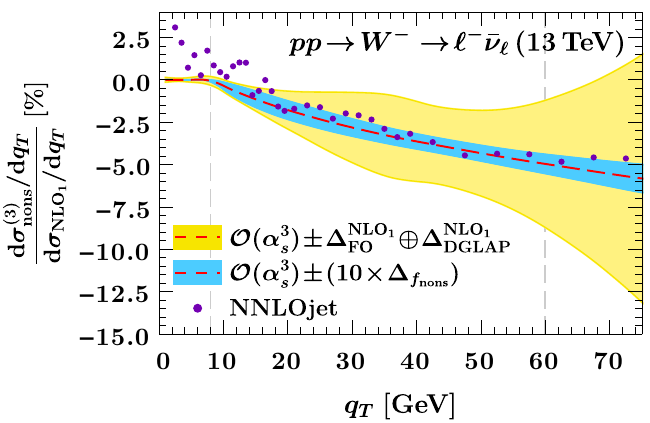}%
\caption{
Comparison of the fitted approximate $\ord{\as^3}$ nonsingular coefficient
against \texttt{NNLOjet} data~\cite{Chen:2022cgv}
for the neutral (top left) and the charged-current (bottom) Drell-Yan process.
We normalize to the total $\ord{\as^2}$ (NLO$_1$) fixed-order $q_T$ spectrum,
as also done in the fit functional form in \eq{ratio_nons_to_nnlo}.
The red and blue bands indicate the error
due to our approximate treatment of the $\ord{\as^3}$ nonsingular coefficient,
which is estimated by conservatively scaling up
the error obtained from fitting to the nonsingular model function for the $Z$ case,
as described in the text.
As a point of reference, the yellow bands indicate the total perturbative uncertainty
on the NLO$_1$ nonsingular cross section, see \sec{uncerts_pert}.
The top right panel shows the same estimated approximation error
as the red and blue bands in the other plots, respectively,
but normalized to show the relative impact of the error
on the final matched prediction at N$^3$LL$^\prime+\ord{\as^3}$.
Vertical dashed gray lines indicate $q_T^* = 8 \GeV$
and the upper edge $65 \GeV$ ($60 \GeV$) of the range
on which we provide predictions for the $Z$ (the $W^\pm$).
}
\label{fig:as3_mocked_nons}
\end{figure*}

For the extraction of $c_{1, 2}$ in \eq{nons_model_func_def}
we perform a two-parameter fit at fixed $q_T^\ast = 8 \GeV$
to the data for the $\ord{\as^3}$ nonsingular coefficient at $q_T > q_T^\ast$.
While the lower end of the fit window in $q_T$ is thus fixed,
it is still important to assess the impact of the choice of upper endpoint
in the fitting procedure.
This is because the FO data, and hence also the corresponding nonsingular,
are more precise for larger values of $q_T$
since they carry smaller relative statistical uncertainties.
On the other hand, increasing the range to $q_T$ values
that are too large carries the risk of overfitting and exceeding
the physical validity range of the model.
To address this, we perform multiple fits of $c_{1, 2}$ by gradually extending
the $q_T$ fit window while inspecting the values of the $R^2$ statistic,
which provides a convenient goodness-of-fit measure.
We find that the optimal fitting range is $8 \GeV\leq q_T \leq 77.5\GeV$,
resulting in
\begin{align}
c_1 = -0.02097 \pm 0.00149
\,, \quad
c_2 = -0.005737 \pm 0.000804
\,, \quad
R^2 = 0.994
\,.\end{align}
The best-fit model function and the original data are shown in the top left panel of \fig{as3_mocked_nons},
where we also indicate the perturbative uncertainty
on the nonsingular cross section at NLO$_1$
(estimated as described in \sec{uncerts_pert}) for comparison.

We stress that the fit we perform here
does not meet the level of rigor found in state-of-the-art nonsingular fits
that can actually \emph{guarantee} a given fixed-order accuracy
on the total cross section in the context of $q_T$ or $\Tau_N$ subtractions.
These fits in particular allow one to thoroughly
assess the uncertainty from extrapolating the nonsingular down to $q_T \to 0$,
see \refscite{Moult:2016fqy, Moult:2017jsg, Billis:2021ecs, Chen:2022cgv},
but necessarily involve a fit function
whose functional form contains \emph{all} logarithmic terms present
in the $\ord{q_T^2/Q^2}$ and $\ord{q_T^4/Q^4}$ terms of the nonsingular at a given order,
possibly using analytic knowledge of the highest logarithms
to reduce the number of free parameters.
Instead, our goal here is to obtain
an approximate form of the nonsingular
that is convenient to use,
appropriately captures the physical few-percent effect for $q_T = 20$-$60 \GeV$,
and smoothens out statistical fluctuations at the level of the finely-binned $q_T$ spectrum,
while still roughly maintaining the fixed-order accuracy of the total integral.

Ready-to-use fixed-order data at $\ord{\as^3}$ of similar quality
are not yet available for $W^\pm$ production,
inclusive $Z$ production,
or fiducial $Z$ production with CMS $13\TeV$ cuts.%
\footnote{We note that a public implementation
of $Z+\mathrm{jet}$ production at $\ord{\as^3}$ has
in the meantime become available~\cite{Neumann:2022lft},
but requires substantial computing resources.}
With the same goal for the nonsingular in mind as above,
and in line with the overall focus of this paper
on the state-of-the-art resummed physics,
we continue to make use of our model function in \eq{nons_model_func_def}
to obtain an approximate $\ord{\as^3}$ nonsingular cross sections also for these processes.
Specifically, in these cases we apply $f_\nons(q_T)$,
which we take to be universal,
to the NLO$_1$ full cross section for the respective process
obtained from our in-house
implementation at $\ord{\as}$ and from \texttt{MCFM} at $\ord{\as^2}$.
For $W^\pm$ production, we are able to qualitatively compare
our approximate $\ord{\as^3}$ nonsingular coefficient
to the data provided in the ancillary files of \refcite{Bizon:2019zgf} as a cross check,
using the fiducial cuts given in that reference.%
\footnote{
We did not attempt a dedicated fit to these data for $W^\pm$ production
because no statistical uncertainties were provided with \refcite{Bizon:2019zgf}.
Note that to produce the bottom two panels of \fig{as3_mocked_nons},
we also extracted the overall normalization of the full fixed-order data in \refcite{Bizon:2019zgf}
by adjusting it to recover the expected power suppression of the nonsingular as $q_T \to 0$.
}
The comparison is shown in the bottom two panels \fig{as3_mocked_nons}.
We observe reasonable agreement on the shape, sign, and magnitude
of the nonsingular cross section, in particular at $q_T > 20 \GeV$,
giving us confidence that the dominant physics
of the nonsingular cross section for $V+\mathrm{jet}$ production
are indeed captured by \eq{ratio_nons_to_nnlo}.

To conclude this section, we estimate the error
associated with this approximate treatment
of the $\ord{\as^3}$ nonsingular cross section coefficient,
specifically the error
affecting the shape of the spectrum at $q_T \geq q_T^*$,
and propagate it into our final predictions.
To do so, we propagate the fit uncertainty on the parameters $c_1$ and $c_2$
of the nonsingular model function $f_\nons$ into an uncertainty $\Delta_{f_\nons}$
on the $\ord{\as^3}$ nonsingular cross section coefficient.
We then conservatively scale this uncertainty up by a factor of $3$
in the case of the $Z$ boson to reflect the fact
that the functional form of the model only captures
the nonsingular coefficient in a simplified way.
The result is shown as a red band in the top left panel of \fig{as3_mocked_nons}.
In the case of the $W^\pm$, we scale $\Delta_{f_\nons}$ up
by a factor of $10$ to in addition reflect the fact
that it was ported from one (related) process to another.
The result is shown as a blue band in the bottom two panels of \fig{as3_mocked_nons}.
We find that with these scaling factors,
our estimate of the approximation error
covers both the scatter of the nonsingular data and its residual deviations
from our approximate model for the $W^\pm$ to a reasonable degree
on the range of $20 < q_T < 60 \GeV$,
where both the data and the model can be considered reliable.
The top right panel shows the same two bands,
but instead normalized to our final prediction at N$^3$LL$^\prime+\ord{\as^3}$.
It is instructive to already at this point compare
the estimated approximation error
to the final perturbative uncertainty $\Delta_\pert$,
see \figss{Z_qT_incl_predictions}{Z_qT_fid_predictions}{W_qT_fid_predictions}.
Focussing on the range of $10 < q_T < 20 \GeV$,
where $\Delta_\pert$ reaches the level of $1 \%$,
we find that the approximation is fully justified for the $Z$ boson,
with the relative approximation error (red) reaching at most a level of $0.2 \%$,
and indeed mainly serves to smoothen out the nonsingular data.
For the $W^\pm$, the estimated approximation error of $\approx 0.6 \%$ (blue)
does not constitute a bottleneck at the present time,
but dedicated high-statistics Monte-Carlo campaigns for the nonsingular cross section,
or more efficient techniques to evaluate it, would certainly be desirable in the future.

\subsection{Matching and profile scale choices}
\label{sec:profile_scales}

\begin{figure*}
\centering
\includegraphics[width=\WidthTwoSubfigs]{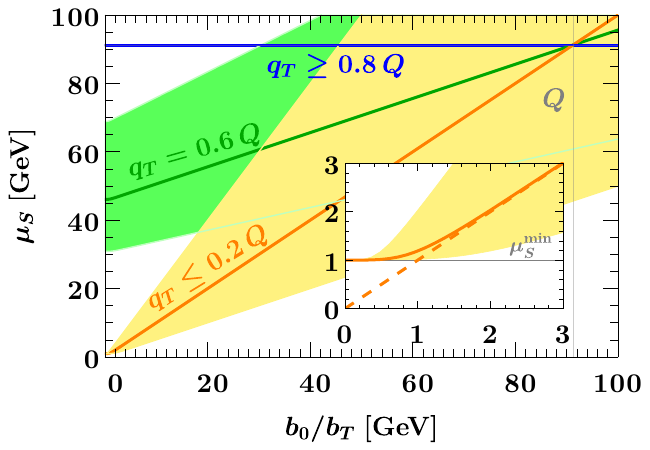}%
\hfill
\includegraphics[width=\WidthTwoSubfigs]{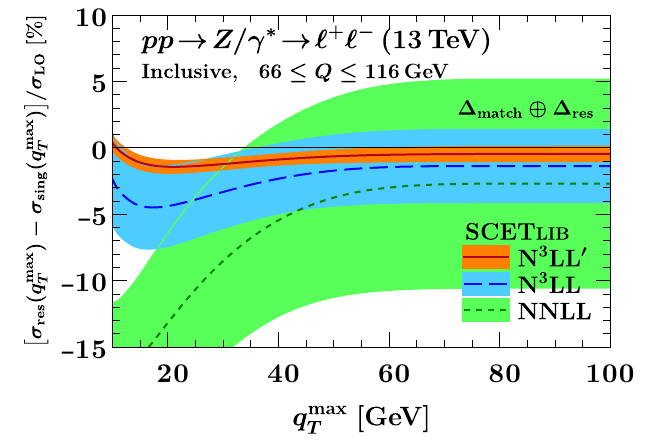}%
\caption{
Left: Illustration of the hybrid profile scales
and the Landau pole treatment,
using the soft virtuality scale $\mu_S$ as an example.
The bands indicate $\mu_S$ variations
entering the resummation uncertainty estimate
as described in \sec{uncerts_pert}.
Right: Difference of the resummed and the fixed-order singular cumulative
inclusive cross section (i.e., without fiducial cuts) at various orders.
The difference is compatible with zero as $q_T^\mathrm{max} \to Q$
within the relevant combined resummation
and matching uncertainty $\Delta_\res \oplus \Delta_\match$,
which is estimated as described in \sec{uncerts_pert}.
}
\label{fig:profiles}
\end{figure*}

The final matched prediction is obtained as
\begin{align} \label{eq:matching}
\frac{\df\sigma_\match}{\df q_T}
= \frac{\df\sigma_\res}{\df q_T}
+ \frac{\df\sigma_\nons}{\df q_T}
= \frac{\df\sigma_\res}{\df q_T}
+ \frac{\df \sigma_{\FO}}{\df q_T} - \frac{\df\sigma_\sing}{\df q_T}
\,,\end{align}
where $\df\sigma_\res$ is the differential resummed cross section
obtained from a numerical Bessel transform,
as given by the first line of the right-hand side of \eq{tmd_factorization_fid_spectrum},
and $\df \sigma_\nons$ is the nonsingular cross section
evaluated at fixed order, as defined in \eq{nons_def}.
At large $q_T$, the resummation in $\df\sigma_\res$ must be turned off
to ensure that it cancels the fixed-order singular cross section
$\df\sigma_\sing$ and the total fixed-order prediction is recovered.%
\footnote{On a practical note, recovering the fixed-order spectrum exactly
also requires expanding
the different fixed-order boundary conditions in the resummed cross section
against each other, and truncating in the overall number
of powers of $\as$ (also at different scales).
}\textsuperscript{,}%
\footnote{
Note that the resummed cross section also contains all leading nonperturbative
effects of $\ordsq{\lqcd^{n+m}/(q_T^n Q^m)}$ with $m = 0, 1$,
as discussed in \sec{np_model}.
These corrections are absent in the nonsingular cross section,
which is calculated using strict leading-twist collinear factorization,
and could thus potentially also upset the cancellation
between the resummed and singular cross sections at large $q_T$.
However, the scaling of even the most severe $\ord{\lqcd^2/q_T^2}$ correction,
which is maintained by the Bessel transform~\cite{Ebert:2022cku},
is such that their numerical effect already at $q_T \sim 30 \GeV$ is negligible.
For performance reasons
we in practice stop using the numerical Bessel transform altogether
when the resummation is fully turned off for all scale
and matching parameter variations and for all points in the considered $Q$ intervals,
which amounts to $q_T \geq 125 \GeV$ for the numerical results presented in this paper.
}
Thus in practice, the scales in \eqs{canonical_scales_bbs}{canonical_scales_ff}
need to be modified to perform the fixed-order matching at large $q_T$.
We achieve this using hybrid profile scales~\cite{Lustermans:2019plv},
\begin{align} \label{eq:profile_scales_central}
\mu_B &= \mu_\FO \, f_\run\biggl(
   \frac{q_T}{Q},
   \frac{1}{Q}\mu_*\Bigl(\frac{b_0}{b_T}, \muBmin \Bigr)
\biggr)
\,, \quad &
\nu_B &= \mu_\FO
\,, \nn \\[0.4em]
\mu_S &= \mu_\FO \, f_\run\biggl(
   \frac{q_T}{Q},
   \frac{1}{Q}\mu_*\Bigl(\frac{b_0}{b_T}, \muSmin \Bigr)
\biggr)
\,, \quad &
\nu_S &= \mu_\FO \, f_\run\biggl(
   \frac{q_T}{Q},
   \frac{1}{Q}\mu_*\Bigl(\frac{b_0}{b_T}, \nuSmin \Bigr)
\biggr)
\,, \nn \\[0.4em]
\mu_0
&= \mu_*\Bigl( \frac{b_0}{b_T}, \muZeromin \Bigr)
\,, \quad &
\mu_f &= \mu_F \,  f_\run\biggl(
   \frac{q_T}{Q},
   \frac{1}{Q}\mu_*\Bigl(\frac{b_0}{b_T}, \mufmin \Bigr)
\biggr)
\end{align}
where $\mu_\FO = \mu_F = \mu_H = Q$ is the central fixed-order scale used in our predictions,
and
\begin{align} \label{eq:f_run}
f_\run(x,y) &= 1 + g_\run(x)(y-1)
\,, \\
g_\run(x) &= \begin{cases}
1
\,, &
0 < x \leq x_1
\,, \\
1 - \frac{(x-x_1)^2}{(x_2-x_1)(x_3-x_1)}
\,, &
x_1 < x \leq x_2
\,, \\
\frac{(x-x_3)^2}{(x_3-x_1)(x_3-x_2)}
\,, &
x_2 < x \leq x_3
\,, \\
0
\,, &
x_3 \leq x
\nn \,,\end{cases}
\end{align}
ensures that the scales evaluate to the canonical $b_T$-space scales
for $x = q_T/Q < x_1$, and smoothly transition to constant fixed values $\mu_\FO$
as $x \to x_3$, as illustrated in the left panel of \fig{profiles}.%
\footnote{
An alternative way of implementing the transition
is by raising $y$ to a power as $f_\run^\mathrm{alt}(x, y) = y^{g_\run(x)}$,
which directly translates to controlling the size of the resummed logarithms
since $y$ ultimately appears as a ratio of scales in the resummed cross section.
We have verified that this alternative choice
leads to predictions that show quantitatively similar behavior
in the transition region and are completely compatible with the above
within the matching uncertainty defined in \sec{uncerts_pert}.
All numerical predictions given in this paper are obtained using \eq{f_run}.
}
Based on the observed size of the nonsingular cross section, we choose
\begin{align} \label{eq:x1x2x3_central}
(x_1, x_2, x_3) = (0.3, 0.6, 0.9)
\end{align}
for our central predictions.
This choice of matching points has the important property
that the integral of the resummed and matched inclusive spectrum
indeed recovers the fixed-order result,
as it must because fiducial resummation effects~\cite{Ebert:2020dfc}
are absent in that case.
We verify this property in the right panel of \fig{profiles},
where we show the difference
\begin{align}
&\int^{q_T^\mathrm{max}} \! \df q_T \, \frac{\df \sigma_\match^\incl}{\df q_T}
- \int^{q_T^\mathrm{max}} \! \df q_T \, \frac{\df \sigma_\FO^\incl}{\df q_T}
\nn \\
&= \int^{q_T^\mathrm{max}} \! \df q_T \, \frac{\df \sigma_\res^\incl}{\df q_T}
- \int^{q_T^\mathrm{max}} \! \df q_T \, \frac{\df \sigma_\sing^\incl}{\df q_T}
\equiv \sigma_\mathrm{res}(q_T^\mathrm{max})
- \sigma_\mathrm{sing}(q_T^\mathrm{max})
\end{align}
of the resummed and
the fixed-order singular cumulative inclusive cross section
as a function of the upper integral cutoff $q_T^\mathrm{max}$
at different orders.
To give a sense of the effect size,
we normalize the difference to the leading-order inclusive Drell-Yan cross section,
which evaluates to $\sigma_\mathrm{LO}^\incl = 1182.8 \pb$ for our settings.
(The nonsingular cross section exactly drops out in the difference,
and therefore need not be included.)
We find that within the relevant combined resummation
and matching uncertainty $\Delta_\res \oplus \Delta_\match$
estimated as described in \sec{uncerts_pert},
the integral of the resummed cross section
indeed recovers the fixed-order result.
While more complicated implementations
of the matching such as cumulant scales or the so-called Bolzano algorithm~\cite{Bertolini:2017eui}
exist that would make the agreement exact,
we conclude that the hybrid (spectrum) scales we use here
already ensure the fixed-order integral property
to a sufficient degree.
This behavior of spectrum scales
strongly differs e.g.\ from that of the thrust observable
considered in \refcite{Bertolini:2017eui}, and is due to the fact
that for the bulk of the cross section in the peak,
the $b_T$-space scales are in fact canonical and independent of $q_T$
in our construction, such that up to fiducial resummation effects,
the cumulant integral can directly be taken
at the level of the Bessel integral kernel in \eq{tmd_factorization}
without having to integrate the scales by parts.

The central profile scales in \eq{profile_scales_central} serve the additional purpose
of freezing out the scales
at which the perturbative components in \eq{profile_scales_central}
are evaluated before $b_0/b_T$ reaches the Landau pole of QCD.
This is implemented through the function $\mu_*$ appearing
in the second argument of $f_\run$ in \eq{profile_scales_central}
and controlled by, in general, four additional
parameters $\muBmin, \muSmin, \nuSmin, \mufmin \sim 1 \GeV$.
The function $\mu_*(a, b)$ must be a monotonic function of $a$
and in addition have the following asymptotic behavior,
\begin{alignat}{3} \label{eq:mu_star_asymptotic_behavior}
a &\gg b
\,: \qquad &
\mu_*(a, b)
&= a \Bigl[ 1 + \ORd{\frac{b^p}{a^p}} \Bigr]
\,, \nn \\
a &\ll b
\,: \qquad &
\mu_*(a, b)
&= b \Bigl[ 1 + \ORd{\frac{a^q}{b^q}} \Bigr]
\,,\end{alignat}
where $p, q > 0$ are positive powers.
This ensures that at perturbative $b_0/b_T \gg \lqcd$
(and $q_T \ll Q$, specifically $q_T \leq x_1 Q$),
the profile scales in \eq{profile_scales_central} indeed evaluate
to the canonical scales in \eq{canonical_scales_bbs}
up to power corrections of $(\lqcd b_T)^p$.
In the opposite limit $b_0/b_T \lesssim \lqcd$ it ensures
that the profile scales remain bounded from below
by the respective minimum scale.

Various choices for $\mu_*$ have been considered in the literature.
One common choice going back to Collins and Soper~\cite{Collins:1981va} is to use
\begin{align} \label{eq:mu_star_cs}
\mu_*^\mathrm{CS}(a, b) = \sqrt{a^2 + b^2}
\,,\end{align}
which results in $p = q = 2$.
Here, $p \geq 2$ is desirable
because it ensures that the Landau pole prescription
does not introduce spurious power corrections
with stronger scaling than the genuine $\lqcd^2 b_T^2$ power corrections
to the leading term of the OPEs in \eq{ope_beam_soft_tmd}.
As discussed in \refcite{Ebert:2022cku}, this scheme may be extended
to ensure that also higher powers in the OPE are not contaminated
by the Landau pole prescription.
Two simple options are

\begin{subequations} \label{eq:mu_star_cs4_and_exp}
\begin{align} \label{eq:mu_star_cs4}
\mu_*^\mathrm{CS4}(a, b) &= (a^4 + b^4)^\frac{1}{4}
\,, \\[0.4em]
\label{eq:mu_star_exp}
\mu_*^\mathrm{exp}(a, b)
&= \begin{cases}
   a \Bigl[ 1 - \exp\Bigl( - \frac{2}{2b/a - 1}\Bigr) \Bigr]^{-1}
   \,, \quad &
   a < 2b
   \,, \\
   a
   \,, \quad &
   a \geq 2b
\,.\end{cases}
\,,\end{align}
\end{subequations}
leading to $p = q = 4$ and $p = q = \infty$, respectively.
The emphasis of our analysis is on the treatment of the $\ord{\lqcd^2 b_T^2}$
power corrections, so the $\mu_*^\mathrm{CS4}$ prescription
that guarantees their direct field-theoretic interpretation
in terms of higher-twist matrix elements
is sufficient for our purposes,
and we thus adopt it throughout our numerical results.%
\footnote{We note that the Pavia prescription
$\mu_*^\mathrm{Pavia}(a, b) = b \bigl[ 1 - \exp\bigl( - b^4/a^4 )]^{-\frac{1}{4}}$
introduced in \refcite{Bacchetta:2017gcc} also has $p = 4 > 2$ (and $q = \infty$).}
We pick the minimum scales as
\begin{align} \label{eq:choices_min_scales}
\muBmin = \muSmin = \muZeromin = 1 \GeV
\,, \qquad
\nuSmin = 0
\,.\end{align}
For $\mufmin$ we pick the maximum of the $Q_0$ value of the respective PDF set,
i.e., down to which value of $Q_0 \leq \mu$ the \texttt{LHAPDF} interpolation grid is provided,
and the value of the charm quark mass as quoted in the PDF set.
This avoids both extrapolation outside of the \texttt{LHAPDF} grid
and large effects from the charm threshold,
which is effectively screened by this prescription.
At the same time, it allows us to resum Sudakov logarithms to the largest
extent possible by letting the scales relevant for it ($\mu_B$, $\mu_S$, $\nu_S$)
decrease further until $1 \GeV$.

Following this prescription in practice,
we set $\mufmin = 1.65 \GeV$ ($Q_0$) for the NNLO PDF sets by the \texttt{NNPDF} collaboration~\cite{NNPDF:2017mvq, NNPDF:2021njg},
whereas for the aN$^3$LO~\cite{NNPDF:2024nan} set we use $\mufmin = 1.51 \GeV$ (charm mass).
For all sets produced by the \texttt{MSHT} collaboration~\cite{Bailey:2020ooq, McGowan:2022nag} we set $\mufmin = 1.40 \GeV$ (charm mass),
and for sets by the \texttt{CTEQ} collaboration~\cite{Hou:2019efy} $\mufmin = 1.30 \GeV$ (charm mass).

For the convenience of the reader,
we again present the equivalent profile scales
in terms of TMD boundary scales,
which are a ready-to-use implementation of large-$q_T$ fixed-order matching
for general TMD studies (suitably adjusting the $\mu_*$ function
to recover existing scale choices at small $q_T$).
The explicit relations for the central scale choices are
\begin{align} \label{eq:relations_scales_tmd_bbs}
\mu_\init &= \mu_\FO \, f_\run\biggl(
   \frac{q_T}{Q},
   \frac{1}{Q}\mu_*\Bigl(\frac{b_0}{b_T}, \muinitmin \Bigr)
\biggr)
\,, \nn \\
\sqrt{\zeta_\init} &= \mu_\FO \, f_\run\biggl(
   \frac{q_T}{Q},
   \frac{1}{Q}\mu_*\Bigl(\frac{b_0}{b_T}, \sqrt{\zetainitmin} \Bigr)
\biggr)
\,, \nn \\
\mu_\fin &= \sqrt{\zeta_\fin} = \nu_B = \mu_\FO
\,,\end{align}
where $\mu_0$ is common to both approaches,
and identifying a combined TMD PDF requires setting
$\nuSmin = \sqrt{\zetainitmin}$
and $\muinitmin = \muBmin = \muSmin$
such that $\nu_S = \sqrt{\zeta_\init}$ and $\mu_B = \mu_S$ for central scales.
It is also instructive to discuss how the above prescription relates
to standard, so-called $b^*$ prescriptions.
Assuming an identical functional form of $\mu_*$,
a ``local'' $b^*$ prescription acting only
on boundary scales is obtained by setting
\begin{align} \label{eq:relation_local_bstar}
\muBmin = \muSmin = \nuSmin = \mu_0 = \mu_f = \muinitmin = \sqrt{\zetainitmin}
= \frac{b_0}{b_\mathrm{max,local}}
\,.\end{align}
Note that some implementations that only act on virtuality scales
entering the strong coupling or the PDFs
may instead require $\nuSmin = \zetainitmin = 0$,
corresponding to our earlier choice.
By contrast, a ``global'' $b^*$ prescription amounts
to globally replacing $b_T$
by a suitable function $b^*(b_T)$
of the form $b_0/\mu_*$ such that $b^*(b_T) \to b_\mathrm{max,global}$
as $b_T \to \infty$.
Specifically, this replacement is performed
everywhere under the Fourier integral in \eq{tmd_factorization}
except for the integral kernel,
including in particular the $b_T$ arguments of the TMDs themselves,
i.e.,
\begin{align} \label{eq:global_bstar}
&[B_a B_b S](Q^2, x_a, x_b, \qt, \mu_\fin) \bigl\rvert_\mathrm{global}
\nn \\
&= \frac{1}{2\pi} \int_0^\infty \df b_T \, b_T J_0(b_T q_T) \,
   \tf_a\bigl(x_a, b_*, \mu_\init(b^*), \zeta_\init(b^*) \bigr) \,
   \tf_b\bigl(x_b, b_*, \mu_\init(b^*), \zeta_\init(b^*) \bigr)
\nn \\ & \quad \times
   U^2(\mu_\init(b^*), \zeta_\init(b^*), \mu_\fin, \zeta_\fin = \mu_\fin^2)
\,,\end{align}
where $U$ is the combined quark TMD PDF evolution kernel.
The scales $\mu_\init(b^*)$ and $\zeta_\init(b^*)$ may still be chosen as above
to implement the fixed-order matching using hybrid profile scales
(but without an additional Landau prescription at the level of the scales,
which would be redundant in this case).

\section{Nonperturbative TMD dynamics}
\label{sec:np_model}

\subsection{Sources of nonperturbative contributions}

The resummed unpolarized TMD PDFs
entering the factorized cross section in \eq{tmd_factorization_fid_spectrum},
including all possible nonperturbative effects
allowed by the structure of TMD factorization,
take the following form~\cite{Boussarie:2023izj}:
\begin{align} \label{eq:tmd_pdf_evolved}
\tf_i\bigl(x, b_T, \mu, \zeta)
&= \tf_i^\pert\bigl(x, b_T, \mu_\init, \zeta_\init\bigr) \,
U_i(\mu_\init, \mu, \zeta_\init) \, V_i(b_T, \mu, \zeta_\init, \zeta) \,
\tf_i^\np(x, b_T)
\end{align}
Here $\tf_i^\pert$ denotes the leading term
in the OPE in \eq{ope_beam_soft_tmd}
evaluated at boundary scales $\mu_\init$ and $\zeta_\init$
using a Landau pole prescription such as in \eq{profile_scales_central},%
\footnote{
The additional deformation of the scales
to implement the fixed-order matching for $q_T \to Q$
is not relevant here,
since it is of higher order in $q_T/Q$
than the formal accuracy of the TMD factorization itself.
}
$U_i$ denotes the virtuality evolution of the TMD PDF
at a fixed CS scale (which is purely perturbative by consistency
with the hard function),
and $V_i$ denotes the CS evolution kernel at fixed $\mu$,
which contains a nonperturbative component discussed below.
Finally, $\tf_i^\np(x, b_T)$ denotes the nonperturbative contribution
to the TMD PDF boundary condition, which can be interpreted
as the intrinsic transverse momentum spectrum
of partons within the proton.

The CS evolution factor $V_i$ in \eq{tmd_pdf_evolved} is obtained by solving~\cite{Boussarie:2023izj}
\begin{align} \label{eq:tmd_rapidity_anom_dim_relations}
\gamma^i_\zeta(b_T, \mu)
&= 2 \frac{\df}{\df \ln \zeta} \ln \tf_i(x, b_T, \mu, \zeta)
\nn \\
&= - \frac{\df}{\df \ln \nu} \ln \tB_i(b_T, \mu, \nu)
= + \frac{1}{2} \frac{\df}{\df \ln \nu} \ln \tS_i(b_T, \mu, \nu)
= \frac{1}{2} \tgamma^i_\nu(b_T, \mu)
\,,\end{align}
where $\gamma^i_\zeta$ is known as the Collins-Soper kernel,
and we have included its relation to the beam and soft rapidity anomalous
dimensions in the rapidity renormalization group approach~\cite{Chiu:2011qc, Chiu:2012ir}
for reference. (Note that as in \refcite{Billis:2019vxg},
we reserve the symbol $\tgamma^i_\nu$ for the \emph{soft}
rapidity anomalous dimension, in contrast to Sec.~4.5 of \refcite{Boussarie:2023izj}.)
Similar to the TMD PDFs in \eq{tmd_pdf_evolved}, the CS kernel contains
an evolution term resumming logarithms of $\mu b_T$,
a fixed-order boundary condition evaluated at $\mu_0 \sim b_T$ for $b_0/b_T \gg \lqcd$,
and a nonperturbative contribution, which in this case are all additive~\cite{Boussarie:2023izj},
\begin{align} \label{eq:tmd_rapidity_anom_dim_evolved}
\tgamma_\zeta^i(b_T, \mu)
= -2 \eta^i_\Gamma(\mu_0, \mu) + \tgamma^i_{\zeta}(b_T, \mu_0) + \tgamma^{i, \np}_{\zeta}(b_T)
\,.\end{align}
Expressions for the fixed-order boundary condition $\tgamma^i_{\zeta}(b_T, \mu_0)$
and a recursive analytic solution for $\eta^i_\Gamma$
to N$^4$LL are given in \app{rge_solutions_n4ll},
while the exact analytic solution for the latter to N$^3$LL is given in \refcite{Ebert:2021aoo}.
Following the split of perturbative and nonperturbative physics in $\tgamma_\zeta^i$,
we accordingly define their exponentials as
\begin{align} \label{eq:cs_evolution_factor_pert_np_def}
V_i(b_T, \mu, \zeta_\init, \zeta)
= \exp \Bigl[ \frac{1}{2} \tgamma_\zeta(b_T, \mu) \, \ln \Bigl(\frac{\zeta}{\zeta_\init} \Bigr) \Bigr]
\equiv V_i^\pert(b_T, \mu, \zeta_\init, \zeta) \, V_i^\np(b_T, \zeta_\init, \zeta)
\,,\end{align}
which together make up the total CS kernel.
For later convenience we also define
\begin{align} \label{eq:tmd_pdf_evolved_pert_np_zeta_def}
\tf_i\bigl(x, b_T, \mu, \zeta)
&\equiv \tf_i^\pert \bigl(x, b_T, \mu, \zeta) \, V_i^\np(b_T, \zeta_\init, \zeta) \, \tf_i^\np(x, b_T)
\nn \\[0.4em]
&\equiv \tf_i^{\pert + \np_\zeta} \bigl(x, b_T, \mu, \zeta) \, \tf_i^\np(x, b_T)
\,,\end{align}
i.e., the purely perturbative TMD PDF $\tf_i^\pert$ including the $U_i$ evolution to $\mu$
and the perturbative component of the CS evolution from $\zeta_\init$ to $\zeta$,
and the TMD PDF $\tf_i^{\pert + \np_\zeta}$ including in addition
the nonperturbative part of the CS evolution (but still excluding
the nonperturbative TMD boundary condition).

We remind the reader of the well-known fact
that $\tf_i^\np(x, b_T)$, $\gamma_\zeta^{q,\np}(b_T)$,
and the entire notion of ``splitting''
perturbative and nonperturbative physics
are defined with respect to a given choice of boundary scales
(i.e.\ in particular, a given Landau pole prescription),
for which many choices are available in the literature,
see e.g.\ \refscite{Collins:1981va, Bacchetta:2017gcc, Scimemi:2018xaf}.
Meaningful comparisons and conversions between them
must therefore be performed
at the level of the complete TMD PDF in \eq{tmd_pdf_evolved}.
Nevertheless, the interpretation of the nonperturbative functions
can be made more straightforward
if the Landau pole prescription in $\mu_\init$ and $\mu_0$
(and possibly $\zeta_\init$)
is of $\ord{\lqcd^p b_T^p}$ with $p > 2$~\cite{Ebert:2022cku}.
In this case the power expansions of the model functions
are in direct correspondence to the first subleading terms in the OPEs
of the TMD PDF and the CS kernel in \eqs{ope_beam_soft_tmd}{resummed_rapidity_anom_dim}.
Combining them, they read as follows,
\begin{align} \label{eq:tmd_opes_subleading}
f_i(x, b_T, \mu, \zeta)
&= \Bigl[
   \sum_j \int\! \frac{\df z}{z} \, \tilde C_{ij}(z, b_T, \mu, \zeta) \, f_{j}\Bigl( \frac{x}{z}, \mu \Bigr)
\Bigr]
\nn \\ & \quad \times \Bigl[
   1
   +  b_T^2 \Bigl( \Lambda_i^{(2)}(x) + \frac{1}{2} \gamma^{(2)}_{\zeta,i} \ln \frac{b_T^2 \zeta}{b_0} \Bigr)
+ \ord{\lqcd^4 b_T^4}
\Bigr]
\,,\end{align}
and relate to the nonperturbative functions introduced above as
\begin{align} \label{eq:relation_np_func_ope}
\tgamma^{i, \np}_{\zeta}(b_T)
= \gamma_{\zeta,i}^{(2)} \, b_T^2 + \ord{\lqcd^4 b_T^4}
\,, \qquad
\tf_i^\np(x, b_T) = 1 + \Lambda_i^{(2)} \, b_T^2 + \ord{\lqcd^4 b_T^4}
\,.\end{align}
Here $\gamma_{\zeta,i}^{(2)}$
is a single number
given by a gluon vacuum condensate~\cite{Vladimirov:2020umg},
and $\Lambda_i^{(2)}(x)$
is a function of $x$ and the flavor $i$
given by a normalized combination of twist-4 collinear matrix elements.
If the Landau pole prescription instead starts at $p = 2$,
the relation must include total derivatives of the perturbative piece
with respect to $b_T$ in addition,
and analogously for higher terms in the OPE
and Landau pole prescriptions of higher order.
Here we ignore -- as is common in the literature --
the possible presence of anomalous powers of $\lqcd$
from the RG evolution of the condensate and the higher-twist matrix elements
that is not captured by that of the twist-2 ingredients.
I.e., taking the nonperturbative function to be analytic at $b_T = 0$
appropriately captures the higher-twist contributions for $\lqcd \, b_T \ll 1$
when working to leading order for their Wilson coefficient
and ignoring their running.
Finally, we note that while the leading $\ord{\lqcd^2 b_T^2}$ renormalons
in the CS kernel and TMD PDFs are known to be nonzero~\cite{Scimemi:2016ffw},
we do not include any renormalon subtractions,
which is motivated \emph{a posteriori} by the excellent convergence
we observe in our predictions towards higher perturbative orders,
see \secss{uncerts}{results_Z_cumulants}{results_W}.

We note that while we mostly choose to work
in TMD notation in this section for definiteness
and to make contact with the literature,
one can indeed express all results in a completely equivalent way
by introducing nonperturbative functions
$\tB^\np_i(x, b_T)$ and $\sqrt{S_i^\np}(b_T)$
for the beam and soft functions at their respective boundary scales,
where one may choose $S_i^\np = 1$ and $\tf_i^\np(x, b_T) = \tB^\np_i(x, b_T)$
without loss of generality through a suitable rapidity renormalization scheme.
In our final predictions
based on \eqs{np_rap_model_combined_formula}{np_fid_model_combined_formula}
(which in practice will use effective models,
but the same could be done with a more detailed model
for the underlying $\tf_i^\np$),
we evaluate the evolved perturbative TMD PDF
dressing the nonperturbative model functions
in the more general form of the rapidity renormalization framework
as in \eq{RGevolution}, explicitly,
\begin{align} \label{eq:tmd_pdf_evolved_pert_np_zeta_as_used_in_practice}
&\tf_i^{\pert + \np_\zeta} \bigl(x, b_T, \mu, \zeta)
\\
&= \tB_i^\pert\Bigl(x,b_T,\mu_B,\frac{\nu_B}{\sqrt{\zeta}}\Bigr)
   \exp\biggl[-\frac12 \ln\frac{\nu}{\nu_B} \tgamma^i_\nu(b_T,\mu_B) \biggr]
   \exp\biggl[\int_{\mu_B}^\mu \frac{\df\mu'}{\mu'} \tgamma_B^i\Bigl(\mu',\frac{\nu}{\sqrt{\zeta}}\Bigr) \biggr]
\nn \\ & \quad \times
\sqrt{\tS_i^\pert}(b_T,\mu_S,\nu_S)
   \exp\biggl[\frac12 \ln\frac{\nu}{\nu_S} \tgamma^i_\nu(b_T,\mu_S) \biggr]
   \exp\biggl[\frac12 \int_{\mu_S}^\mu \frac{\df\mu'}{\mu'} \tgamma_S^i(\mu',\nu) \biggr]
\nn \,,\end{align}
which in particular may involve independent $\mu_{B,S}$ and/or $\nu_{B,S}$
variations, as described in \sec{uncerts_pert}.
Here $\tB_i^\pert$ and $\tS_i^\pert$
are defined as the respective leading term of the OPEs in \eq{ope_beam_soft_tmd},
and the nonperturbative component of the resummed CS kernel
at its (implicit) boundary scale $\mu_0$ in \eq{tmd_rapidity_anom_dim_evolved}
is, of course, exactly equivalent to that
of the rapidity anomalous dimension in \eq{tmd_pdf_evolved_pert_np_zeta_as_used_in_practice},
see \eq{tmd_rapidity_anom_dim_explicit_np_model} below.

\subsection{Modeling the CS kernel}
\label{sec:np_cs_kernel}

We consider the following simple nonperturbative model
for the quark rapidity anomalous dimension,
i.e., the CS kernel in the fundamental representation,
\begin{align} \label{eq:tmd_rapidity_anom_dim_explicit_np_model}
\gamma_\zeta^{q,\np}(b_T)
= \frac{1}{2} \gamma_{\nu\,\np}^{q}(b_T) = c_\nu^q \tanh\Bigl( \frac{\omega_{\nu,q}^2}{\abs{c_\nu}} b_T^2 \Bigr)
= \operatorname{sgn}(c_\nu^q) \, \omega_{\nu,q}^2 b_T^2 + \ord{\lqcd^6 b_T^6}
\,,\end{align}
defined with respect to our preferred Landau-pole prescription
in \eqs{mu_star_cs4}{choices_min_scales}.
The above two-parameter model is useful for two reasons.
First,
it asymptotes to a constant $c_\nu$ as $b_T \to \infty$,
so the total CS kernel
also becomes flat (up to logarithmic terms) in the large-$b_T$ limit,
as suggested in \refcite{Collins:2014jpa}.%
\footnote{
We note that these potentially large logarithms of $\muZeromin b_T$
as $b_T \to \infty$ from having $\mu_0 \to \muZeromin$
in our ``local'' Landau-pole prescription,
like the renormalons of the TMD ingredients,
are another possible source of unstable behavior
as the perturbative logarithmic order increases.
While we find no drastic lack of convergence (again \emph{a posteriori})
even at very small $q_T$,
removing these terms explicitly provides
one way of improving the stability of the fit
between different orders
when model parameters are extracted from data,
which we plan to address in the future.
}
The absence of at least linear power-law growth
has in the meantime also been confirmed
by first-principles calculations
on the lattice~\cite{Schlemmer:2021aij, LatticePartonLPC:2022eev, Shu:2023cot, LatticePartonLPC:2023pdv, Avkhadiev:2024mgd},
and we will give an illustrative range
over which our model parameters are varied
based on the spread of these lattice results in \sec{uncerts_np}.
In addition, the most recent global TMD fits~\cite{Bury:2022czx, Bacchetta:2022awv, Moos:2023yfa, Bacchetta:2024qre}
also point to a functional form of this kind.
(However, since these fits use LHC Drell-Yan data,
we cannot include them in an a priori estimate of the nonperturbative impact.)
Second, the sign of the whole model
and the height of the plateau at large $b_T$
are controlled through $c_\nu$
in a way independent of the $\omega_\nu$ parameter
that governs the strength of the quadratic term
and thus the onset of nonperturbative physics.
For ease of illustrating the impact of variations around it, we take
\begin{align} \label{eq:gamma_nu_np_central_params}
 (c_\nu^i, \omega_{\nu,q})
 = (-0.05, 0.25 \GeV)
\end{align}
as our default values used in all our numerical results
unless otherwise noted.

\subsection{Effective rapidity-dependent models for multi-differential Drell-Yan}
\label{sec:np_rap_model}

We first consider the impact of nonperturbative TMD physics
on a generic multi-differential Drell-Yan measurement,
which can be expressed in terms of their contributions
to the helicity cross sections $i = -1, 4$
that factorize in terms of leading-power unpolarized TMD PDFs
as in \eq{tmd_factorization}.
We may rewrite the latter as
\begin{align} \label{eq:np_rap_model_rewritten_fnp_explicit}
\frac{\df \sigma_i}{\df^4 q}
= \frac{1}{2\pi} \int_0^\infty \! \df b_T \, b_T \, J_0(b_T q_T) \,
\sum_{a,b} \tsigma_{i,ab}^{\pert + \np_\zeta}(Q, Y, b_T) \,
\tf_a^\np(x_a, b_T) \, \tf_b^\np(x_b, b_T)
\,,\end{align}
where the nonperturbative TMD boundary conditions are weighted
by a function
\begin{align} \label{eq:np_rap_model_weight_func}
\tsigma_{i,ab}^{\pert + \np_\zeta}(Q, Y, b_T)
&\equiv \frac{1}{2\Ecm^2} \sum_{V,V'} L_{\pm (i) VV'}(Q^2) \,
H_{i\,VV'\,ab}(Q^2, \mu)
\nn \\[0.4em]
& \quad \times
\tf_a^{\pert + \np_\zeta} \bigl(x_a, b_T, \mu, Q^2) \,
\tf_b^{\pert + \np_\zeta} \bigl(x_b, b_T, \mu, Q^2)
\end{align}
that contains the electroweak couplings, hard functions,
and all other parts of the evolved TMD PDFs,
i.e., all its perturbative components
and the nonperturbative CS evolution
as in \eq{tmd_pdf_evolved_pert_np_zeta_def}.
Note that this weight function (like the definitions derived from it below)
also depend on the collider beams and center-of-mass energy $\Ecm$
as well as the type of boson ($Z/\gamma^*$, $W^+$, or $W^-$),
all of which we suppress in the following.
We now simply let
\begin{align} \label{eq:np_rap_model_average_np_func_def}
\overline{F}_i^\np(Q, Y, b_T)
\equiv
\frac{
   \sum_{a,b} \tsigma_{i,ab}^{\pert + \np_\zeta}(Q, Y, b_T) \,
   \tf_a^\np(x_a, b_T) \, \tf_b^\np(x_b, b_T)
}{
   \sum_{a,b} \tsigma_{i,ab}^{\pert + \np_\zeta}(Q, Y, b_T)
}
\end{align}
which satisfies
$\overline{F}_i^\np(Q, Y, b_T) = 1 + \ord{\lqcd^2 b_T^2}$
by construction.
Note that $i = -1,4$ in this case labels helicity cross sections,
not parton species $u, d, \bar{u}, \dots$
as in the TMD PDF nonperturbative function in \eq{tmd_pdf_evolved}.
Inserting \eq{np_rap_model_average_np_func_def} into \eq{np_rap_model_rewritten_fnp_explicit}, we have
\begin{align} \label{eq:np_rap_model_rewritten_with_average}
\frac{\df \sigma_i}{\df^4 q}
= \frac{1}{2\pi} \int_0^\infty \! \df b_T \, b_T \, J_0(b_T q_T) \,
\Bigl[ \sum_{a,b} \tsigma_{i,ab}^{\pert + \np_\zeta}(Q, Y, b_T) \Bigr] \,
\overline{F}_i^\np(Q, Y, b_T)
\,.\end{align}
This is precisely in the spirit of \refcite{Ebert:2022cku},
where an approach was proposed to extract effective, flavor-averaged
nonperturbative parameters at $\ord{\lqcd^2 b_T^2}$,
but generalized to all powers in $\lqcd$.
Specifically, the -- in general complicated -- flavor and $x$ dependence
of the underlying TMD PDFs is transformed
into effective, flavor-averaged model functions
that are both minimal and sufficient
to describe the nonperturbative TMD physics
contained in a given process at a given collider,
as relevant when using one process at a time
to e.g.\ extract fundamental parameters like $\as(m_Z)$ or $m_W$.

The above result can be simplified further for resonant Drell-Yan
(integrated over $Q$)
by noting that the weight factor in \eq{np_rap_model_rewritten_with_average}
is strongly dominated by $Q \sim m_V$, with $m_V = m_Z$ or $m_W$,
whereas the resonant enhancement cancels in \eq{np_rap_model_average_np_func_def},
leaving behind only the slowly varying $x$
and logarithmic $Q$ dependence of the resummed perturbative TMD PDFs.
It is then reasonable to approximate
\begin{align} \label{eq:np_rap_model_average_np_func_approx}
\overline{F}_i^\np(Q, Y, b_T)
\approx \overline{F}_i^\np(m_V, Y, b_T)
\equiv \overline{F}_i^\np(Y, b_T)
\end{align}
by a simpler, two-dimensional function of just $Y$ and $b_T$.
We stress that cross terms between the finite vector boson width
and the nonperturbative TMD boundary conditions encoded in $\overline{F}_i^\np(Y, b_T)$
are in fact retained and treated exactly in our numerics,
i.e., the width is kept exact in $\tilde{\sigma}_{i,ab}^{\pert+\np_\zeta}$
in our final \eq{np_rap_model_combined_formula}.
The above manipulations using the narrow width thus only limit the degree
to which our model predictions \emph{deviate from the most general form}
of nonperturbative dynamics allowed by TMD factorization
(similar to the impact of picking any given functional form for the model),
with the residual model dependence starting
at $\ordsq{(\lqcd^2 \Gamma_V)/(q_T^2 m_V)}$ in this case.

Fixing $Q = m_V$ also simplifies the averaging procedure,
since in this case
the perturbative $\mu$ evolution
and the (non)perturbative CS evolution
cancel exactly between numerator and denominator,
\begin{align} \label{eq:np_rap_model_average_np_func_approx_simplify}
\overline{F}_i^\np(Y, b_T)
= \frac{
   \sum_{a,b} \tsigma_{i,ab}^{\pert}(m_Z, Y, b_T) \,
   \tf_a(x_a, b_T, \mu, Q^2) \, \tf_b(x_b, b_T, \mu, Q^2)
}{
   \sum_{a,b} \tsigma_{i,ab}^{\pert}(m_Z, Y, b_T)
}
\end{align}
i.e., the nonperturbative model for $\gamma_\zeta^q$ decouples
from the averaging procedure,
allowing us to drop the superscript $\np_\zeta$.
If we further ignore the numerically tiny effect
of singlet hard Wilson coefficients, see \sec{singlet_coeffs},
the higher-order corrections to the hard function also cancel,
leaving behind the simple formula
\begin{align} \label{eq:np_rap_model_average_np_func_final}
\overline{F}_i^\np(Y, b_T)
= \frac{
   \sum_{a,b} \sigma_{i,ab}^B(m_Z) \,
   \tf_a^{\pert} \,
   \tf_b^{\pert}
   \tf_a^\np(x_a, b_T) \, \tf_b^\np(x_b, b_T)
}{
   \sum_{a,b} \sigma_{i,ab}^B(m_Z) \,
   \tf_a^{\pert} \,
   \tf_b^{\pert}
}
\,,\end{align}
where
\begin{align} \label{eq:helicity_born_xsec}
\sigma_{i,ab}^B(Q)
\equiv \frac{1}{2 Q^2} \sum_{V,V'} L_{\pm (i) VV'}(Q^2) \,
H_{i\,VV'\,ab}^\mathrm{LO}
\end{align}
is the Born cross section
in the given helicity channel, containing the quark and lepton electroweak charges
(see appendix~A in \refcite{Ebert:2020dfc}),
and the perturbative TMD PDFs are evaluated at
$x_{a,b}, b_T, \mu_\init, \zeta_\init$ with $x_{a,b} = m_Z/\Ecm \, e^{\pm Y}$.
To low perturbative accuracy one may even approximate
$\tf_i(x, b_T, \mu_\init, \zeta_\init)$
by a collinear PDF $f_i(x, \mu_\init)$,
yielding a very simple pocket formula
that can be evaluated quickly
to convert between effective and general nonperturbative models.

Based on the above discussion, our final, recommended nonperturbative TMD model
for multi-differential resonant Drell-Yan studies reads
\begin{align} \label{eq:np_rap_model_combined_formula}
\frac{\df \sigma_i}{\df^4 q}
= \frac{1}{2\pi} \int_0^\infty \! \df b_T \, b_T \, J_0(b_T q_T) \,
\Bigl[ \sum_{a,b} \tsigma_{i,ab}^{\pert + \np_\zeta}(Q, Y, b_T) \Bigr] \,
\overline{F}_i^\np(Y, b_T)
\,,\end{align}
where $\tsigma_{i,ab}^{\pert+\np_\zeta}$ includes the contribution
from the nonperturbative Collins-Soper kernel
modeled e.g.\ as in \eq{tmd_rapidity_anom_dim_explicit_np_model}.
We stress that for $Z$ boson production,
the $\overline{F}_i^\np(Y, b_T)$
for $i = -1, 4$ (i.e., the inclusive cross section
and the forward-backward asymmetry)
will in general differ from each other due to the different ways
flavors are weighted with electroweak charges.
Nevertheless, the two functions exactly coincide for $W^+$ and $W^-$ production
due to the $V-A$ structure of the weak current,
leading to a large degree of universality.
(Note that the effective model functions for each of $W^+$ and $W^-$
are not equal to each other in general
due to e.g.\ a different $u$ vs.\ $d$ valence TMD PDF contributing to each).
To predict their degree of correlation, or their correlation(s)
across different colliders and bosons,
one necessarily has to resort to a more complete
model of the TMD PDF boundary condition
and perform the conversion using \eq{np_rap_model_average_np_func_final}.%
\footnote{
\label{ftn:dbm}%
We note that in principle,
the result in \eq{np_rap_model_combined_formula}
for the case of the $i = 2$ helicity cross section
of the $Z$ boson should be extended
by another effective function of $Y$ and $b_T$ starting at $\ord{\lqcd^2 b_T^2}$
that encodes the flavor-weighted contribution
from the product of two Boer-Mulders functions,
possibly supplemented by their known twist-3 matching~\cite{Scimemi:2018mmi}.
(The Double Boer-Mulders (DBM) contribution is suppressed by the width
and/or the size of singlet contributions for $i = 5$ in the $Z$ case,
and vanishes altogether for $i = 2,5$ in the case of $W^\pm$
due to the $V-A$ structure of the current~\cite{Boer:1999mm, Gao:2024xxx},
such that none of this affects direct $m_W$ measurements
that do not rely on tuning to the $Z$.)
As shown in \refcite{Ebert:2020dfc},
this DBM contribution can in fact become leading
in general multi-differential Drell-Yan studies
when leptonic observables $p_L$ approach $p_L \sim q_T \sim \lqcd$.
Examples with nonzero DBM contributions in this regime
analyzed in \refcite{Ebert:2020dfc}
are $p_L = \phi^*_\eta Q$ and, indeed,
the distance $p_L = Q - 2 p_T^\ell$ to the Jacobian peak
in the $p_T^\ell$ spectrum as relevant for $m_W$-like $m_Z$ measurements.

At this point we like to remark that our statements in this section
about the \emph{dimensionality} of the effective function space
are actually unaffected by the presence of the DBM effect
because the approximate leptonic tensors for $p_L \sim q_T \ll Q$ in \refcite{Ebert:2020dfc}
can be recast as effective integral functionals
acting directly on the $b_T$ space integrand
upon performing the $q_T$ integral,
following the approach of \refcite{Gao:2022bzi},
which in turn preserve the power counting
of terms of $\ord{\lqcd^n b_T^n}$ as $\ord{\lqcd^n/p_L^n}$.
Thus one still only requires one nonperturbative function
of $Y$ to describe the total net nonperturbative effect
at each power in $\lqcd^2/p_L^2$,
and a single number if no rapidity-differential information
($\eta_\ell$ or $Y$) is retained.
Details on this point will be given elsewhere.

Finally, we note that introducing an explicit DBM model function,
just like decorrelating the nonperturbative models
between $i = -1$ and $i = 4$ for the $Z$,
is not easy to reconcile with the commonly chosen approach in $m_W$ measurements,
where the angular coefficients (as \emph{ratios} of helicity cross sections)
are evaluated at fixed order,
and would first require introducing dedicated resummed components
into the description of at least
the leading-power helicity cross sections $i = -1, 2, 4$
(and $i = 5$ for full generality).
}

On the other hand, based on the argument above,
we find that one can make $\bar{F}_i^\np(Y, b_T)$
the primary target of the TMD nonperturbative modelling effort
when considering multi-differential Drell-Yan distributions
for one boson at a time,
with the understanding that they are defined
with respect to a definite choice of boundary scales,
just like the underlying flavor and $x$-dependent model would be.
For definiteness, one may pick
the following flexible model,
\begin{align} \label{eq:np_rap_model_explicit_form}
\overline{F}_i^\np(Y, b_T)
= \bigl[ 1 + \overline{\Lambda}_{i,2}(Y) \, b_T^2\bigr]^2
\exp\bigl[-2 \overline{\Lambda}_{i,4}(Y) \, b_T^4 \bigr]
\end{align}
with power expansion
\begin{align} \label{eq:np_rap_model_expansion}
\overline{F}_i^\np(Y, b_T)
= 1
+ \overline{\Lambda}_{i,2}(Y) \, b_T^2
+ \bigl[ \overline{\Lambda}_{i,2}^2(Y) - \overline{\Lambda}_{i,4}(Y) \bigr] b_T^4
+\ord{\lqcd^6 b_T^6}
\,,\end{align}
where the coefficient of the quadratic term $\overline{\Lambda}_{i,2}(Y) \sim \lqcd^2$
may take either sign as long as $\overline{\Lambda}_{i,4}(Y) \sim \lqcd^4$ is positive.
The functions $\overline{\Lambda}_{i,n}(Y)$ must be even functions of $Y$
for $pp$ collisions and may, again for definiteness, be chosen as
\begin{align} \label{eq:np_rap_model_explicit_Y_dep}
\overline{\Lambda}_{i,n}(Y) = \Lambda_{i,n} + \Delta \Lambda_{i,n} \, Y^2
\,.\end{align}
We stress that the parameters of the model upon expansion have
a direct field-theoretic interpretation
in terms of (averaged and normalized) entries in the OPE
of the TMD PDFs, accounting for the effect of the Landau-pole prescription
starting at $\ord{\lqcd^4 b_T^4}$.
Since the $x$ dependence of the collinear twist-2 PDFs has been divided out,
we generally expect a slowly varying $Y$ dependence
and thus no need to include terms of $\ord{Y^4}$
or choose a more complicated functional form for the rapidity dependence.

\subsection{Effective models for fiducial \texorpdfstring{$q_T$}{qT} spectra}
\label{sec:np_fid_model}

We can take the averaging procedure further
by also including the rapidity integral in it,
as relevant for the single-differential fiducial $q_T$ spectra
that are the main focus of this paper.
To do so, it is necessary to also consider the action of fiducial cuts $\Theta$,
since they shape the contributing rapidity region in this case.
Up to corrections of $\ordsq{\lqcd^2/(q_T Q)}$,
which are small for $q_T$ spectra in resonant Drell-Yan production,
we may simply ignore the $q_T$ dependence of the leptonic tensors
when acting on the nonperturbative model,
i.e., we evaluate $L_{i\,VV'}(q, \Theta)$
in \eq{tmd_factorization_fid_spectrum} at $q_T = 0$.
This results in a formula similar to \eq{np_rap_model_combined_formula},
\begin{align} \label{eq:np_fid_model_combined_strict_lp}
&\frac{1}{\pi q_T} \frac{\df \sigma^\lp(\Theta)}{\df q_T}
\\[0.4em]
&= \frac{1}{2\pi} \int_0^\infty \! \df b_T \, b_T \, J_0(b_T q_T) \,
\Bigl[
\sum_{a,b} \int \! \df Q^2 \, \df Y \,
\tsigma_{ab}^{\lp, \pert + \np_\zeta}(Q, Y, \Theta, b_T)
\Bigr]
\overline{F}^\np(\Theta, b_T)
\nn \,,\end{align}
where the leading-power fiducial perturbative cross section in $b_T$ space is defined as
\begin{align} \label{eq:np_fid_model_weight}
\tsigma_{ab}^{\lp, \pert + \np_\zeta}(Q, Y, \Theta, b_T)
&= \frac{3}{8} \sum_{i = -1, 4}
\tsigma_{i,ab}^{\pert + \np_\zeta}(Q, Y, b_T)
\int_{-1}^{1} \! \df \cos \theta \,
g_i(\theta) \,
\hat{\Theta}^\lp(Q, Y, \theta)
\,,\end{align}
with $\hat{\Theta}^\lp(Q, Y, \theta)$
the acceptance function $\hat{\Theta}(q, \theta, \varphi)$
of the fiducial cuts $\Theta$ evaluated for $q_T = 0$,
which implies that the $\varphi$ dependence also drops out~\cite{Ebert:2020dfc}.
The corresponding averaged (effective) nonperturbative model is given by
\begin{align} \label{eq:np_fid_model_average_np_func_def}
\overline{F}^\np(\Theta, b_T)
= \frac{
\sum_{a,b} \int \! \df Y \,
\sigma_{ab}^B(m_V, Y, \Theta) \,
\tf_a^{\pert} \,
\tf_b^{\pert} \,
\tf_a^\np(x_a, b_T) \, \tf_b^\np(x_b, b_T)
}{
\sum_{a,b} \int \! \df Y \,
\sigma_{ab}^B(m_V, Y, \Theta) \,
\tf_a^{\pert} \,
\tf_b^{\pert}
}
\end{align}
where the fiducial Born cross section is defined in terms of \eq{helicity_born_xsec} as
\begin{align} \label{eq:fiducial_born_xsec}
\sigma_{ab}^B(Q, Y, \Theta)
= \frac{3}{8} \sum_{i=-1,4}
\sigma_{i,ab}^B(Q)
\int_{-1}^{1} \! \df \cos \theta \,
g_i(\theta) \,
\hat{\Theta}^\lp(Q, Y, \theta)
\,,\end{align}
the perturbative TMD PDFs are again evaluated at $x_{a,b},b_T, \mu_\init, \zeta_\init$,
and we have made use of $Q \approx m_V$ as before
such that the (non)perturbative CS evolution, hard functions,
and $\mu$ evolution drop out.

Our final predictions for the resummed contribution
to fiducial $q_T$ spectra are then obtained
by inserting this model back
as an overall common factor to all factorized helicity cross sections
into the factorization formula with exact leptonic kinematics,
\begin{align} \label{eq:np_fid_model_combined_formula}
\frac{1}{\pi q_T} \frac{\df\sigma_\res(\Theta)}{\df q_T}
&= \frac{1}{4\Ecm^2} \int \! \df Q^2 \, \df Y
\sum_{i = -1, 4} \sum_{V, V'} L_{i\,VV'}(q, \Theta)
\frac{1}{2\pi} \int_0^\infty \df b_T \, b_T J_0(b_T q_T)
\nn \\ & \qquad \times
\Bigl[ \sum_{a,b} H_{i\,VV'\,ab}(Q^2, \mu) \,
   \tf_a^{\pert + \np_\zeta}(x_a, b_T, \mu, Q^2) \,
   \tf_b^{\pert + \np_\zeta}(x_b, b_T, \mu, Q^2)
\Bigr]
\nn \\ & \qquad \times
\overline{F}^\np(\Theta, b_T)
\,,\end{align}
which also follows e.g.\
from \eqss{np_rap_model_combined_formula}{np_rap_model_explicit_form}{np_rap_model_explicit_Y_dep}
by setting $\Lambda_{-1,n} = \Lambda_{4,n}$ and $\Delta \Lambda_{-1,n} = 0$,
i.e., dropping the $Y$ dependence.
We stress that as for the narrow-width approximation already used above,
cross terms of $\ordsq{\lqcd^2/(q_T Q)}$
between nonperturbative and linear fiducial power corrections
are indeed retained exactly in our final predictions
through the exact kinematic dependence of $L_{i\,VV'}(q, \Theta)$,
and it is only the degree of model (in)dependence
that is qualified by the approximations we make.
In particular, the final formula in \eq{np_fid_model_combined_formula}
still retains all the exact linear power corrections
dressing the perturbative resummation.

In order to illustrate the effect of the nonperturbative TMD boundary condition
on the predictions at the highest perturbative orders
that we present in this paper,
we make the following, simple Gaussian choice for the model function,
\begin{align} \label{eq:np_fid_model_explicit_form}
\overline{F}^\np(\Theta, b_T)
= \exp \bigl( - 2 \Omega_V b_T \bigr)
= 1 - 2 \Omega_V^2 b_T^2 + 2 \Omega_V^4 b_T^4 + \ord{\lqcd^6 b_T^6}
\,,\end{align}
and we include a subscript $V = Z, W^+, W^-$ to remind the reader
that the parameter is decorrelated between bosons
(and their associated fiducial phase-space volumes $\Theta$).
We note that while this is numerically equivalent
to assuming a common Gaussian width $\Omega_V$ for all quark and antiquark flavors,
it is much more rigorously justified
through our effective model approach,
where the existence of the model is well-defined, minimal, and sufficient,
and only in a second step does one pick a specific form for it.
Again, in order to be able to easily showcase the impact
of variations around it, we will take
\begin{align}
\Omega_V = 0.5 \GeV
\end{align}
as our default central value.

\paragraph{Discussion.}

\Eqs{np_rap_model_combined_formula}{np_fid_model_combined_formula}
make precise the intuitive notion
that a one-dimensional $q_T$ (two-dimensional $Y$ and $q_T$) spectrum
contains only information about a single effective
one-dimensional (two-dimensional)
nonperturbative model function of $b_T$ ($Y$ and $b_T$),
independent of the underlying complexity
of the flavor and $x$-dependent TMD PDF nonperturbative boundary conditions.
In particular, upon expansion in $\lqcd \, b_T$ (or $\lqcd/q_T$),
the results imply that only a single number
(or a single one-dimensional function of $Y$)
is required in addition to the first correction to the CS kernel
in order to completely describe the leading nonperturbative effect.
\Eq{np_fid_model_average_np_func_def} generalizes the formula
derived in \refcite{Ebert:2022cku}
for the effective parameter $\overline{\Lambda}^{(2)}$
appearing in the $\ord{\lqcd^2/q_T^2}$ term in the fiducial spectrum
to all powers in $\lqcd$.%
\footnote{
Note that eq.~(4.11) in \refcite{Ebert:2022cku}
misses a factor of $2$ in the denominator).
}
(As a downside of extending to all powers in $\lqcd$, the manifestly linear impact
of the leading nonperturbative term in \refcite{Ebert:2022cku} is lost,
and dedicated numerics like in this paper are required to obtain fit templates
for each set of parameters describing $\overline{F}^\np$.)
Conversely, no additional information about the underlying
TMD nonperturbative structure beyond $\overline{F}_i^\np(Y, b_T)$
or $\overline{F}^\np(\Theta, b_T)$
can be extracted from resonant Drell-Yan spectra.
This formally explains the observations of \refcite{Bacchetta:2024qre},
which found that the combination of Drell-Yan and semi-inclusive
deep-inelastic scattering data for different hadron species
was crucial to achieve flavor separation of the TMDs.
Of course, in the context of a global, flavor-dependent TMD fit
or if a more granular model is desired and e.g.\ explicit
flavor ratios are known from the lattice,
one can always revert
to instead using the product $\tilde{f}_a^\np(x_a, b_T) \, \tilde{f}_b^\np(x_b, b_T)$
on the last line of \eq{np_fid_model_combined_formula}
and pulling it back under the flavor sums
and $(Q, Y)$ integral.

\section{Results for the inclusive and fiducial \texorpdfstring{$p_T^Z$}{pTZ} spectrum}
\label{sec:uncerts}

\begin{figure*}
\includegraphics[width=\WidthTwoSubfigs]{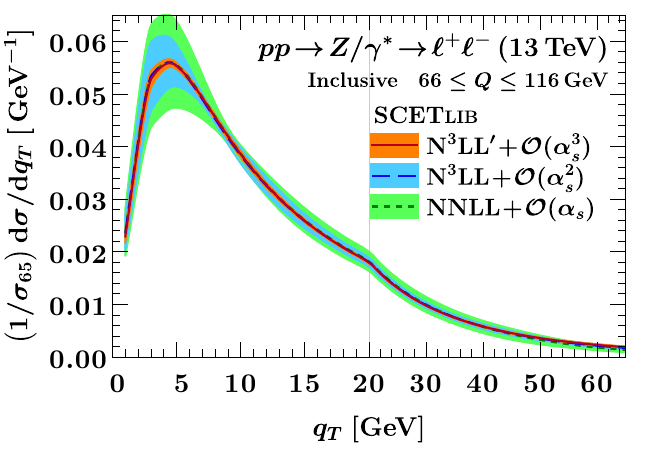}%
\hfill%
\includegraphics[width=\WidthTwoSubfigs]{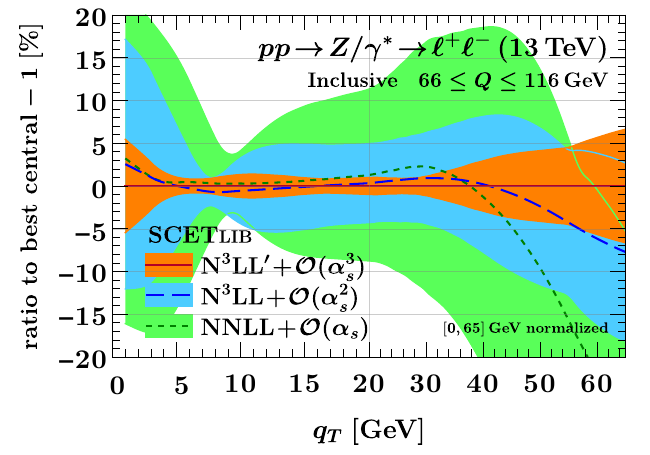}%
\\%
\centering%
\includegraphics[width=\WidthTwoSubfigs]{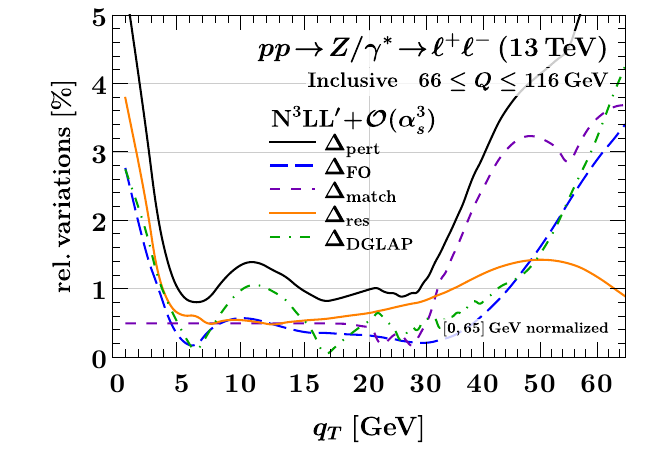}%
\caption{Predictions for the normalized inclusive $p_T^Z$ spectrum at the $13 \TeV$ LHC (top left),
its relative difference to the highest-order prediction (top right),
and the complete perturbative uncertainty breakdown
of the N$^3$LL$^\prime+\ord{\as^3}$ result
in terms of each contributing source (bottom).}
\label{fig:Z_qT_incl_predictions}
\end{figure*}

\begin{table}
\begin{center}
\begin{tabular}{c | c | c | c}
& NNLL$+\ord{\as}$ & N$^3$LL$+\ord{\as^2}$ & N$^3$LL$^\prime\!+\!\ord{\as^3}$
\\
\hline
$\sigma_{65}~[\mathrm{pb}]$ & 1813.5 & 1806.9 & 1787.3
\end{tabular}
\caption{Normalization factors for the inclusive $p_T^Z$ spectra with \texttt{MSHT20nnlo}
shown in \fig{Z_qT_incl_predictions}.}
\label{tab:Z_qT_incl_predictions_norm}
\end{center}
\end{table}

\begin{figure*}
\includegraphics[width=\WidthTwoSubfigs]{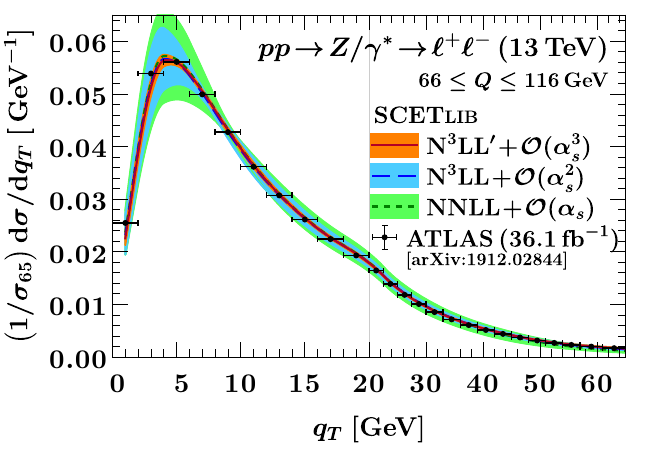}%
\hfill%
\includegraphics[width=\WidthTwoSubfigs]{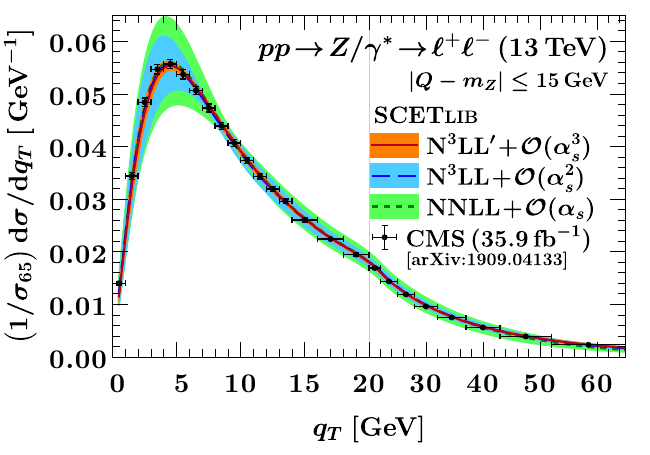}%
\\%
\includegraphics[width=\WidthTwoSubfigs]{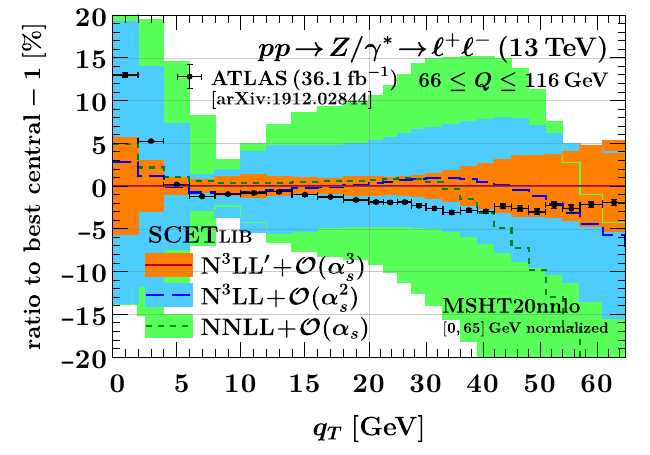}%
\hfill%
\includegraphics[width=\WidthTwoSubfigs]{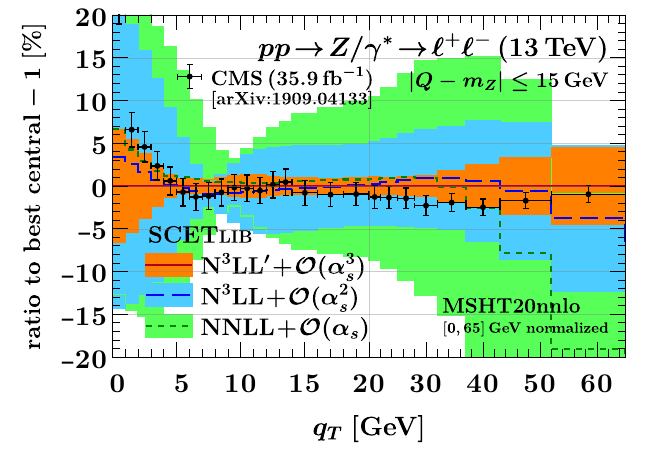}%
\\%
\includegraphics[width=\WidthTwoSubfigs]{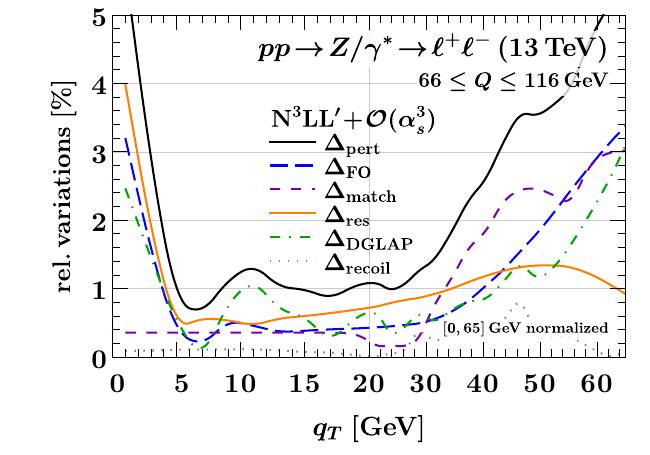}%
\hfill%
\includegraphics[width=\WidthTwoSubfigs]{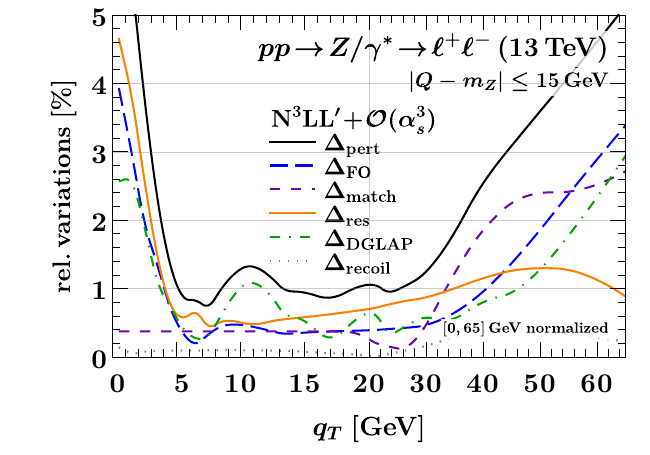}%
\caption{Predictions for the normalized fiducial $p_T^Z$ spectrum
at the $13 \TeV$ LHC (top row),
its relative difference to the highest-order prediction (middle row),
and the complete perturbative uncertainty breakdown
of the N$^3$LL$^\prime+\ord{\as^3}$ result
in terms of each contributing source (bottom row).
In the left (right) column, we compare to the ATLAS $13\TeV$~\cite{Aad:2019wmn}
(CMS $13\TeV$~\cite{Sirunyan:2019bzr}) measurement,
using the respective set of fiducial cuts.
}
\label{fig:Z_qT_fid_predictions}
\end{figure*}

\begin{table}
\begin{center}
\begin{tabular}{c | c | c | c}& NNLL$+\ord{\as}$ & N$^3$LL$+\ord{\as^2}$ & N$^3$LL$^\prime\!+\!\ord{\as^3}$
\\
\hline
$\sigma_{65}~[\mathrm{pb}]$ \quad \multrow{ATLAS cuts~\cite{Aad:2019wmn} \\ CMS cuts~\cite{Sirunyan:2019bzr}} & \multrow{682.1 \\ 673.1} & \multrow{682.6 \\ 673.9} & \multrow{673.4\\ 664.9}
\end{tabular}
\caption{Normalization factors for the fiducial $p_T^Z$ spectra with \texttt{MSHT20nnlo} shown in \fig{Z_qT_fid_predictions}.}
\label{tab:norm_factors_fid_pTZ}
\label{tab:Z_qT_fid_predictions_norm}
\end{center}
\end{table}

In \figs{Z_qT_incl_predictions}{Z_qT_fid_predictions}
we present our predictions for the inclusive
and fiducial $Z$-boson transverse-momentum spectrum
at different orders in resummed and matched perturbation theory
up to N$^3$LL$'$$+$$\ord{\as^3}$.
We use the \texttt{MSHT20nnlo} PDF set~\cite{Bailey:2020ooq} as our default
with the attendant value of $\as(m_Z) = 0.118$.
We consider two sets of fiducial cuts on the dilepton pair,
as used in the ATLAS~\cite{Aad:2019wmn}
and CMS~\cite{Sirunyan:2019bzr} $13 \TeV$ measurements, respectively,
\begin{alignat}{5} \label{eq:fiducial_cuts}
&\text{ATLAS cuts:} \qquad
&& p_T^\ell > 27 \GeV
\,, \qquad
&& \abs{\eta_\ell} < 2.5
\nn \\
&\text{CMS cuts:} \qquad
&& p_T^\ell > 25 \GeV
\,, \qquad
&& \abs{\eta_\ell} < 2.4
\,,\end{alignat}
which are applied to both the lepton and the antilepton.
For \refcite{Aad:2019wmn}, we compare to the result obtained for Born leptons
after combining the electron and muon channels.
For \refcite{Sirunyan:2019bzr},
we compare to the result obtained using dressed leptons,
again from a combination of electron and muon channels.%
\footnote{
We note that at least as of \texttt{version 2},
the \texttt{HEPData} entry for \refcite{Sirunyan:2019bzr}
had an incorrect total integral ($\neq 1$) for the normalized $p_T^Z$ spectrum,
while relative uncertainties in each bin were correct.
The correct result is obtained by dividing by the total integral once more.
We thank Aram Apyan for help with this issue.
}

Our electroweak inputs are as follows, see \refcite{Ebert:2020dfc} for details,
\begin{alignat}{5}
m_Z &= 91.1535\GeV
\,, \qquad
\Gamma_Z = 2.4943 \GeV
\,,\nn\\
m_W &= 80.3580 \GeV
\,,\qquad
\Gamma_W = 2.0843 \GeV
\,, \\[0.4em]
G_F &= 1.1663787 \times 10^{-5} \GeV^{-2}
\,,\nn\\[0.4em]
V_{\rm CKM}
&= \begin{pmatrix}
    V_{ud} \quad V_{us} \quad V_{ub} \\
    V_{cd} \quad V_{cs} \quad V_{cb} \\
    V_{td} \quad V_{ts} \quad V_{tb} \\
   \end{pmatrix}
 = \begin{pmatrix}
    0.97446 \quad 0.22452 \quad 0.00365 \\
    0.22438 \quad 0.97359 \quad 0.04214 \\
  ~0.00896 \quad 0.04133 \quad 0.999105 \\
   \end{pmatrix}
\,,\nn\\[0.4em]
\sin^2\theta_w
&= 0.22284
\,,\qquad
\aem
= \frac{1}{132.357}
\,.\end{alignat}
The numerical treatment of the oscillatory Bessel integral
and our semi-analytic evaluation of the leptonic phase-space integrals
are also described in section~(4.1) of \refcite{Ebert:2020dfc}.

To easily study the uncertainties in the $p_T^Z$ spectrum
and compare to the (most often normalized) experimental data,
we find it convenient to normalize the spectrum
on the range $0\leq q_T\leq 65\GeV$,%
\footnote{
Experimental data are re-normalized on this range
by dividing by the central value of the norm
and maintaining the relative uncertainties in each bin.
We ignore possible correlations between individual bins and the norm
that are negligible compared to the original relative uncertainties
in each (narrow) bin.
}
where we empirically found that nonsingular contributions are small.
For reference, the total integrals of our predictions
as used for the normalization,
which we denote by $\sigma_{65}$,
are reported in \tabs{Z_qT_incl_predictions_norm}{Z_qT_fid_predictions_norm}.
From the data comparisons in \fig{Z_qT_fid_predictions}
we observe that our predictions
with the default PDF set and default nonperturbative parameters
closely track the fiducial ATLAS~\cite{Aad:2019wmn} and CMS~\cite{Sirunyan:2019bzr} data
over a wide range of $q_T$,
but residual differences are well visible.
In the following subsections,
we assess the impact of various sources of uncertainties
and different physical effects on our predictions.

\subsection{Scale variations, convergence, and estimated matching uncertainties}
\label{sec:uncerts_pert}

Predictions in either fixed-order or resummation-improved perturbation theory
are necessarily subject to an uncertainty from missing higher-order terms.
A commonly used approach to assess these
is to vary the scales at which various objects appearing
in the prediction are renormalized.
Since the dependence on any individual scale
is beyond the working order in a factorized cross section,
these variations can indicate the typical size of missing higher-order terms.
While they do not offer any handle on the correlations
between perturbative uncertainties e.g.\ between different bins,
scale variations nevertheless can serve, and are heavily used,
as a crude estimate of the truncation uncertainty
based on the available lower-order data.
For this reason, we do present estimates of the perturbative uncertainty
within the scale variation paradigm in this section.
(For an alternative that overcomes all the shortcomings,
see \refscite{Tackmann:2024xxx, Tackmann:2024kci}.)
To do so in the most reliable and granular fashion possible
within the conventional paradigm,
we refine the profile scale variation setup of \refcite{Ebert:2020dfc},
which in turn was based on \refcite{Stewart:2013faa}.
We here give a self-contained description
of the profile scale variations we perform,
commenting on the improvements and differences
to \refcite{Ebert:2020dfc} along the way.
The profile scales and variations originally designed
in the course of preparing the present manuscript
have recently been applied already in \refcite{Cal:2023mib},
where the ability to separately freeze out the PDF scale $\mu_f$
proved particularly useful for heavy flavor-induced processes like $b \bar{b} \to H$.

For the resummed cross section, the task is to vary
the central boundary scales given in \eq{profile_scales_central}
around their central values by (up to) the conventional factor of $2$
in all possible ways
such that the variations (a)~do not induce undue sensitivity to the Landau pole
and (b)~smoothly transition into a set of conventional fixed-order scale variations
in the tail at $q_T \sim Q$.
We achieve this task by varying them as
\begin{align} \label{eq:profile_scales_variations}
\mu_H &= \mu_\FO^\central \, 2^{w_\FO}
\,, \nn \\[0.4em]
\mu_B &= \mu_\FO^\central \, 2^{w_\FO} \, f_\vary^{v_{\mu_B}} \, f_\run\biggl(
   \frac{q_T}{Q},
   \frac{1}{Q}\mu_*\Bigl(\frac{b_0}{b_T}, \frac{\muBmin}{2^{w_\FO} \, f_\vary^{v_{\mu_B}}} \Bigr)
\biggr)
\,, \nn \\[0.4em]
\nu_B &= \mu_\FO^\central \, 2^{w_\FO} \, f_\vary^{v_{\nu_B}}
\,, \nn \\[0.4em]
\mu_S &= \mu_\FO^\central \, 2^{w_\FO} \, f_\vary^{v_{\mu_S}} \, f_\run\biggl(
   \frac{q_T}{Q},
   \frac{1}{Q}\mu_*\Bigl(\frac{b_0}{b_T}, \frac{\muSmin}{2^{w_\FO} \, f_\vary^{v_{\mu_S}}} \Bigr)
\biggr)
\,, \nn \\
\nu_S &= \mu_\FO^\central \, 2^{w_\FO} \, f_\vary^{v_{\nu_S}} \, f_\run\biggl(
   \frac{q_T}{Q},
   \frac{1}{Q}\mu_*\Bigl(\frac{b_0}{b_T}, \frac{\nuSmin}{2^{w_\FO} \, f_\vary^{v_{\nu_S}}} \Bigr)
\biggr)
\,, \nn \\[0.4em]
\mu_f &= \mu_F^\central \, 2^{v_{\mu_f}} f_\run\biggl(
   \frac{q_T}{Q},
   \frac{1}{Q}\mu_*\Bigl(\frac{b_0}{b_T}, \frac{\mufmin}{2^{v_{\mu_f}}} \Bigr)
\biggr)
\,,\end{align}
where $\mu_\FO^\central = \mu_F^\central = Q$
are the central fixed-order renormalization
and factorization scales.
The central scales are recovered for
\begin{align}
v_{\mu_B} = v_{\nu_B} = v_{\mu_S} = v_{\nu_S} = w_{\FO} = v_{\mu_f} = 0
\,.\end{align}
The boundary scale $\mu_0$ of the rapidity anomalous dimension
is held fixed at its central value since its variations
would be double counted with the explicit $\nu_{B,S}$ variations~\cite{Stewart:2013faa}.
Here $f_\vary \equiv f_\vary(q_T/Q)$ governs the strength of
those variations that must turn off as $q_T/Q \to 1$, $f_\vary \to 1$,
where we choose
\begin{align} \label{eq:f_vary_def}
f_\vary(x) = \begin{cases}
2(1 - x^2/x_3^2)
\,, &
0 \leq x < x_3/2
\,, \\
1 + 2(1 - x/x_3)^2 \
\,, &
x_3/2 \leq x < x_3
\,, \\
1
\,, &
x_3 \leq x
\,.\end{cases}
\end{align}
For $q_T \geq x_3 Q$, the above variations thus all reduce to
\begin{align}
\mu_B = \nu_B = \mu_S = \nu_S = \mu_\FO^\central \, 2^{w_\FO}
\,, \qquad
\mu_f = \mu_F^\central \, 2^{v_{\mu_f}}
\,,\end{align}
which exactly matches the variations
we perform on the renormalization and factorization scale
in the nonsingular cross section (and thus on the whole prediction
in the far tail),
\begin{align} \label{eq:scale_variations_nons}
\mu_R = \mu_\FO^\central \, 2^{w_\FO}
\,, \qquad
\mu_F = \mu_F^\central \, 2^{v_{\mu_f}}
\,.\end{align}
We note the appearance of factors like $2^{w_\FO}$, $2^{v_{\mu_f}}$,
or $f_\vary^{v_{\mu_X}}$ in the argument of the $\mu_*$ functions
in \eq{profile_scales_variations}.
These ensure that scale variations are also frozen out
at long distances as $b_T \to 1/\lqcd$,
which in turns ensures that no scales $\lesssim 1 \GeV$
are probed by the variations, see the bands in the left panel of \fig{profiles}.
Like the addition of the minimum scales implemented through the $\mu_*$
functions themselves, this change compared to \refcite{Ebert:2020dfc}
is necessary because unlike \refcite{Ebert:2020dfc} (which simply froze out the coupling and PDFs),
we aim to maintain a straightforward OPE interpretation
of the nonperturbative functions we introduced in \sec{np_model}.

We now describe how the above variation flags are grouped
into subsets of joint and individual variations that each estimate
a distinct source of uncertainty,
as shown in the bottom rows of \figs{Z_qT_incl_predictions}{Z_qT_fid_predictions}.

\paragraph{Resummation uncertainty $\Delta_\res$.}

A first set of variations concerns those independent scales that emerge
in the canonical resummation region $q_T \ll Q$, and estimates
the perturbative uncertainty on (mainly) the shape of
the resummed Sudakov spectrum.
In \eq{profile_scales_variations},
they are controlled by $v_{\mu_B}$, $v_{\nu_B}$, $v_{\mu_S}$, $v_{\nu_S}$,
which we vary as
\begin{align} \label{eq:def_v_res}
(v_{\mu_B}, v_{\nu_B}, v_{\mu_S}, v_{\nu_S})
\in V_\res = \bigl\{&
(\uparrow,-,-,-),
(-,\uparrow,-,-),
(-,-,\uparrow,-),
(-,-,-,\uparrow),
\nn \\ &
(-,-,-,\downarrow),
(-,-,\downarrow,-),
(-,\downarrow,-,-),
(\downarrow,-,-,-),
\nn \\ &
(\uparrow,\uparrow,-,-),
(\uparrow,-,\uparrow,-),
(\uparrow,-,-,\uparrow),
(\uparrow,-,-,\downarrow),
(\uparrow,\downarrow,-,-),
\nn \\ &
(-,\uparrow,\uparrow,-),
(-,\uparrow,-,\uparrow),
(-,\uparrow,\downarrow,-),
(-,-,\uparrow,\uparrow),
\nn \\ &
(-,-,\downarrow,\downarrow),
(-,\downarrow,\uparrow,-),
(-,\downarrow,-,\downarrow),
(-,\downarrow,\downarrow,-),
(\downarrow,-,-,\uparrow),
\nn \\ &
(\downarrow,-,\downarrow,-),
(\downarrow,\downarrow,-,-),
(\uparrow,\uparrow,\uparrow,-),
(\uparrow,\uparrow,-,\uparrow),
(\uparrow,-,\uparrow,\uparrow),
\nn \\ &
(\uparrow,\downarrow,\uparrow,-),
(\uparrow,\downarrow,-,\downarrow),
(-,\uparrow,\uparrow,\uparrow),
(-,\downarrow,\downarrow,\downarrow),
(\downarrow,\uparrow,-,\uparrow),
\nn \\ &
(\downarrow,\downarrow,-,\downarrow),
(\downarrow,\downarrow,\downarrow,-),
(\uparrow,\uparrow,\uparrow,\uparrow),
(\downarrow,\downarrow,\downarrow,\downarrow)
\bigr\}
\,,\end{align}
where we write $\uparrow$ ($\downarrow$, $-$) for $v = +1$ ($v = -1$, $v = 0$) for brevity.
The set of variations in \eq{def_v_res} is defined
by considering all $3^4 - 1 = 80$ possible variations of the four $v$,
and removing all the ones where the argument of a logarithm exponentiated
by the renormalization group evolution between terms
would be varied by a factor of four~\cite{Stewart:2013faa}.
The final uncertainty estimate is obtained
by taking the symmetrized envelope of all tuples $\vec{v}$ of variations above,
\begin{align}
\Delta_\res = \max_{\vec{v} \in V_\res} \Abs{\df \sigma_{\vec{v}} - \df \sigma_\central}
\,.\end{align}
It has been noted in \refscite{Stewart:2013faa, Ebert:2020dfc}
that while the set of variations in \eq{def_v_res} appears large,
keeping all combinations under the envelope
is often important to prevent accidental underestimates
when several scale variations at once cross the central value
at some point in $q_T$.
For definiteness, we take each individual variation to be fully
correlated between each bin and the total integral
when normalizing the spectrum.
This effect on the total integral
is responsible for the nonzero (but subdominant) contribution of $\Delta_\res$
also at large $q_T \geq 60 \GeV$ in \figs{Z_qT_incl_predictions}{Z_qT_fid_predictions}
where the underlying variations are largely turned off already
at the level of the unnormalized spectrum.
We note at this point that while \eq{tmd_factorization_bbs}
with separately finite soft and beam functions
can easily be translated into suitably evolved TMD PDFs
for central scales, see \eq{tmd_pdf_evolved_pert_np_zeta_as_used_in_practice},
the rapidity renormalization group does provide
a larger set of possible nontrivial scale variations,
which we here make use of for our estimate of the perturbative uncertainties.

\paragraph{Fixed-order uncertainty $\Delta_\FO$.}

A second set of variations, which affects the spectrum everywhere
and can be thought of as (mainly) varying its normalization,
concerns the overall choice of fixed-order renormalization scale
entering in particular the strong coupling,
\begin{align}
\mu_\FO = \mu_\FO^\central \, 2^{w_\FO}
\,,\end{align}
which we vary as follows, taking the uncertainty estimate to be the symmetrized envelope,
\begin{align}
w_\FO \in V_\FO = \{ +1, -1 \}
\,, \qquad
\Delta_\FO = \max_{w_\FO  \in V_\FO} \Abs{\df \sigma_{w_\FO } - \df \sigma_\central}
\,.\end{align}
Variations of $w_\FO$,
which we take to be fully correlated when normalizing the spectrum,
are consistently propagated into the resummed and nonsingular cross sections
through \eqs{profile_scales_variations}{scale_variations_nons},
which involve factors of $2^{w_\FO}$
in front of all renormalization-like scales.
Compared to the most common approaches
to estimate the total truncation uncertainty
in fixed-order perturbation theory for hadronic processes,
we yet have to include an uncertainty component
from variations of the \emph{factorization scale},
which we turn to next.

\paragraph{DGLAP (factorization scale) uncertainty $\Delta_\dglap$.}
While it is challenging in general to identify common underlying sources
of uncertainty for different regions of the spectrum,
one source can in fact be readily identified
as the truncation uncertainty in the DGLAP evolution
of the twist-2 collinear PDF that feature as a common ingredient everywhere.
To estimate it, we perform a common variation
of the scales entering the PDFs everywhere in the prediction as follows,
\begin{align} \label{eq:def_delta_dglap}
v_{\mu_f} \in V_\dglap = \{ +1, -1 \}
\,, \qquad
\Delta_\dglap = \max_{v_{\mu_f}  \in V_\dglap} \Abs{\df \sigma_{v_{\mu_f} } - \df \sigma_\central}
\,.\end{align}
Variations of $v_{\mu_f}$,
which we take to be fully correlated when normalizing the spectrum,
are again consistently propagated into the resummed and nonsingular cross sections
through \eqs{profile_scales_variations}{scale_variations_nons}.
We recall that in the resummed cross section,
only the beam function matching coefficient is evaluated at $\mu_B$,
while the underlying PDF is evaluated (in general) at $\mu_f \neq \mu_B$
using our dedicated implementation of \eq{ope_beam_muf_neq_mu},
which precisely induces a set of compensating logarithmic terms in $\mu_f/\mu_B$
whose difference to the central value probes the residual dependence on $\mu_f$.
Note that in contrast to $\mu_B$, whose variations
are damped by $f_\vary$ at large $q_T$ in \eq{profile_scales_variations},
the variations of $\mu_f$ stay fully turned on all the way into the tail,
as is required to recover a standard $\mu_F$ variation
of the fixed-order result in that region.
Compared to \refscite{Stewart:2013faa, Ebert:2020dfc},
having a dedicated uncertainty component $\Delta_\dglap$
is new for a resummed and matched prediction;
both references effectively only considered common diagonal variations
of $\mu_\FO$ and $\mu_F^\central$, which then were inherited
by the respective $\mu_B = \mu_f$ in the resummed cross section
when computing the total $\Delta_\FO$ quoted in those references.
We would also like to point out
that in our present setup, $\mu_R$ and $\mu_F$ variations
are effectively combined in quadrature in the final perturbative uncertainty estimate
also in the fixed-order region, see \eq{def_delta_pert}.
In our case, this choice of separating out $\Delta_\dglap$
(rather than moving it under a common envelope)
is motivated by the special role of PDF evolution
affecting the spectrum everywhere.
We point out that the mildly oscillatory behavior of $\Delta_\dglap$
in \figs{Z_qT_incl_predictions}{Z_qT_fid_predictions}
is induced by the discontinuity of the PDF at the $m_b$ threshold,
which in turn spreads over the entire $q_T$ range by the inverse $b_T$-integral
with a period of $\Delta q_T \sim m_b$.

\paragraph{Matching uncertainty $\Delta_\match$.}

The profile scales that implement
the transition from the canonical to the fixed-order region
involve three transition points $x_i$
that determine the onset, midpoint, and endpoint of the transition.
While the values of these transition points are based
on a quantitative assessment of the size of nonsingular terms~\cite{Ebert:2020dfc},
they are not uniquely determined by this criterion,
and the associated uncertainty must be assessed.
As in \refscite{Stewart:2013faa, Ebert:2020dfc},
we thus compute a matching uncertainty $\Delta_\match$ as follows,
\begin{align} \label{eq:def_v_match}
(x_1, x_2, x_3)
&\in V_\match
= \{
   (0.4, 0.75, 1.1),
   (0.2, 0.45, 0.7),
   (0.4, 0.55, 0.7),
   (0.2, 0.65, 1.1)
\}
\,, \nn \\
\Delta_\match &= \max_{\vec{x}  \in V_\match} \Abs{\sigma_{\vec{x}} - \sigma_\central}
\,,\end{align}
where we take the variations to be fully correlated when normalizing the spectrum.
At the level of the unnormalized spectrum,
the matching uncertainty vanishes for $q_T \leq x_1^\mathrm{min} Q$
and $q_T \geq x_3^\mathrm{max} Q$,
where $x_1^\mathrm{min}$ and $x_3^\mathrm{max}$ are the minimum
or maximum value taken by $x_1$ and $x_3$, respectively.
The $0.5\%$ effect visible at small $q_T$ in the normalized spectra
in \figs{Z_qT_incl_predictions}{Z_qT_fid_predictions}, on the other hand,
is precisely due to the residual effect
of the matching variations on the total integral,
which is not guaranteed to be preserved by either the central $x_i$
or their variations in \eq{def_v_match}.
We note that if a symmetric impact of variations entering
the matching uncertainty is desired, an alternative scheme
is to simply vary the midpoint $x_2$ of the profile
up and down by a suitable amount~\cite{Cal:2023mib}.

\paragraph{Recoil uncertainty $\Delta_\recoil$.}

A final contribution to our total perturbative uncertainty estimate
is not related to any specific scale choice,
but -- like the matching uncertainty -- still concerns
the split between resummed and fixed-order terms.

Specifically, when decomposing the fiducial $q_T$ spectrum
(or any observable sensitive to the differential lepton-antilepton distribution)
as in \eq{tmd_factorization_fid_spectrum},
a choice is made in which frame to decompose
the hadronic tensor in terms of helicity cross sections,
where subsequently the leading-power ones are resummed to all orders
while dressing them with the exact lepton kinematics
to ensure the exact treatment of linear or leptonic (``fiducial'') power corrections.
Making these choices explicit, we compute the resummed and fixed-order singular
cross sections using the following formula,
\begin{align}
\frac{1}{\pi q_T} \frac{\df\sigma_X(\Theta)}{\df q_T}
&=  \frac{1}{2\Ecm^2} \int \! \df Q^2 \, \df Y
\sum_{i = -1,4} \sum_{V, V'} L_{i\,VV'}^X(q, \Theta)\,W_{i\,VV'}^{\lp,X}(q, P_a, P_b)
\,,\end{align}
where $X$, affecting both the leptonic prefactors and the structure functions,
indicates the choice of frame, and $X = \mathrm{CS}$ (the Collins-Soper frame)
for our central prediction.
While these frame choices are subject to the constraint
that their $z^\mu$ axes coincide with the beam axis as $q_T \to 0$,
this does not uniquely specify them,
and different choices for the structure function decomposition
(i.e., the choice of $Z$ rest frame, i.e., the choice of a specific recoil prescription)
differ by moving a set of terms of $\ord{q_T^2/Q^2}$~\cite{Ebert:2020dfc}
between the leading-power and subleading-power structure functions.
This means that as soon as the leading-power resummation is performed,
they will differ by whether these terms are resummed to all orders or evaluated at fixed order.
In order to assess the potential presence of an ambiguity due to this choice,
we here perform -- to our knowledge, for the first time --
an explicit variation of the choice of structure function decomposition
in a resummed and matched prediction.%
\footnote{A distinct issue is how the choice of frame
for computing the singular cross section affects the efficiency
of differential $q_T$ subtractions, see \refcite{Ebert:2020dfc}.
The two issues are related because they both hinge
on the size of the remaining fixed-order nonsingular,
but our study here concerns the potential scheme dependence
of fiducial resummation effects
that remain present even after taking all technical cutoffs to zero.
}
Specifically, we consider the two extreme cases
\begin{align} \label{eq:def_v_recoil}
X
\in V_\recoil
= \{
   \mathrm{GJ}
   \,,
   \overline{\mathrm{GJ}}
\}
\,.\end{align}
Here the decomposition is performed in either the Gottfried-Jackson ($X = \mathrm{GJ}$)
frame~\cite{Lam:1978pu}, where the $z^\mu$ axis is aligned with the hadron incoming
along the positive $z^\mu_\mathrm{lab}$ direction,
or performed in the ``anti-Gottfried Jackson'' frame ($X = \overline{\mathrm{GJ}}$)
where the opposite hadron is chosen as reference;
our default value $X = \mathrm{CS}$ amounts to a certain symmetric choice
between the two that has compact covariant expressions~\cite{Ebert:2020dfc}
and a particularly simple power expansion~\cite{Gao:2024xxx}.
The associated uncertainty is then computed by taking the envelope,
\begin{align} \label{eq:def_delta_recoil}
\Delta_\match = \max_{X  \in V_\recoil} \Abs{\df \sigma_{X} - \df \sigma_\central}
\,.\end{align}
Like e.g.\ the matching uncertainty,
$\Delta_\recoil$ vanishes by construction at the level of the unnormalized spectrum
since it amounts to reexpanding certain $\ord{q_T^2/Q^2}$ terms.
We nevertheless point out that for the normalized fiducial spectra
shown in \fig{Z_qT_fid_predictions},
its impact instead again tends to a constant as $q_T \to 0$
due to its residual effect on the total matched fiducial cross section.
From this observation, we find it interesting to note that this uncertainty
on the total fiducial cross section is in fact under excellent control,
where we may read off its relative size from the uncertainty breakdown plots
in \fig{Z_qT_fid_predictions} at $q_T \to 0$ as roughly a permille.

\paragraph{Combined perturbative uncertainty $\Delta_\pert$.}

The final perturbative uncertainty estimate $\Delta_\pert$,
which provides the bands at different orders in the ratio plots
in \figs{Z_qT_incl_predictions}{Z_qT_fid_predictions}
and is indicated by a solid black line in the uncertainty breakdown plots,
is computed by adding the uncertainties from all individual sources in quadrature,
\begin{align} \label{eq:def_delta_pert}
\Delta_\pert = \Delta_\res \oplus \Delta_\FO \oplus \Delta_\dglap \oplus \Delta_\match \oplus \Delta_\recoil
\,.\end{align}
As can be seen from the top right panel
in \fig{Z_qT_incl_predictions}
and the center row of panels in \fig{Z_qT_fid_predictions},
where we show ratios of predictions at different orders to the highest order,
the resulting uncertainty estimate features
excellent perturbative coverage,
i.e., higher orders are well contained within the lower-order uncertainty estimates.
We note that while the convergence at the level of the central value
seems to be more rapid than the uncertainty estimates would allow for,
we see no justification for reducing the estimate after the fact.

\subsection{Impact of nonperturbative TMD physics}
\label{sec:uncerts_np}

\begin{figure*}
\centering
\includegraphics[width=\WidthTwoSubfigs]{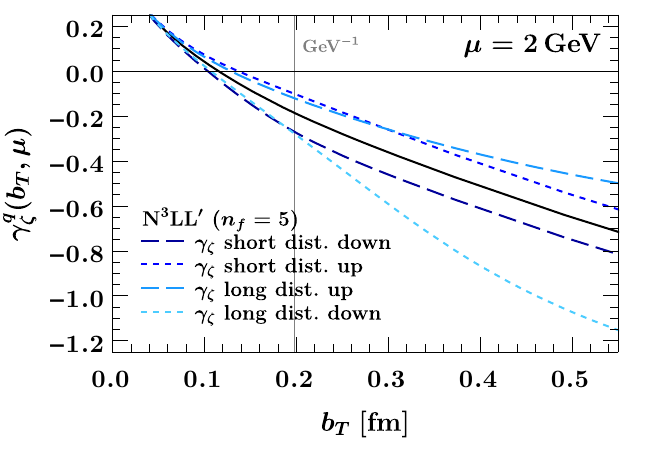}%
\hfill%
\includegraphics[width=\WidthTwoSubfigs]{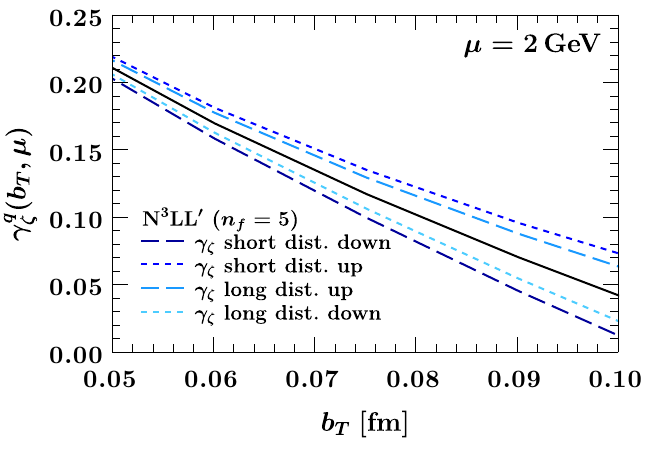}%
\caption{
Illustration of the model parameter spread we consider for the Collins-Soper kernel,
i.e., the rapidity anomalous dimension.
The black solid line indicates our default choice of model parameters.
The gray vertical line indicates a distance
of $b_T = 1 \GeV^{-1} \approx 0.197 \fm$ for reference.
The right panel shows a close-up in the small-$b_T$ region.
}
\label{fig:gamma_nu_np_model}
\end{figure*}

\begin{figure*}
\includegraphics[width=\WidthTwoSubfigs]{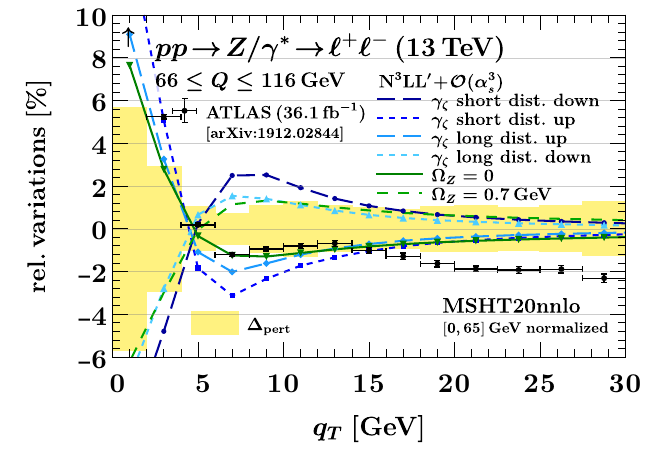}%
\hfill%
\includegraphics[width=\WidthTwoSubfigs]{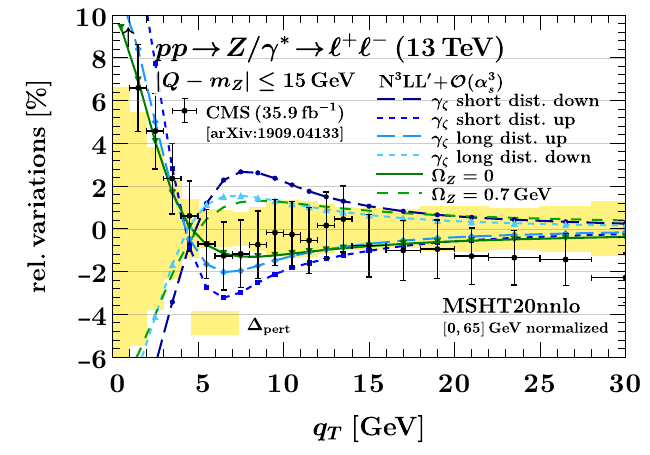}%
\caption{
Impact of parameter variations
in the nonperturbative TMD models at N$^3$LL$^\prime+\ord{\as^3}$
compared to ATLAS $13\TeV$~\cite{Aad:2019wmn} (left) and CMS $13\TeV$~\cite{Sirunyan:2019bzr} (right) measurements.
For illustration, we also compare to the estimated size $\Delta_\pert$ of the perturbative uncertainty.
}
\label{fig:Z_qT_np_uncertainties}
\end{figure*}

We next consider the impact
of the nonperturbative model functions introduced in \sec{np_model}
on our predictions.
We do so by performing illustrative variations of the model parameters
to determine which regions in the spectrum
are affected by nonperturbative physics, and to which extent.
While this already helps to assess whether residual differences to the data
may be of nonperturbative origin,
we leave a dedicated fit of the nonperturbative contributions to future work.

We vary the parameters entering the model for the Collins-Soper kernel
(i.e., the rapidity anomalous dimension)
in \eq{tmd_rapidity_anom_dim_explicit_np_model} as follows,
\begin{align} \label{eq:gamma_nu_np_variations}
(c_\nu^q, \omega_{\nu,q}) \in \{ &
   (-0.15, 0.433 \GeV),
   (0.05,  0.25 \GeV),
\nn \\ &
   (0.5,   0.15 \GeV),
   (-0.5,  0.37 \GeV)
\}
\,.\end{align}
The result for the total CS kernel at N$^3$LL$'$
(i.e., retaining the complete three-loop boundary condition
and four-loop cusp evolution)
is shown in \fig{gamma_nu_np_model}
with the default set of model parameters
$(c_\nu^q, \omega_{\nu,q}) = (-0.05, 0.25 \GeV)$
shown as a solid black line.
The first two variations (indicated by long and short-dashed dark blue lines)
are chosen such that they predominantly affect
the CS kernel at short transverse distances $b_T \lesssim 0.197 \fm \sim 1 \GeV^{-1}$
by varying the quadratic coefficient in the OPE down or up,
as illustrated in the right panel of \fig{gamma_nu_np_model}.
Since we expect the impact of the quadratic coefficient to be linear
at $q_T \gtrsim 5 \GeV$,
we choose the magnitude of $\w_{\nu,q}$ and the sign of $c_\nu^q$
such that the variations they induce on the leading quadratic coefficient
$\operatorname{sgn}(c_\nu^q) \, \omega_{\nu,q}^2 \in \{ -0.1875 \GeV^2, +0.0625 \GeV^2\}$
are symmetric around the central value of $-0.0625 \GeV^2$.

By contrast, the third and fourth variation in \eq{gamma_nu_np_variations}
are chosen such that they only change the quadratic coefficient in the CS kernel
by a smaller amount, and instead predominantly
affect the behavior of the CS kernel at long distances
by varying the sign $\operatorname{sgn}(c_\nu^q)$
and height $\abs{c_\nu^q}$ of the plateau in the model function,
as illustrated in the left panel of \fig{gamma_nu_np_model}.
Here we choose the magnitude of the variation
such that it covers the spread of available lattice results
as reviewed in \refcite{Shanahan:2021tst}.%
\footnote{While high-precision lattice results at physical pion masses
have recently become available~\cite{Avkhadiev:2024mgd},
we leave it to future work to incorporate them directly in our prediction.
Doing so in particular requires one
to include and smoothly match the known perturbative bottom quark mass effects
in the CS kernel~\cite{Pietrulewicz:2017gxc}
in order to correctly transition from the $n_f = 5$ massless limit
to the precision results of \refcite{Avkhadiev:2024mgd},
which feature three light and one massive (charm) flavor.}

For the effective nonperturbative one-parameter model
for the TMD boundary conditions in \eq{np_fid_model_explicit_form}
we choose the following illustrative variations,
\begin{align}
\Omega_Z \in \{ 0, 0.707 \GeV\}
\,.\end{align}
Compared to our default central value of $\Omega_Z = 0.5 \GeV$,
these again amount to a symmetric variation of the leading quadratic coefficient
$\Omega_Z^2 \in \{ 0, 0.5 \GeV^2 \}$ around the central value of $0.25 \GeV^2$.
We thus again expect the impact of these variations
to be linear for $q_T \gtrsim 5 \GeV$~\cite{Ebert:2022cku}.

Our results for the impact of the above nonperturbative parameter variations
on the fiducial $p_T^Z$ spectrum N$^3$LL$^\prime+\ord{\as^3}$
are shown in \fig{Z_qT_np_uncertainties}.
We indeed observe the expected symmetric opposite-sign impact
of the short-distance down/up variations in the CS kernel
and the variations of $\Omega_Z$, respectively, at $q_T \gtrsim 5 \GeV$.
We furthermore observe that the impact of the short-distance variations
of the CS kernel dominates over the long-distance variation
for all bins at the LHC (including the very first one, which we clipped for readability),
suggesting that the strong Sudakov suppression of the $b_T$-space cross section
for resonant Drell-Yan makes it challenging to access the genuine long-distance behavior.
We also note that the impact of the nonperturbative parameter variations,
which we chose for illustration,
is comparable to the estimated total perturbative uncertainty $\Delta_\pert$,
which is shown as the yellow band in \fig{Z_qT_np_uncertainties} for reference.
It is interesting to note that the corresponding variations
in the underlying model functions, see e.g.\ \fig{gamma_nu_np_model},
exceed the typical uncertainties obtained in TMD global fits~\cite{Moos:2023yfa, Bacchetta:2024qre},
which however do not account for perturbative uncertainties in the fit.

Comparing to the experimental data,
it seems likely -- based on the first few bins --
that the data in fact prefer weaker
nonperturbative effects than our default choices
(e.g.\ $\Omega_Z = 0$, solid green).
However, the fact that the prediction for our default PDF set overshoots
the data at $q_T\gtrsim 20\GeV$
cannot be addressed by the nonperturbative model
since this region lies well outside of its effective range,
i.e., because of the fact that the leading nonperturbative effects in the OPE
have to fall off as $1/q_T^2$ upon Fourier transform~\cite{Ebert:2022cku}.

\subsection{Parametric \texorpdfstring{$\as$}{alphas} and PDF uncertainties}
\label{sec:uncerts_as_pdf}

\begin{figure*}
\includegraphics[width=\WidthTwoSubfigs]{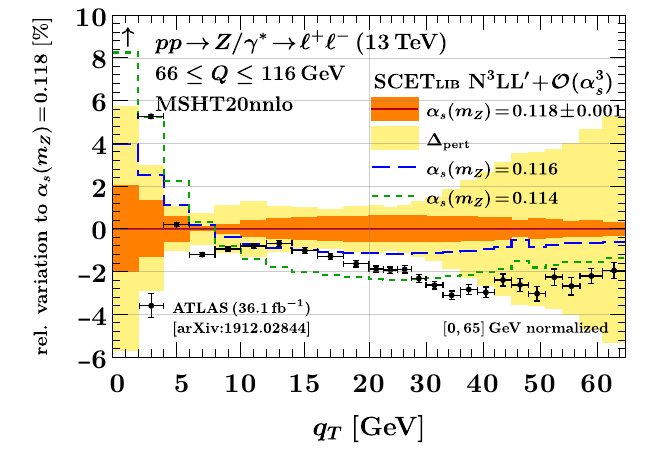}%
\hfill%
\includegraphics[width=\WidthTwoSubfigs]{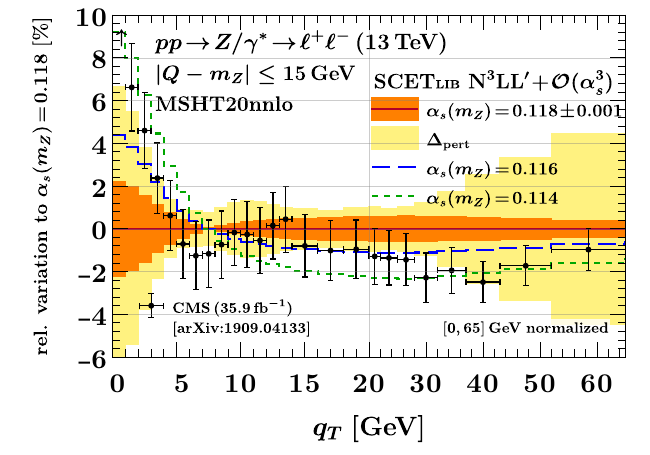}%
\\%
\includegraphics[width=\WidthTwoSubfigs]{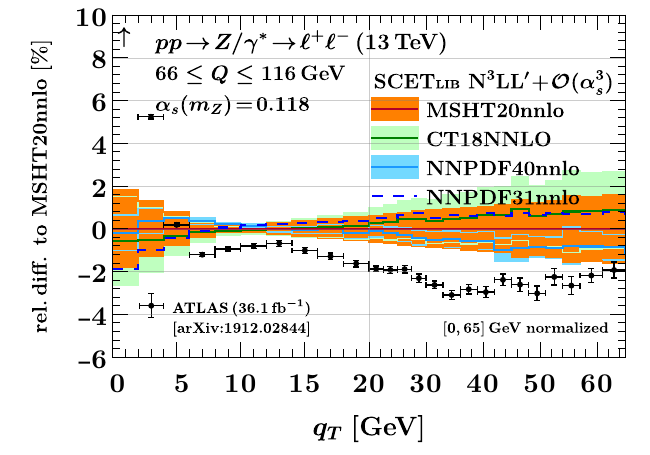}%
\hfill%
\includegraphics[width=\WidthTwoSubfigs]{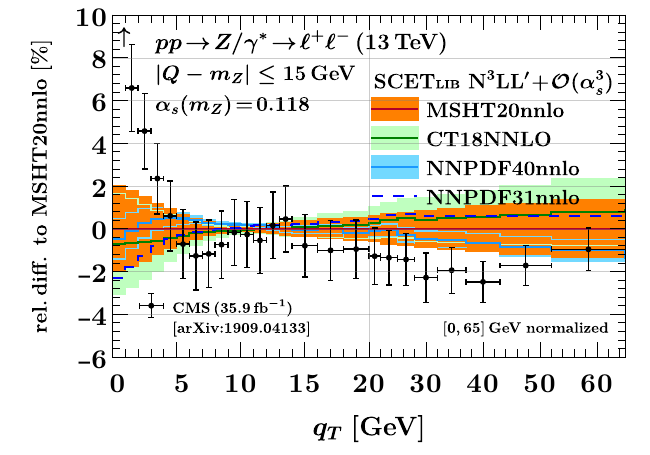}%
\caption{
Impact of parametric strong coupling variations (top)
and parametric PDF variations (bottom),
as well as alternate PDF sets, on the normalized $p_T^Z$ spectrum
at N$^3$LL$^\prime+\ord{\as^3}$
compared to the ATLAS $13\TeV$~\cite{Aad:2019wmn} (left)
and CMS $13\TeV$~\cite{Sirunyan:2019bzr} (right) measurements.
}
\label{fig:Z_qT_as_pdf_uncertainties}
\end{figure*}

\begin{table}
\begin{center}
\begin{tabular}{c | c | c | c}
& \texttt{CT18NNLO} & \texttt{NNPDF31nnlo} & \texttt{NNPDF40nnlo}
\\
\hline
$\sigma_{65}~[\mathrm{pb}]$ \quad \multrow{ ATLAS cuts~\cite{Aad:2019wmn} \\ CMS cuts~\cite{Sirunyan:2019bzr}} & \multrow{659.2 \\ 650.1} & \multrow{678.1 \\ 669.9} & \multrow{691.3 \\ 682.9}
\end{tabular}
\label{tab:norm_factors_fid_pTZ_pdf_sets}
\caption{Normalization factors for the central N$^3$LL$^\prime\!+\!\ord{\as^3}$ fiducial $p_T^Z$ spectra
shown in \fig{Z_qT_as_pdf_uncertainties}. The corresponding one for \texttt{MSHT20nnlo} is given in \tab{norm_factors_fid_pTZ}.}
\end{center}
\end{table}

We next estimate the parametric uncertainties
related to the strong coupling constant and the collinear PDFs on our prediction
using standard methods, closely following those described
in \refcite{Butterworth:2015oua}.
In practical terms, both variations are performed only in the resummed singular cross section
and the $\ord{\as}$ nonsingular coefficient while keeping the $\ord{\as^2}$ and $\ord{\as^3}$
nonsingular coefficients at central values and at the reference PDF set,
as already discussed in \sec{nons_cross_section}.
This strategy for the uncertainty estimation
avoids performing CPU-intensive runs for the FO nonsingular coefficients
and is justified by the smallness
of the $\ord{\as^2}$ and the $\ord{\as^3}$ nonsingular coefficients for $q_T\lesssim 65\GeV$.

For the uncertainty related to $\as$, we perform variations of
\begin{align} \label{eq:alphas_variations}
\as(m_Z) = 0.118 \pm 0.001
\,,\end{align}
properly taking into account its effect on the DGLAP running
and on PDF determinations by using the PDF set that
consistently employs the corresponding value of $\as(m_Z)$.
\Eq{alphas_variations} roughly corresponds
to the currently quoted PDG 2023 uncertainty~\cite{ParticleDataGroup:2024cfk}.
The associated parametric $\as$ uncertainty is computed as
\begin{align} \label{eq:as_unc_formula}
\Delta_{\as}
= \frac{1}{\sqrt{2}}
\Bigl[
\bigl(\df\sigma_{\as,\up} - \df\sigma_\central \bigr)^2
+ \bigl(\df\sigma_{\as,\down} - \df\sigma_\central \bigr)^2
\Bigr]^{1/2}
\,.\end{align}

The resulting uncertainty on the normalized fiducial $p_T^Z$ spectrum
is shown in the top row of \fig{Z_qT_as_pdf_uncertainties} (orange band).
On the range $5 \leq q_T \leq 30 \GeV$, the uncertainty
is on par with $\Delta_\pert$ (light yellow band) at this order.
Neither of them is able to remedy
the overshoot of the prediction compared to the data at ($q_T\gtrsim 15\GeV$).
Curiously, we find that a lower value of the strong coupling,
$\as(m_Z)=0.114$ (short-dashed green), captures the data trend
much more closely over a wide range of $q_T$,
including in particular the region $q_T \geq 20 \GeV$
where nonperturbative effects are already negligible, see \sec{uncerts_np}.
While this observation did \emph{not} result from an $\as$ fit,
it is interesting to note since similarly low $\as$ values
have been extracted in the past~\cite{Abbate:2010xh, Abbate:2012jh}
from other resummation-sensitive observables, specifically $e^+ e^-$ event shapes.
We stress, however, that a complete fit of $\as$ to hadron-collider $p_T^Z$ data
must include profiling over the PDF set,
as done e.g.\ in \refscite{Camarda:2022qdg, ATLAS:2023lhg},
and in \sec{approximate_n4ll} we will indeed confirm that the inclusion of approximate N$^3$LO effects
in the PDF set in fact yields much closer agreement with the data
for $\as(m_Z) = 0.118$ from the start.

For the uncertainty related to the collinear PDFs,
we show results for \texttt{CT18NNLO}~\cite{Hou:2019efy},
\texttt{NNPDF31nnlo}~\cite{NNPDF:2021njg}, \texttt{NNPDF40nnlo}~\cite{NNPDF:2021njg},
and for our default set \texttt{MSHT20nnlo}~\cite{Bailey:2020ooq}.
Since these collaborations in general employ different
methodologies in estimating an uncertainty from the fit,
we follow the respective collaboration's nominal formula
for calculating the corresponding uncertainty on our prediction.
For Hessian sets such as \texttt{CT18NNLO} and \texttt{MSHT20nnlo} we use
\begin{align} \label{eq:pdf_unc_formula}
\Delta^{\mathrm{Hessian}}_{\mathrm{PDF}}
= \frac{1}{\sqrt{2}}
\Biggl[ \sum_i^{n_{\mathrm{rep}}} (\df\sigma_i - \df\sigma_{\mathrm{central}})^2 \Biggr]^{1/2}
\,.\end{align}
The index $i$ runs over the set of $n_{\mathrm{rep}}$ different replicas provided by each
collaboration and $\df\sigma_{\mathrm{central}}$ ($\df\sigma_i$)
denotes the prediction evaluated using
the central member ($i$\textsuperscript{th} member) of the PDF set.
Compared to eq.~(20) of \refcite{Butterworth:2015oua},
the relative factor of $1/\sqrt{2}$ in \eq{pdf_unc_formula}
avoids the possible double counting
induced by replicas of opposite eigenvectors in the symmetric limit.
We note that we in addition rescale the resulting uncertainty
for the \texttt{CT18NNLO} set
to account for the difference in their quoted confidence level (CL, 90\%)
to our common target CL of 68\% for all PDF sets.
For Monte-Carlo (MC) sets such as \texttt{NNPDF31nnlo} and \texttt{NNPD40nnlo}, we use
\begin{align}
\Delta^{\mathrm{MC}}_{\mathrm{PDF}}
= \frac{1}{\sqrt{n_{\mathrm{rep}}-1}}
\Biggl[ \sum_i^{n_{\mathrm{rep}}}  (\df\sigma_i - \df\sigma_{\mathrm{central}})^2 \Biggr]^{1/2}
\,,\end{align}
where for $\df\sigma_{\mathrm{central}}$ we use the averaged PDF replica provided by each set.

Turning to the bottom row of \fig{Z_qT_as_pdf_uncertainties},
we show the respective central values
and parametric PDF uncertainties for \texttt{CT18NNLO}~\cite{Hou:2019efy}
(solid green), \texttt{NNPDF31nnlo}~\cite{NNPDF:2017mvq} (dashed blue)
and \texttt{NNPDF40nnlo}~\cite{NNPDF:2021njg} (solid blue)
for the normalized $p_T^Z$ spectrum,
with the associated normalization factors given in \tab{norm_factors_fid_pTZ_pdf_sets}.
Specifically, we show relative deviations
from our default choice, \texttt{MSHT20nnlo}~\cite{Bailey:2020ooq} (red).
It is worth noting that all sets agree at the level
of the normalized $p_T^Z$ spectrum well within the uncertainties
of the \texttt{CT18NNLO} and \texttt{MSHT20nnlo} predictions,
while the \texttt{MSHT20nnlo} and \texttt{CT18NNLO} central values
fall on the upper edge or outside of the \texttt{NNPDF40nnlo}, respectively.
In particular, none of the PDF parametric uncertainties,
and no alternate choice of central set at this PDF order (NNLO),
is able to capture the discrepancy to the data at $q_T \geq 15 \GeV$.
We have checked that this behavior of NNLO PDF sets,
with a characteristic overshoot above the data at $q_T \geq 15 \GeV$,
in fact persists for the $p_T^Z$ spectrum
as measured in bins of the $Z$ boson rapidity at $8 \TeV$
by the ATLAS collaboration~\cite{Aaboud:2017ffb}.

Concerning the nonperturbative region of $q_T \leq 5 \GeV$,
it has been pointed out in \refcite{Bury:2022czx}
that fits of nonperturbative TMD physics
are subject to a sizable bias due to the choice
of reference collinear PDF set.
We find this surprising in light of the small ($\leq 1\%$) relative differences
between modern PDF sets that we observe in this region for common nonperturbative inputs
at the level of the self-consistently normalized $p_T^Z$ spectrum.
We hasten to add, however, that while the TMD fit of \refcite{Bury:2022czx}
like other fits heavily relies on LHC data for resonant Drell-Yan,
the latter are, of course, not the only data set entering the fit,
and by our own universality arguments
in \sec{np_fid_model} insufficient to determine the individual TMDs
or, indeed, fully assess the presence of bias due to the collinear PDFs.

\begin{figure*}
\includegraphics[width=\WidthTwoSubfigs]{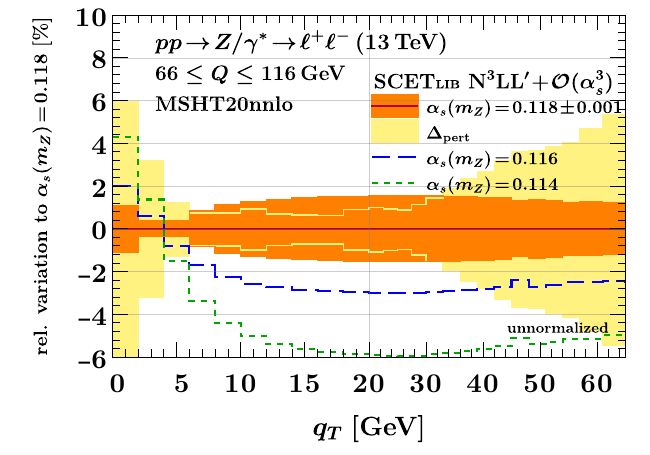}%
\hfill%
\includegraphics[width=\WidthTwoSubfigs]{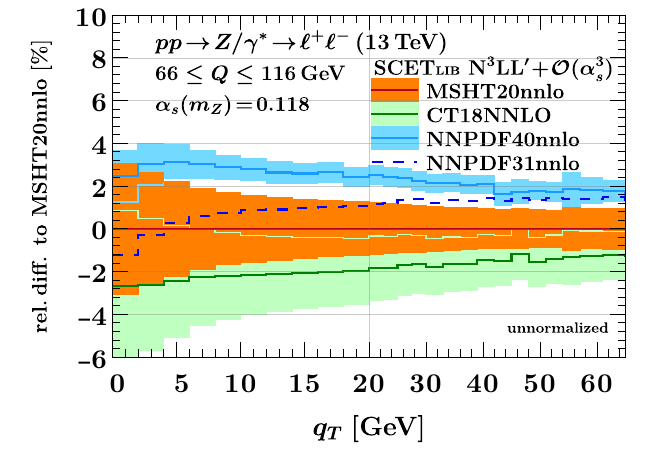}%
\caption{
Impact of parametric strong coupling variations (left)
and parametric PDF variations (right),
as well as alternate PDF sets, on the \emph{unnormalized} $p_T^Z$ spectrum
at N$^3$LL$^\prime+\ord{\as^3}$.
}
\label{fig:Z_qT_as_pdf_uncertainties_unnormalized}
\end{figure*}

We find it interesting to also consider the impact of the parametric
uncertainties on the \emph{unnormalized} $p_T^Z$ spectrum,
which we show in \fig{Z_qT_as_pdf_uncertainties_unnormalized} for reference.
Since the data sets for unnormalized spectra are more limited,
we simply show our predictions on their own here,
using the ATLAS cuts for definiteness;
in \sec{results_Z_cumulants} we will perform direct comparisons to available data
at the level of the unnormalized cumulative cross section,
i.e., the piece of information orthogonal to the normalized spectrum.
Interestingly, while the various PDF sets were in close agreement
on the shape of the normalized spectrum, the agreement worsens
at the level of the unnormalized spectrum,
see the right panel of \fig{Z_qT_as_pdf_uncertainties_unnormalized}.
Together with our findings on the impact of nonperturbative physics
and the size of the nonsingular cross section,
this suggests an attractive strategy to distinguish between PDF sets
at complete three-loop accuracy using the cumulative cross section,
see \sec{results_Z_cumulants},
which precisely retain the PDF sensitivity that usually
cancels to a large extent in normalized spectra.

\subsection{Impact of \texorpdfstring{N$^4$LL}{N4LL} Sudakov effects
and approximate \texorpdfstring{N$^3$LO}{N3LO} PDFs}
\label{sec:approximate_n4ll}

\begin{figure*}
\includegraphics[width=\WidthTwoSubfigs]{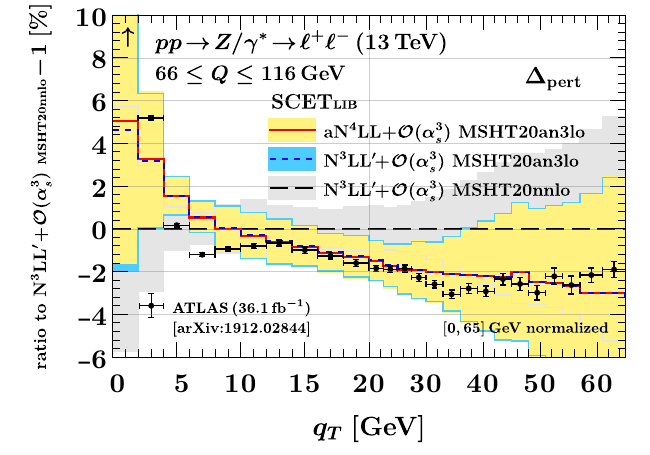}%
\hfill%
\includegraphics[width=\WidthTwoSubfigs]{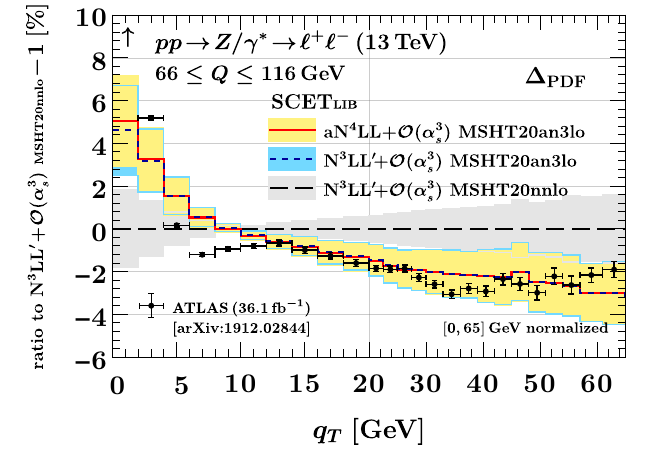}%
\\
\includegraphics[width=\WidthTwoSubfigs]{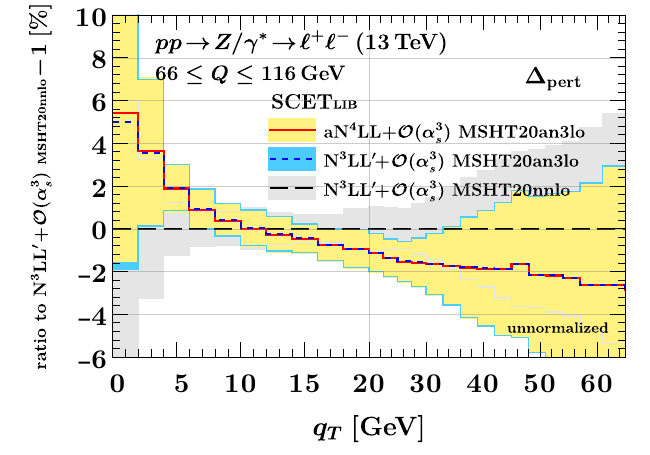}%
\hfill%
\includegraphics[width=\WidthTwoSubfigs]{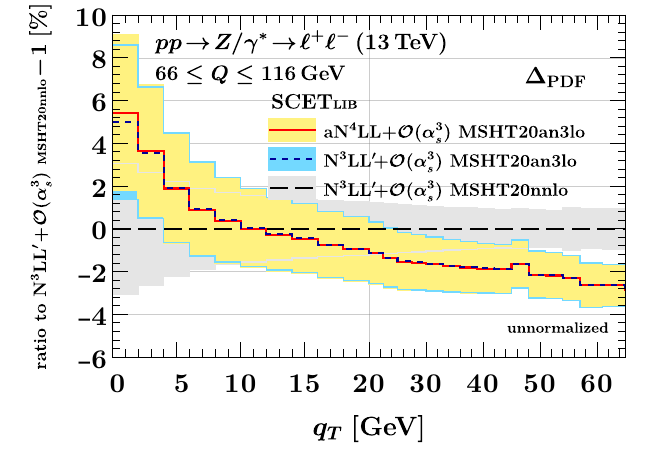}%
\caption{
Impact of aN$^3$LO PDFs (blue)
and of N$^4$LL Sudakov effects (red/yellow,
with overall aN$^4$LL$+ \ord{\as^3}$ accuracy)
on the fiducial $p_T^Z$ spectrum
relative to our baseline prediction (black/gray)
using an NNLO PDF set
at N$^3$LL$^\prime + \ord{\as^3}$
compared to the ATLAS $13\TeV$ measurement~\cite{Aad:2019wmn}.
Here we compare the \texttt{MSHT20nnlo} and \texttt{MSHT20an3lo} PDF sets
produced by the \texttt{MSHT} collaboration~\cite{Bailey:2020ooq, McGowan:2022nag}.
Predictions on the left (right) are dressed
with the perturbative (parametric PDF) uncertainty.
The top (bottom) row shows the impact on the normalized (unnormalized) $p_T^Z$ spectrum.
}
\label{fig:Z_an3lopdf_ratio_msht20}
\end{figure*}

\begin{figure*}
\includegraphics[width=\WidthTwoSubfigs]{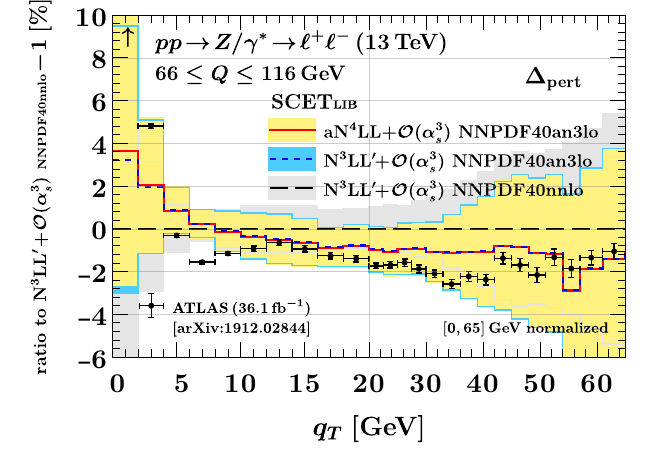}%
\hfill%
\includegraphics[width=\WidthTwoSubfigs]{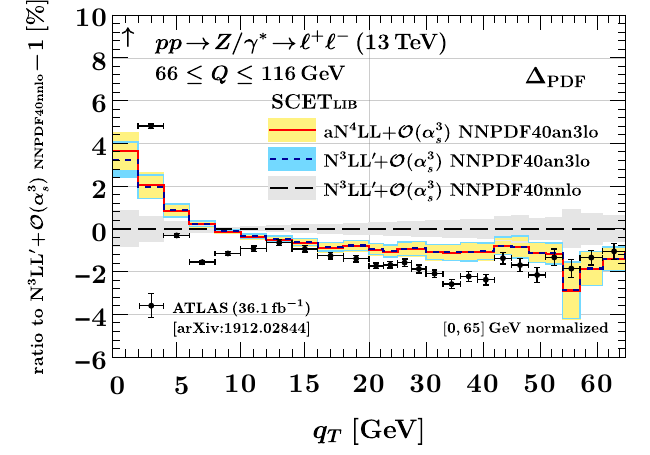}%
\\
\includegraphics[width=\WidthTwoSubfigs]{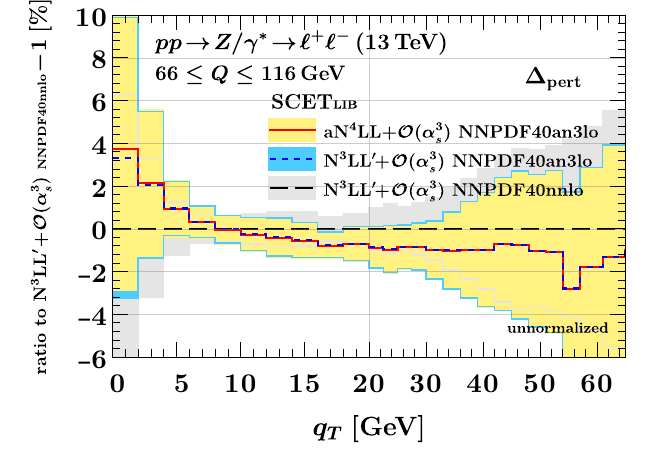}%
\hfill%
\includegraphics[width=\WidthTwoSubfigs]{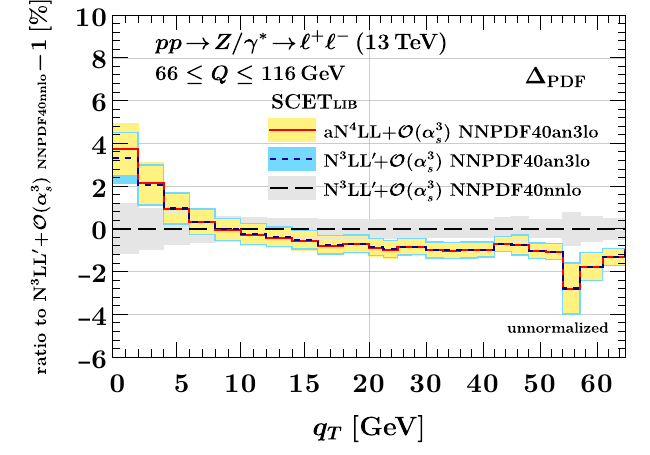}%
\caption{
Same as \fig{Z_an3lopdf_ratio_msht20}, but comparing the
\texttt{NNPDF40nnlo} and \texttt{NNPDF40an3lo} PDF sets
produced by the \texttt{NNPDF} collaboration~\cite{NNPDF:2021njg, NNPDF:2024nan}.
}
\label{fig:Z_an3lopdf_ratio_nnpdf40}
\end{figure*}

\begin{table}
\begin{center}
\begin{tabular}{c | c | c}
& \texttt{MSHT20an3lo} & \texttt{NNPDF40an3lo}
\\
\hline
$\sigma_{65}~[\mathrm{pb}]$ \quad \multrow{ aN$^4$LL$+\ord{\as^3}$ \\ \,\,\,\,N$^3$LL$^\prime+\ord{\as^3}$ } & \multrow{675.975 \\ 675.979} & \multrow{ 691.963 \\ 691.969}
\end{tabular}
\label{tab:Z_an3lopdf_norm}
\caption{
Normalization factors for the additional fiducial $p_T^Z$ spectra
shown in \figs{Z_an3lopdf_ratio_msht20}{Z_an3lopdf_ratio_nnpdf40}.
Normalization factors for the respective baseline NNLO PDF sets
are given in \tabs{Z_qT_fid_predictions_norm}{norm_factors_fid_pTZ_pdf_sets}.
}
\end{center}
\end{table}

\begin{figure*}
\centering
\includegraphics[width=\WidthTwoSubfigs]{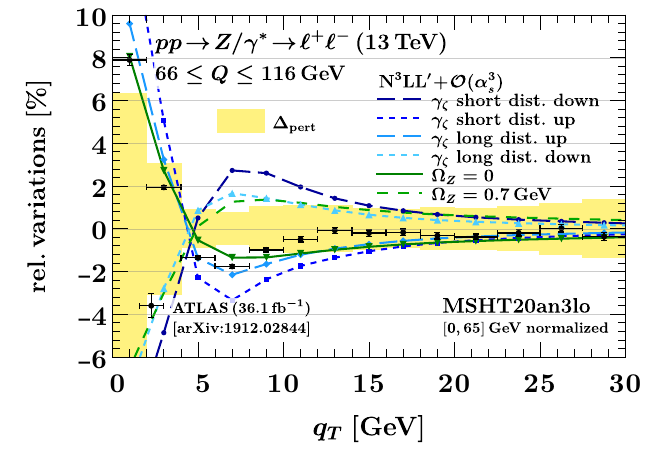}%
\hfill%
\includegraphics[width=\WidthTwoSubfigs]{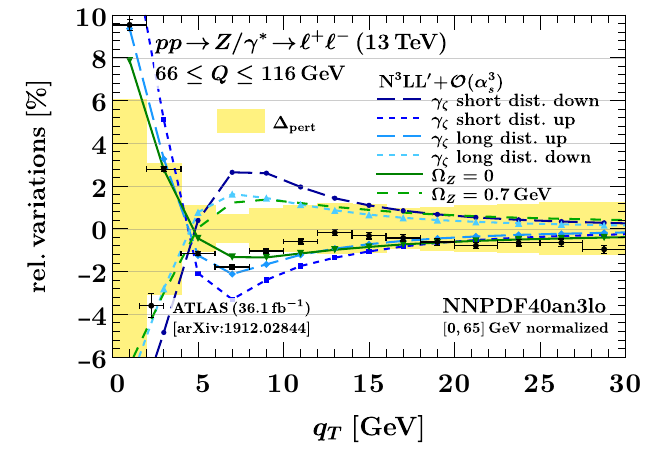}%
\caption{Impact of parameter variations in the nonperturbative TMD models at N$^3$LL$^\prime+\ord{\as^3}$
for the \texttt{MSHT20an3lo} PDF set~\cite{McGowan:2022nag} (left) and for the \texttt{NNPDF40an3lo} PDF set~\cite{NNPDF:2024nan} (right)
compared to the ATLAS $13\TeV$ measurement~\cite{Aad:2019wmn}.
For illustration, we also compare to the estimated size $\Delta_\pert$ of the perturbative uncertainty.
}
\label{fig:Z_an3lopdf_np_unc}
\end{figure*}

While all of our highest-order predictions so far involved partonic cross sections at complete three-loop accuracy,
they employed NNLO PDFs determined using NNLO (three-loop) DGLAP evolution and NNLO theory predictions in the fit,
which until recently constituted the state-of-the-art.
Lately, both the \texttt{MSHT} and the \texttt{NNPDF} collaboration
exploited the rapidly growing body of
knowledge~\cite{Falcioni:2023luc, Falcioni:2023tzp, Moch:2023tdj, Falcioni:2024xyt, Falcioni:2024qpd}
on the N$^3$LO DGLAP and mass decoupling kernels
in order to produce approximate N$^3$LO (aN$^3$LO)
PDF sets~\cite{McGowan:2022nag, NNPDF:2024nan, Cridge:2024exf, Cooper-Sarkar:2024crx, MSHT:2024tdn}.
Notably, in \refcite{McGowan:2022nag}
both the set of unknown N$^3$LO $K$ factors for the relevant physical processes
and the few remaining unknown N$^3$LO DGLAP ingredients
are promoted to nuisance parameters~\cite{Tackmann:2024xxx, Tackmann:2024kci}.
In addition, this allows for an improved uncertainty estimation of the PDF set,
in principle also addressing common obstacles such as correlations
between missing higher-order uncertainties in the fit and in predictions that use the PDF set.

Importantly, the DGLAP kernels also enter as a key set of noncusp anomalous dimensions
in resummed predictions for the $q_T$ spectrum, since they govern the evolution
of the collinear PDFs from their input scale to the beam function scale $\mu_B \sim q_T \sim 1/b_T$.
Specifically, since the input scale is effectively set by the weak scale $m_{W,Z} \sim Q$ in modern PDF sets
due to the wealth of LHC data entering the fits,
this evolution indeed resums a large single logarithm
and cannot be bypassed when attempting to achieve N$^4$LL accuracy
in any TMD or resummed prediction that uses collinear PDFs as input.
Therefore, achieving (approximate) N$^3$LO DGLAP evolution
provides the last missing ingredients for achieving (approximate) N$^4$LL accuracy at the level of the resummed spectrum.
This is thanks to dedicated previous efforts to calculate (or numerically estimate) the five-loop $\beta$
function~\cite{Herzog:2017ohr},
the five-loop cusp anomalous dimension~\cite{Herzog:2018kwj},
the four-loop rapidity anomalous dimension~\cite{Duhr:2022yyp, Moult:2022xzt},
and the four-loop virtuality anomalous dimensions~\cite{vonManteuffel:2020vjv, Agarwal:2021zft},
all of which are also necessary ingredients at this order.
We collectively refer to these as ``N$^4$LL Sudakov effects'',
as opposed to the purely single-logarithmic DGLAP evolution.
Conversely, since all of
these contributions
are known either fully analytically
or, in the case of the five-loop cusp anomalous dimension, estimated
to sufficient precision given its small overall impact
a resulting approximate N$^4$LL (aN$^4$LL) resummed prediction
will be ``approximate'' precisely in the sense that the underlying PDF set is aN$^3$LO.

It is clearly of key interest to determine the size of the effect
that N$^3$LO evolution and N$^3$LO fixed-order contributions in PDF fits
have on the extremely precisely measured $p_T^Z$ spectrum at the LHC,
and similarly for N$^4$LL Sudakov effects.
To do so, we have performed a complete implementation
of the additional N$^4$LL Sudakov effects in \texttt{SCETlib}
and interfaced them with the recent aN$^3$LO PDF sets
to achieve aN$^4$LL accuracy.
Numerical results produced with this further upgraded setup are shown
in \figs{Z_an3lopdf_ratio_msht20}{Z_an3lopdf_ratio_nnpdf40},
with normalization factors reported in \tab{Z_an3lopdf_norm}.
Specifically, beginning with the \texttt{MSHT} sets,
we show predictions at N$^4$LL$+\ord{\as^3}$ (red)
and N$^3$LL$^\prime+\ord{\as^3}$ (dashed blue) with the \texttt{MSHT20an3lo} set,
normalized to the N$^3$LL$^\prime+\ord{\as^3}$ results with the \texttt{MSHT20nnlo} set (dotted black)
in \fig{Z_an3lopdf_ratio_msht20}.
All predictions consistently employ the approximate unexpanded analytic RGE solutions~\cite{Billis:2019evv} at the corresponding order,
which are summarized in \app{rge_solutions_n4ll}.

The effect of the approximate N$^3$LO PDF set is quite significant, leading to differences
of up to $\sim 5\%$ to the corresponding NNLO PDF set
in the low-$q_T$ region and, crucially, a decrease of the prediction by $1-2\%$ in the region
at $q_T \geq 15 \GeV$ where we previously observed an irreconcilable disagreement with the data.
This implies that the three-loop ingredients that were included during the fitting procedure have a nontrivial impact,
even at this high perturbative order. The much improved agreement in the region
$q_T \geq 15 \GeV$ is particularly remarkable in light of the fact
that the shape of the $p_T^Z$ spectrum at values of $q_T \leq 30 \GeV$
has never been used as an input for any PDF fits, cf.\ \refcite{Boughezal:2017nla}
for its uses in the large $p_T^Z$ fixed-order region.
These findings are confirmed almost entirely
by performing the analogous comparison between the two relevant \texttt{NNPDF} sets,
as shown in \fig{Z_an3lopdf_ratio_nnpdf40},
with only a slightly reduced change going from NNLO to aN$^3$LO
in the region $q_T \geq 15 \GeV$ compared to \texttt{MSHT}.
A similarly striking change from including aN$^3$LO PDF information
into the prediction for the $q_T$ spectrum
has previously been reported in \refcite{Neumann:2022lft},
and the \texttt{MSHT20an3lo} has in the meantime also been adopted
as the reference set for the $\as(m_Z)$ determination in \refcite{ATLAS:2023lhg}.

Turning to the N$^4$LL Sudakov effects,
we find that their additional effect on top of the aN$^3$LO PDFs at aN$^4$LL
is completely marginal in comparison.
Specifically, in \fig{Z_an3lopdf_ratio_msht20}
the resummed and matched N$^4$LL prediction (red) only exhibits marginal differences
to N$^3$LL$^\prime$ (dashed blue) for the same PDF set (\texttt{MSHT20an3lo}).
As expected, the uncertainty band of the former is contained within that of the latter,
while their central values mildly differ ($\lesssim 0.5\%$) only in the first bin, indicating that the impact of the four-loop rapidity boundary term is not large.

Returning to the comparison to the data in the region $q_T \lesssim 10\GeV$
we find that the agreement of even the aN$^3$LO PDF results
(or, equivalently, the predictions at overall aN$^4$LL)
with the data in this region is not optimal for our default nonperturbative model parameters.
Since our nonperturbative parameters were chosen mainly for illustration,
it is interesting to ask whether varying them further improves the agreement with the data
and allows one to resolve the residual differences
in the region where the nonperturbative model is effective.
In \fig{Z_an3lopdf_np_unc} we thus show the same nonperturbative variations
as in \sec{uncerts_np},
again at N$^3$LL$^\prime+\ord{\as^3}$,
but in this case using the \texttt{MSHT20an3lo} (left) and \texttt{NNPDF40an3lo} sets (right).
Taken at face value, no single parameter variation follows the entire data trend,
but the typical size of the effect of short-distance variations
in the Collins-Soper kernel and the TMD PDF effective model
easily accounts for the remaining differences.

\section{Results for cumulative fiducial \texorpdfstring{$p_T^Z$}{pTZ} cross sections}
\label{sec:results_Z_cumulants}

\begin{figure*}
\includegraphics[width=\WidthTwoSubfigs]{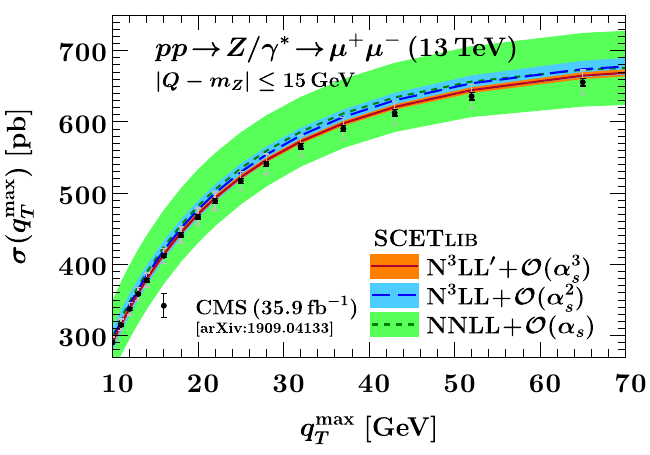}%
\hfill%
\includegraphics[width=\WidthTwoSubfigs]{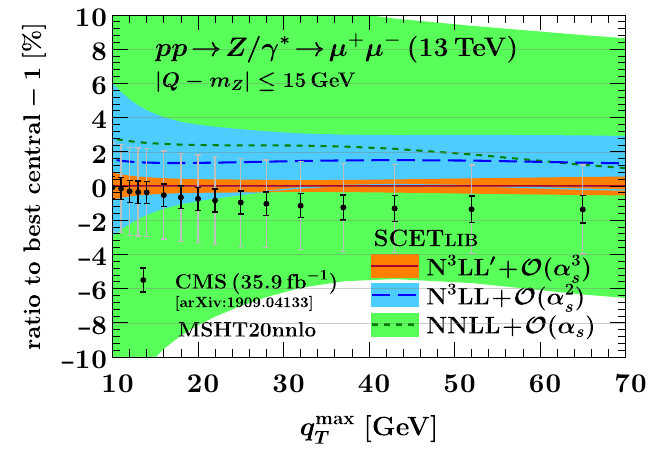}%
\caption{
Predictions for the unnormalized cumulative fiducial cross section (left)
and the relative difference to our prediction at the highest order (right)
compared to the CMS $13\TeV$ measurement~\cite{Sirunyan:2019bzr}.
The black (gray) bars indicate the experimental uncertainty
excluding (including) the luminosity uncertainty.
}
\label{fig:Z_qT_cms_cumulant_ratio}
\end{figure*}

Our earlier observation of differences between PDF sets
at the level of the unnormalized spectrum
leads us to consider the cumulative $p_T^Z$ cross section,
defined in terms of the fiducial $p_T^Z$ spectrum as follows,
\begin{align}
\sigma(\qTmax) = \int^{\qTmax} \!\!\!\! \df q_T \, \frac{\df\sigma}{\df q_T}
\,,\end{align}
which can readily be computed in our framework
by accumulating over bins in our previous predictions.
To estimate perturbative uncertainties in this case,
we perform the envelopes defined in \sec{uncerts_pert}
after accumulating over bins,
i.e., we treat the individual variations entering
the envelope as fully correlated across bins.
We stress that while this amounts to an ad-hoc assumption
on the profile scale variations of \sec{uncerts_pert},
it is, of course, a well-defined procedure
for the parametric nonperturbative, $\as$, and collinear PDF variations
described in \secs{uncerts_np}{uncerts_as_pdf}.
The correct correlations of the perturbative uncertainties in $q_T$
can be fully accounted for using the approach of \refscite{Tackmann:2024xxx, Tackmann:2024kci}.

We compare our predictions to the CMS $13 \TeV$ measurement
in the $\mu^+ \mu^-$ channel~\cite{Sirunyan:2019bzr} using dressed muons.
Both the unnormalized fiducial spectrum
and its complete experimental covariance matrix were reported for this channel,
allowing us to fully reconstruct the experimental uncertainty
on the cumulative cross section as a function of $\qTmax$.
Our results are shown in \fig{Z_qT_cms_cumulant_ratio}.
We again observe excellent perturbative coverage and convergence,
this time at the level of the cumulative cross section,
with the total perturbative uncertainty estimate below the percent level
at N$^3$LL$^\prime+\ord{\as^3}$, on par with the experimental uncertainty
if one excludes the common overall luminosity uncertainty.
The prediction for our default \texttt{MSHT20nnlo} PDF set
overshoots the data, but the difference can easily be accounted for,
as we will see below, by the spread between NNLO PDF sets.

\begin{figure*}
\includegraphics[width=\WidthTwoSubfigs]{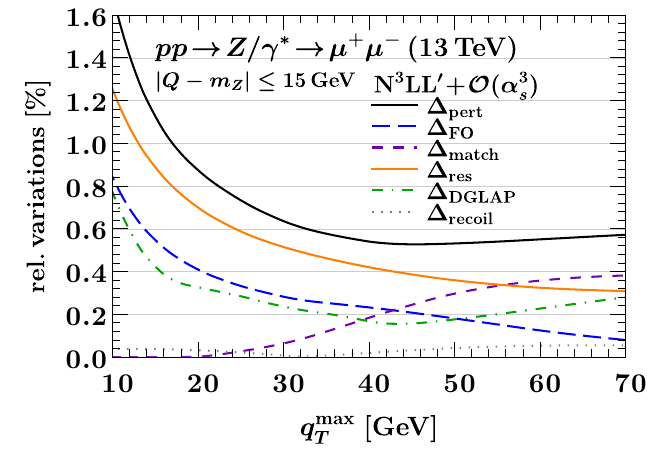}%
\hfill%
\includegraphics[width=\WidthTwoSubfigs]{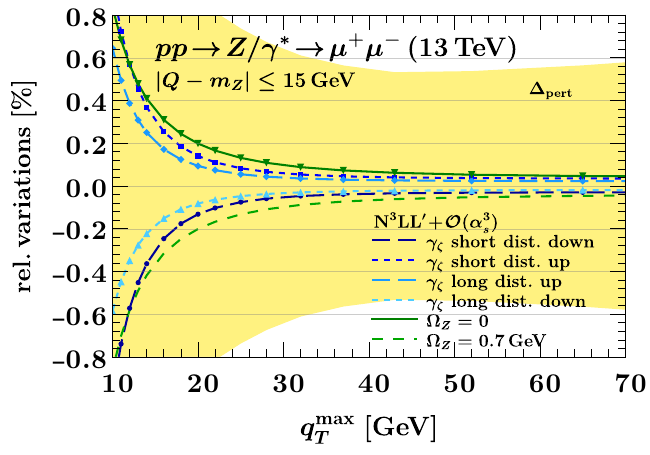}%
\caption{Left: Breakdown of perturbative uncertainties contributing
to the cumulative fiducial cross section at N$^3$LL$^\prime+\ord{\as^3}$.
Right: Impact of parameter variations in the nonperturbative TMD models
on the cumulative fiducial cross section at N$^3$LL$^\prime+\ord{\as^3}$.
For illustration, we also compare to the estimated size $\Delta_\pert$ of the total perturbative uncertainty.
}
\label{fig:Z_qT_cms_uncertainties}
\end{figure*}

To substantiate the percent-level perturbative precision
of the N$^3$LL$^\prime+\ord{\as^3}$ result,
we consider the breakdown of the perturbative uncertainty
in the left panel of \fig{Z_qT_cms_uncertainties}.
We find that the matching uncertainty, which increases with $\qTmax$,
plays a crucial role in stabilizing the uncertainty estimate
towards the fixed-order region.
While an uncertainty of $\leq 0.6 \%$ at $\qTmax \geq 30 \GeV$ may seem aggressive
even for a three-loop prediction,
we believe that our careful estimate derived from a large number of sources
is reliable (and indeed could be considered conservative at the level of the spectrum).
In the right panel of \fig{Z_qT_cms_uncertainties}
we consider the impact of the same model parameter variations
encoding nonperturbative TMD that we previously introduced in \sec{uncerts_np}.
We find that the relative impact of any reasonably sized variations is
at the permille level already for $\qTmax \geq 25 \GeV$,
as expected from their $(1/\qTmax)^2$ falloff.
This suggests that while the cumulative cross section
in this region is still sensitive to the effects of \emph{perturbative} resummation,
it is essentially unaffected by genuinely nonperturbative physics,
implying that it is fully predicted in terms of the strong coupling
and the collinear PDFs.

\begin{figure*}
\centering
\includegraphics[width=\WidthTwoSubfigs]{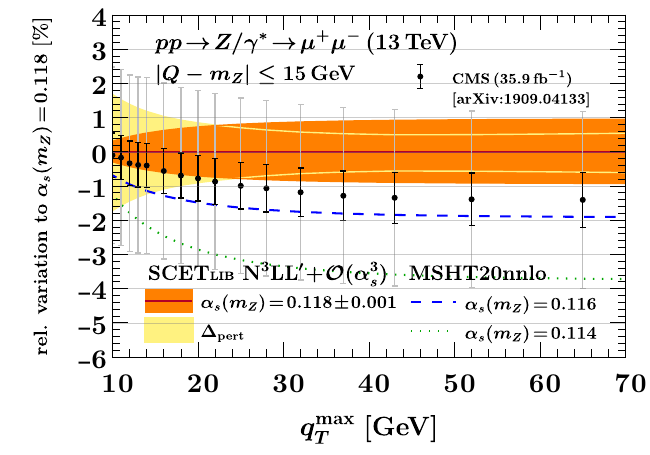}%
\hfill%
\includegraphics[width=\WidthTwoSubfigs]{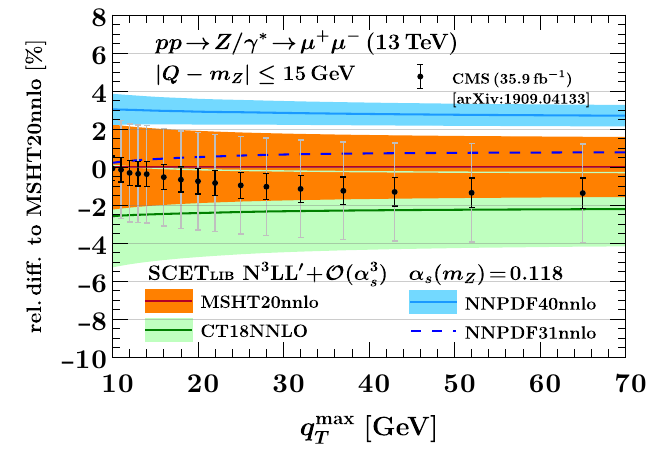}%
\\
\includegraphics[width=\WidthTwoSubfigs]{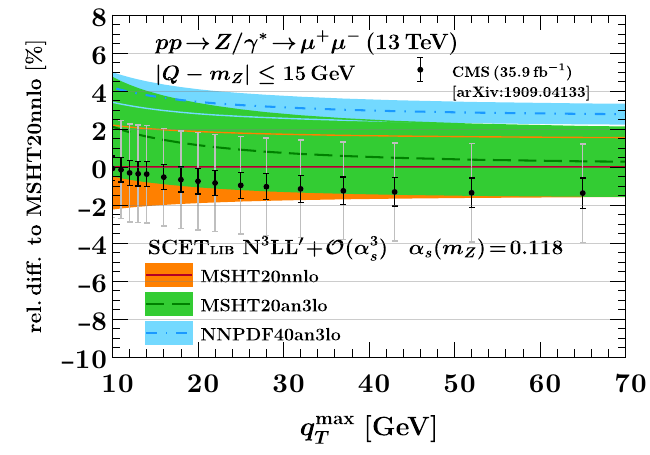}%
\caption{
Impact of parametric strong coupling variations (top left)
and parametric PDF variations
for alternate NNLO PDF sets (top right),
as well as for aN$^3$LO PDF sets (bottom),
on the cumulative fiducial cross section
at N$^3$LL$^\prime+\ord{\as^3}$,
compared to the CMS $13\TeV$ measurement~\cite{Sirunyan:2019bzr}.
The black (gray) bars indicate the experimental uncertainty
excluding (including) the luminosity uncertainty.
}
\label{fig:Z_qT_cms_as_pdf_uncertainties}
\end{figure*}

Turning now to the top left panel of \fig{Z_qT_cms_as_pdf_uncertainties},
we see that the impact of
strong coupling variations of $0.118 \pm 0.001$ on the cumulative cross section
is small and roughly of $\ord{1\%}$,
which should be contrasted with the much stronger impact
of $\as$ on the shape of the spectrum,
see the top left panel of \fig{Z_qT_as_pdf_uncertainties}.
The most interesting observation of this section
is found in the top right and bottom panels of \fig{Z_qT_cms_as_pdf_uncertainties},
where we compare the impact of alternate PDF choices
and PDF parametric uncertainties on the prediction for NNLO PDFs and aN$^3$LO PDFs, respectively.
Keeping in mind the $\leq 1\%$ perturbative accuracy of our predictions,
we find that our prediction together with the similarly
precise experimental data can easily distinguish between both NNLO and aN$^3$LO PDF sets,
and also constrain them further to a fraction of their currently quoted uncertainties.

\paragraph{Discussion:}
Our findings for the cumulative cross section
suggest an appealing strategy to perform PDF fits
at complete three-loop accuracy using our predictions here.
Specifically, one could envision a scheme where the
nonsingular cross sections at $\ord{\as^2}$ and $\ord{\as^3}$,
which are numerically extremely expensive
but small, are computed only once at a reference PDF set
and then treated as a fixed bias correction
while propagating the PDF through the orders of magnitude cheaper
resummed cross section during the fit.
The latter has complete three-loop accuracy,
but much more beneficial scaling of computational cost with the loop order.
Such a scheme is particularly attractive given that PDF fit templates
at complete N$^3$LO accuracy, as required to further improve
on the existing aN$^3$LO PDF fits~\cite{McGowan:2022nag, NNPDF:2024nan, Cridge:2024exf, Cooper-Sarkar:2024crx, MSHT:2024tdn},
are still lacking beyond total inclusive cross sections~\cite{Duhr:2020seh},
and have to be extrapolated to fiducial quantities using $K$ factors.
By contrast, the scheme we propose here could even be
extended easily to the cumulative cross section
differential in rapidity at negligible additional cost (mainly due
to the more differential reference nonsingular).
Importantly, our findings for the impact of nonperturbative TMD physics
confirm that while resummation sensitive, the cumulative cross section
for the $\qTmax$ values of interest is in fact nearly free of nonperturbative effects
and well suited for a perturbative QCD fit at leading twist.

\section{Results for the fiducial \texorpdfstring{$p_T^{W^\pm}$}{pTW} spectrum}
\label{sec:results_W}

\begin{figure*}
\includegraphics[width=\WidthTwoSubfigs]{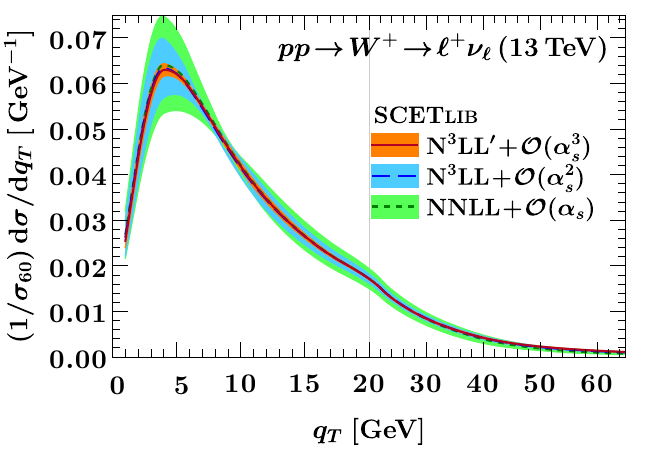}%
\hfill%
\includegraphics[width=\WidthTwoSubfigs]{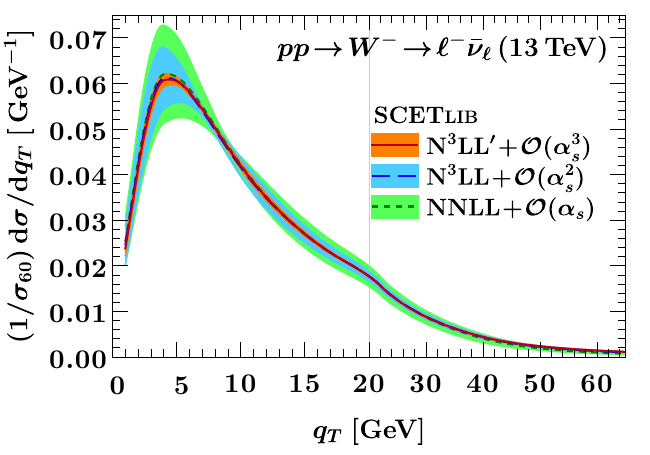}%
\\
\includegraphics[width=\WidthTwoSubfigs]{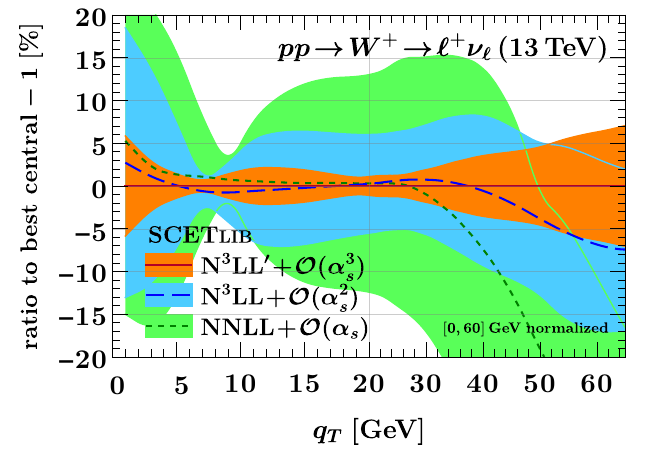}%
\hfill%
\includegraphics[width=\WidthTwoSubfigs]{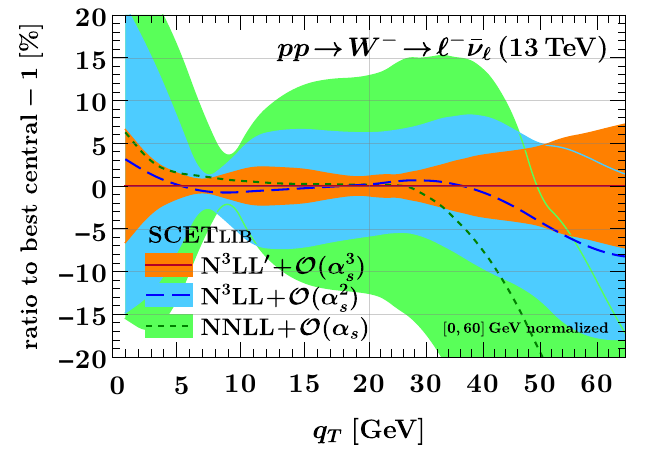}%
\\
\includegraphics[width=\WidthTwoSubfigs]{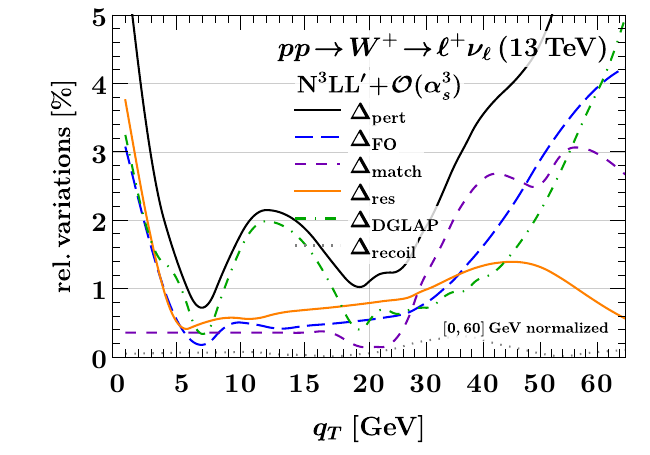}%
\hfill%
\includegraphics[width=\WidthTwoSubfigs]{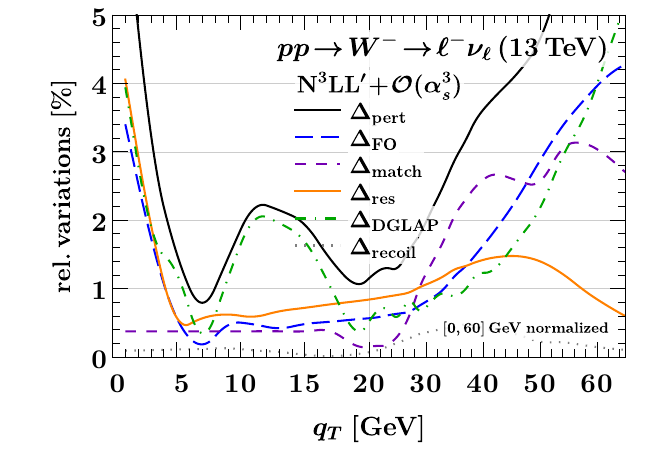}%
\caption{
The normalized transverse momentum spectrum (top), its relative difference to the highest-order prediction (middle),
and the complete perturbative uncertainty decomposition in terms of each conributing source (bottom) for $W^+$ (left)
and $W^-$ (right) production.
}
\label{fig:W_qT_fid_predictions}
\end{figure*}

\begin{table}
\begin{center}
\begin{tabular}{c | c | c | c}& NNLL$+\ord{\as}$ & N$^3$LL$+\ord{\as^2}$ & N$^3$LL$^\prime\!+\!\ord{\as^3}$
\\
\hline
$\sigma_{60}~[\mathrm{pb}]$ \quad \multrow{$W^+$ \\ $W^-$} & \multrow{4294.6 \\ 3281.5} & \multrow{4277.6 \\ 3257.8} & \multrow{4230.3 \\ 3220.1}
\end{tabular}
\caption{Normalization factors for the fiducial $p_T^W$ spectra shown in \fig{W_qT_fid_predictions}.}
\label{tab:norm_factors_fid_pTW}
\end{center}
\end{table}

We finally use our setup to provide predictions
for the transverse momentum spectrum of $W^+$ and $W^-$ bosons at the LHC.
We apply the following set of reference fiducial cuts,
\begin{align}
p_T^\ell > 25 \GeV
\,, \qquad
\abs{\eta_\ell} < 2.5
\,, \qquad
p_T^\nu > 25 \GeV
\,, \qquad
m_{T,W} > 50 \GeV
\end{align}
where the definition of a ``transverse mass''
often used in $W$ analyses reads
\begin{align}
m_{T,W}^2 \equiv \bigl(p_T^\ell + p_T^\nu\bigr)^2 - \bigl(\vec{p}_T^{\,\ell} + \vec{p}_T^{\,\nu}\bigr)^2
\,,\end{align}
and it is understood that the magnitude and direction of $\vec{p}_T^{\,\nu}$
are reconstructed from the missing transverse energy and transverse momentum in the event.
Our results for the normalized $p_T^{W}$ spectrum, its perturbative uncertainty,
and the perturbative uncertainty breakdown,
are shown in \fig{W_qT_fid_predictions}.
To account for the slightly larger nonsingular cross section
due to the lower effective value of $Q$,
we here choose to normalize our predictions to a reference range
of $0 \leq q_T \leq 60 \GeV$,
with normalization factors given in \tab{norm_factors_fid_pTW}.
We find that most features of the prediction,
notably including the perturbative convergence and coverage,
closely resemble those we found for the $p_T^Z$ spectrum.
An interesting difference between the two is found
when considering the ``factorization scale'' uncertainty $\Delta_\dglap$,
see \eq{def_delta_dglap},
which has a much more pronounced peak at $q_T \sim 12 \GeV$ in this case,
reaching a peak height of about $2 \%$.
As previously found for the $Z$, these oscillations are due to the discontinuity
of the PDF $\mu$ dependence at the bottom quark threshold,
and in this case -- to our understanding -- are more pronounced
because the sea and heavy quark channels (strange and charm) relevant here
are more susceptible than valence quarks
to the discontinuous change in the gluon PDF.
We note that an alternate, more direct way of assessing these kinds
of secondary heavy-quark effects would consist
of performing variations of the bottom-quark decoupling scale $\mu_b$
away from the canonical $\mu_b = m_b$.
\enlargethispage{\baselineskip}

\begin{figure*}
\includegraphics[width=\WidthTwoSubfigs]{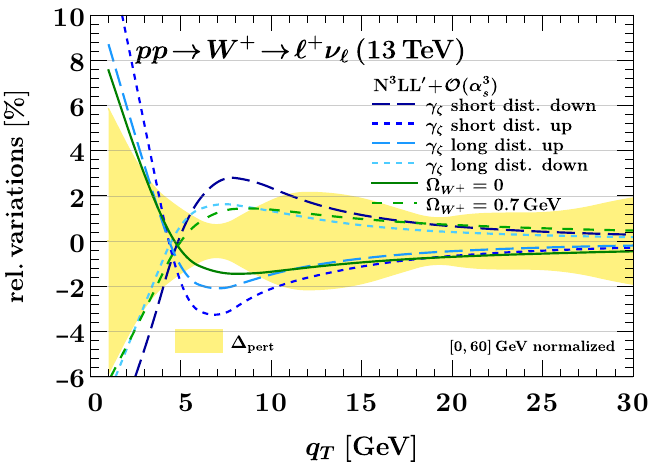}%
\hfill%
\includegraphics[width=\WidthTwoSubfigs]{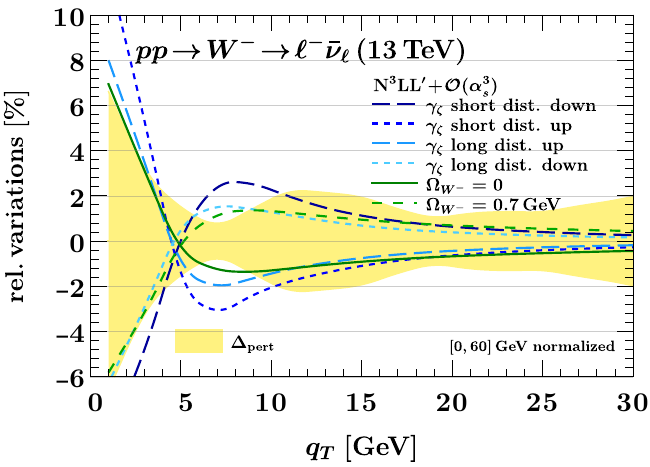}%
\caption{
Impact of parameter variations
in the nonperturbative TMD models
on the normalized $W^+$ (left) and $W^-$ (right) $p_T$ spectrum
at N$^3$LL$^\prime+\ord{\as^3}$.
For illustration, we also compare to the estimated size $\Delta_\pert$ of the perturbative uncertainty.
}
\label{fig:W_qT_np_uncertainties}
\end{figure*}

Finally, the impact of the nonperturbative parameter variations
described in \sec{uncerts_np} on the $p_T^W$ spectra is shown in \fig{W_qT_np_uncertainties}.
While the $W^+$ and $W^-$ cases closely resemble each other
and the $Z$ results in \fig{Z_qT_np_uncertainties},
we remind the reader that the interpretation and value
of the effective parameter $\Omega_V$ encoding
the TMD PDF boundary conditions is in principle different for each of $V = Z, W^+, W^-$,
and correlations between them must necessarily be predicted
from a full TMD flavor model.
By contrast, for an analysis  as in \refcite{CMS:2024lrd} involving \emph{only} $W$ bosons,
but no tuning to the $Z$, effective (rapidity-dependent) nonperturbative functions
are fully sufficient, as we proved in \secs{np_rap_model}{np_fid_model}.

\section{Summary and conclusions}
\label{sec:conclusions}

In this paper, we have provided state-of-the-art QCD precision
predictions for the transverse momentum $(q_T)$ spectra
of electroweak bosons at the LHC up to N$^3$LL$'$ and approximate N$^4$LL
in resummation-improved perturbation theory,
matched to available $\ord{\as^3}$ fixed-order results.
Our predictions fully account for the effect
of realistic fiducial selection cuts on the decay leptons,
incorporate the entire relevant and available perturbative information
at three, four, and five-loop order in QCD,
and feature a rigorously defined field-theoretic description
of the nonperturbative TMD physics at small $q_T \gtrsim \lqcd$.
We have placed particular emphasis on careful estimates
of the magnitude of residual perturbative uncertainties on our predictions,
also assessing in detail the impact of scheme choices made
to perform the matching between nonperturbative, resummed perturbative,
and fixed-order contributions to the spectrum.
This makes our predictions the first of their kind
to consistently incorporate all information
from the nonperturbative region of $q_T \gtrsim \lqcd$
all the way up to the fixed-order tail $q_T \sim m_Z, m_W$
with a complete assessment of the associated matching uncertainties.
In addition, we have studied the parametric strong coupling and PDF uncertainties
in detail, finding that NNLO PDF sets consistently overshoot the data
at intermediate $q_T$, a mismatch that is successfully resolved \emph{a priori}
by the recent aN$^3$LO PDF determinations.
By contrast, the additional N$^4$LL Sudakov resummation ingredients
at overall aN$^4$LL accuracy
only have a negligible impact beyond the baseline N$^3$LL$'$ prediction.
Our predictions are made possible by a fast, modular, and well-tested
implementation of N$^4$LL resummation with fiducial cuts
in the \texttt{SCETlib} numerical \texttt{C++}\ library.
This \texttt{scetlib-qT} module, for which a public release is foreseen,
has in the meantime already
found several applications~\cite{vonKuk:2023jfd, Alekhin:2024mrq, CMS:2024lrd, Tackmann:2024kci},
in particular as a key ingredient of the recent CMS $m_W$ measurement~\cite{CMS:2024lrd}.
\enlargethispage{\baselineskip}

Regarding the treatment of nonperturbative TMD physics, our analysis
for the first time clarifies in detail under what conditions
the -- in general very complicated -- flavor and Bjorken-$x$ dependence
of the nonperturbative quark TMD PDFs can be captured
by a much simpler effective nonperturbative function for a given process.
Indeed, we are able to formally derive the intuitive notion
that all nonperturbative effects
in the $p_T^V$ transverse momentum and rapidity spectrum
of a given resonantly produced electroweak boson $V$
at a given collider configuration
can be captured by a \emph{single effective one-dimensional function}
of the transverse distance $b_T$.
We also showed simple pocket formulas
to convert back to general flavor-dependent TMD models.
A similar result holds for the effective two-dimensional function
if the rapidity of $V$ (or the pseudorapidity of a decay lepton)
is measured in addition.
The complexity reduces further to a single nonperturbative number
(or a one-dimensional function of the rapidity)
if only the leading quadratic correction of $\ord{\lqcd^2 b_T^2}$ is considered,
generalizing the approach of \refcite{Ebert:2022cku}.
While at face value this may seem like a step backward
from the desired goal of TMD universality as assessed by global TMD fits,
our insights in fact have two important uses:
\begin{itemize}
   \item[(a)]
   If a self-contained analysis of a single process is performed,
   e.g.\ when extracting the strong coupling from resonant $Z$ production
   or the $W$ mass from $W$ boson data \emph{alone},
   our analysis clarifies the most general form of the nonperturbative physics
   that must be included in this case.
   In fact, all of our predictions in this paper for single-differential resonant
   $p_T^Z$ and $p_T^W$ spectra make heavy use of this,
   since it allows us to illustrate the most general effect of TMD physics
   (beyond that of the Collins-Soper kernel)
   by varying a single parameter $\Omega_V$ in each case.
   \item[(b)]
   Conversely, our statements about the validity of effective models
   can be read as precisely specifying the maximum information
   on nonperturbative TMD physics
   that can be extracted from a given $p_T^V$ spectrum
   in the context of a global fit.
   Specifically, in the idealized limit
   where the spectrum is measured and perturbatively predicted to perfect precision,
   it reduces the most general TMD nonperturbative parameter space
   by exactly one dimension at each order in $(\lqcd b_T)^{2n}$.
\end{itemize}

As an exciting immediate application of our predictions,
we have identified a strong sensitivity of the unnormalized \emph{cumulative}
$p_T^Z$ cross section to the physics of collinear PDFs,
with differences between recent PDF sets easily resolved
within our three-loop perturbative and the typical experimental uncertainties.
Our framework also allows us to fully assess the impact of nonperturbative physics
on the cumulative cross section, indicating that it is at the permille level
already for a cumulative cut at $\qTmax = 20 \GeV$.
Making use of the power expansion in $(p_T^Z/m_Z)^2$ valid in this region,
we furthermore expect that the small -- but numerically extremely expensive --
nonsingular cross section can be treated as a fixed bias correction
in future PDF fits to this observable,
while the PDF can easily be propagated exactly through the much cheaper
three-loop resummed cross section during the fit, also differential in rapidity.
Fits to the cumulative fiducial $p_T^Z$ cross section
at a value of $\qTmax = 20-30 \GeV$
thus provide a very promising avenue for a numerically inexpensive,
but nevertheless fully three-loop accurate theory template
for future improvements to approximate N$^3$LO PDF sets.

In summary, the predictions and conceptual advances in this paper
constitute another important step towards a rigorous
and comprehensive study of transverse momentum spectra at hadron colliders.
We look forward to future applications of our results
to precision QCD and electroweak physics at the LHC.
\enlargethispage{\baselineskip}

\acknowledgments

We gratefully acknowledge Markus Ebert
for collaboration during the early stages of this work,
by now belonging to a distant past,
and Rebecca von Kuk, Iain Stewart, and Zhiquan Sun
for discussion and collaboration on related work.
We would like to thank many members of the CMS $m_W$ analysis team,
and Josh Bendavid, Kenneth Long, and Simone Amoroso in particular,
for discussion and useful feedback on the \texttt{SCETlib} production code.
J.M.\ would like to thank Artur Avkhadiev, Phiala Shanahan, Michael Wagman,
and Yong Zhao for discussion on lattice extractions of the Collins-Soper kernel.

This work was supported in part by the Office of Nuclear Physics of the U.S.\
Department of Energy under Contract No.\ DE-SC0011090,
and within the framework of the TMD Topical Collaboration.
This work has received funding from the European Research Council (ERC)
under the European Union's Horizon 2020 research and innovation programme
(Grant agreement No. 101002090 COLORFREE).
G.B.\ was supported by MIUR through the FARE grant R18ZRBEAFC.
J.M.\ was supported by the D-ITP consortium, a program of NWO that is funded by the Dutch Ministry of Education, Culture and Science (OCW).

\appendix

\section{Renormalization group solutions at \texorpdfstring{N$^4$LL}{N4LL}}
\label{app:rge_solutions_n4ll}

The four-loop boundary condition for the rapidity anomalous dimension
at a general scale $\mu \sim 1/b_T$ is easily derived from \eq{gamma_nu_RGE}
in terms of the cusp anomalous dimension and the recently calculated
four-loop constant term $\tilde{\gamma}_{\nu\,3}^i$ at $\mu = b_0/b_T$~\cite{Moult:2022xzt, Duhr:2022cob}.
Using the conventions and notation of \refcite{Billis:2019vxg}, it reads
\begin{align}
\tilde{\gamma}_\nu^{i\,(3)}(b_T, \mu)
&= - L_b^4 \frac{1}{2} \beta_0^3 \Gamma_0^i
   + L_b^3 \bigl( \beta_0^3 \tilde{\gamma}_{\nu\,0}^i - \frac{5}{3} \beta_0 \beta_1 \Gamma_0^i - 2 \beta_0^2 \Gamma_1^i \bigr)
   \nn \\ & \quad
   + L_b^2 \Bigl( \frac{5}{2} \beta_0 \beta_1 \tilde{\gamma}_{\nu\,0}^i + 3 \beta_0^2 \tilde{\gamma}_{\nu\,1}^i  - \beta_2 \Gamma_0^i -2 \beta_1 \Gamma_1^i -3 \beta_0 \Gamma_2^i \Bigr)
   \nn \\ & \quad
   + L_b \bigl( \beta_2 \tilde{\gamma}_{\nu\,0}^i + 2\beta_1 \tilde{\gamma}_{\nu\,1}^i + 3 \beta_0 \tilde{\gamma}_{\nu\,2}^i - 2\Gamma_3^i \bigr)
   + \tilde{\gamma}_{\nu\,3}^i
\,,\end{align}
where $L_b = \ln \bigl(b_T^2 \mu^2/b_0^2 \bigr)$.

Next, using the conventions and nomenclature of \refcite{Billis:2019evv},
the N$^4$LL iterative solution for the running of the strong coupling reads
\begin{align}
\frac{\alpha(\mu_0)}{\alpha(\mu)}
&= X + \eps\,\frac{\alpha(\mu_0)}{4\pi}\, b_1 \ln X
+ \eps^2\, \frac{\alpha(\mu_0)^2}{(4\pi)^2}\biggl(b_2\, \frac{X-1}{X} + b_1^2\, \frac{1 -X +\ln X}{X} \biggr)
\nn \\ & \quad
+\eps^3\,\frac{\alpha(\mu_0)^3}{(4\pi)^3}
\biggl[
b_3\,\frac{X^2 - 1}{2X^2} + b_2 b_1\Bigl(\frac{1-X}{X} +\frac{\ln X}{X^2}\Bigr)
+ b_1^3\,\frac{(1-X)^2-\ln^2 X}{2X^2}
\biggr]
\nn \\ & \quad
+\eps^4 \frac{\alpha(\mu_0)^4}{(4\pi)^4} \frac{1}{6X^3}
\biggl[
(1-X) \Bigl( (2X^2-X -1) b_1^4 -6(X^2-1) b_1^2 b_2 + 2(X^2+X-2)b_2^2
\nn \\ & \qquad
+ (4X^2+X+1) b_1 b_3 -2(X^2+X+1) b_4 \Bigr)
+ 6 b_1 \Bigl( (X-1) b_1^3 + (1-X) b_1 b_2 + b_3 \Bigr) \ln X
\nn \\ & \qquad
- 3 (b_1^4 +2 b_1^2 b_2) \ln^2 X + 2 b_1^4 \ln^3 X
\biggr]
\,,\end{align}
where $b_n = \beta_n/\beta_0$, $X = 1 + \frac{\as(\mu_0)}{2\pi} \, \beta_0 \ln (\mu/\mu_0)$,
and $\eps = 1$ is a bookkeeping parameter for the order of the expansion.
Finally, using again the techniques and notation of \refcite{Billis:2019evv},
the N$^4$LL iterative solutions for the building blocks
of the Sudakov evolution kernels are given by
\begin{align}
K_{\Gamma}(\mu_0,\mu)
&= -\frac{\Gamma_0}{4\beta_0^2} \biggl\{
\frac{4\pi}{\alpha(\mu_0)}\Bigl(1-\frac{1}{r}-\ln r\Bigr)
+ \eps \Bigl[(\hat{\Gamma}_1-b_1)(1-r+\ln r)+\frac{b_1}{2}\ln^2 r\Bigr]
\nn \\ & \quad
+ \eps^2\, \frac{\alpha(\mu_0)}{4\pi} \biggl[
   (b_1^2-b_2)\Bigl(\frac{1-r^2}{2} + \ln r\Bigr) + (b_1\hat{\Gamma}_1-b_1^2)(1-r+r\ln r)
\nn \\ & \qquad
- (\hat{\Gamma}_2-b_1\hat{\Gamma}_1)\frac{(1-r)^2}{2} \biggr]
+ \eps^3\, \frac{\alpha(\mu_0)^2}{(4\pi)^2}
\biggl[(b_2 - b_1^2)(\hat{\Gamma}_1-b_1)\frac{(1-r)^2(2+r)}{3}
\nn \\ & \qquad
+ (\hat{\Gamma}_3-b_3-b_1(\hat{\Gamma}_2-b_2))\Bigl(\frac{1-r^3}{3}-\frac{1-r^2}{2}\Bigr)
\nn \\ & \qquad
+ b_1 (\hat{\Gamma}_2-b_2-b_1 (\hat{\Gamma}_1-b_1 ))\Bigl(\frac{1-r^2}{4}+\frac{r^2\ln r}{2}\Bigr)
\nn \\ & \qquad
+ (-b_3+2 b_1 b_2 -b_1^3 )\Bigl(\frac{1-r^2}{4}+\frac{\ln r}{2}\Bigr)
\biggr]
\nn \\ & \quad
+ \eps^4\, \frac{\alpha(\mu_0)^3}{(4\pi)^3}
\biggl[
\frac{b_3\hat{\Gamma}_1}{12} \bigl(5-6r-2r^3 + 3r^4\bigr)
+ \frac{b_2 \hat{\Gamma}_2}{4} (1-2r^2 +r^4 )
- \frac{\hat{\Gamma}_4}{12} (1 - 4r^3 + 3r^4)
\nn \\ & \qquad
- \frac{b_4}{36} \bigl( 1 + 8 r^3 - 9 r^4 + 12 \ln r \bigr)
- \frac{b_2^2}{36} \bigl( 5 - 18 r^2 + 4 r^3 + 9 r^4 - 12 \ln r \bigr)
\nn \\ & \qquad
-\frac{b_1^4}{12} \bigl( 7 - 6 r (1+r) + 2 r^3 + 3 r^4 - 4 \ln r + 4 r^3 \ln r \bigr)
\nn \\ & \qquad
+\frac{b_1^3 \hat{\Gamma}_1}{36} \bigl( 25 - 18r (1+r) + 2 r^3 + 9 r^4 + 12 r^3 \ln r \bigr)
\nn \\ & \qquad
-\frac{b_1^2 \hat{\Gamma}_2}{36} \bigl( 13 - 18 r^2 - 4 r^3 + 9 r^4 + 12 r^3 \ln r \bigr)
\nn \\ & \qquad
+\frac{b_2 b_1^2}{36} \bigl( 41 - 36 r (1+r) + 4 r^3 + 27 r^4 - 36 \ln r + 24 r^3 \ln r \bigr)
\nn \\ & \qquad
-\frac{b_3 b_1}{18} \bigl( 7 - 9 r - 7 r^3 + 9 r^4 - 12 \ln r + 6 r^3 \ln r \bigr)
+ \frac{b_1 \hat{\Gamma}_3}{36} \bigl( 7 - 16 r^3 + 9 r^4 + 12 r^3 \ln r \bigr)
\nn \\ & \qquad
- \frac{b_2 b_1 \hat{\Gamma}_1 }{18} \bigl( 20 - 18 r - 9 r^2 - 2 r^3 + 9 r^4 + 6 r^3 \ln r \bigr)
\biggr]
\biggr\}
\,,\\
K_\gamma (\mu_0,\mu)
&= -\frac{\gamma_0}{2\beta_0} \biggl[\eps\ln r
+ \eps^2\, \frac{\alpha(\mu_0)}{4\pi} (\hat{\gamma}_1-b_1)(r-1)
+ \eps^3\, \frac{\alpha(\mu_0)^2}{(4\pi)^2} (\hat{\gamma}_2
 -b_1\hat{\gamma}_1+b_1^2-b_2)\frac{r^2-1}{2}
\nn \\ & \quad
+ \eps^4\, \frac{\alpha(\mu_0)^3}{(4\pi)^3} \Bigl[\hat{\gamma}_3-b_3-b_1 (\hat{\gamma}_2-b_2)+(b_1^2-b_2
 )(\hat{\gamma}_1-b_1)\Bigr]\frac{r^3-1}{3}
\biggr]
\,,\\
\eta_\Gamma (\mu_0,\mu)
&= -\frac{\Gamma_0}{2\beta_0} \biggl[\ln r
+ \eps\, \frac{\alpha(\mu_0)}{4\pi} (\hat{\Gamma}_1-b_1)(r-1)
+ \eps^2\, \frac{\alpha(\mu_0)^2}{(4\pi)^2} (\hat{\Gamma}_2
 -b_1\hat{\Gamma}_1+b_1^2-b_2)\frac{r^2-1}{2}
\nn \\ & \quad
+ \eps^3\, \frac{\alpha(\mu_0)^3}{(4\pi)^3} \Bigl[\hat{\Gamma}_3-b_3-b_1 (\hat{\Gamma}_2-b_2)+(b_1^2-b_2
 )(\hat{\Gamma}_1-b_1)\Bigr]\frac{r^3-1}{3}
\nn \\ & \quad
+ \eps^4\, \frac{\alpha(\mu_0)^4}{(4\pi)^4}
\Bigl[ b_1^4 +b_2^2 - b_4 -b_1^3 \hat{\Gamma}_1 -b_3 \hat{\Gamma}_1 -b_2 \hat{\Gamma}_2
\nn \\ & \qquad \qquad \qquad \qquad
+ b_1^2 (\hat{\Gamma}_2 - 3 b_2) + b_1 (2b_3 + 2b_2 \hat{\Gamma}_1 - \hat{\Gamma}_3) + \hat{\Gamma}_4 \Bigr] \frac{r^4-1}{4}
\biggr]
\,,\end{align}
where in addition $\hat{\Gamma}_n = \Gamma_n/\Gamma_0$,
$\hat{\gamma}_n = \gamma_n/\gamma_0$,
and $r = \as(\mu)/\as(\mu_0)$.
\enlargethispage{\baselineskip}

\section{Reference results for spectra normalized to the full \texorpdfstring{$q_T$}{qT} range}
\label{app:results_Z_nnlojet_13tev_full_range}

\begin{figure*}
\centering
\includegraphics[width=\WidthTwoSubfigs]{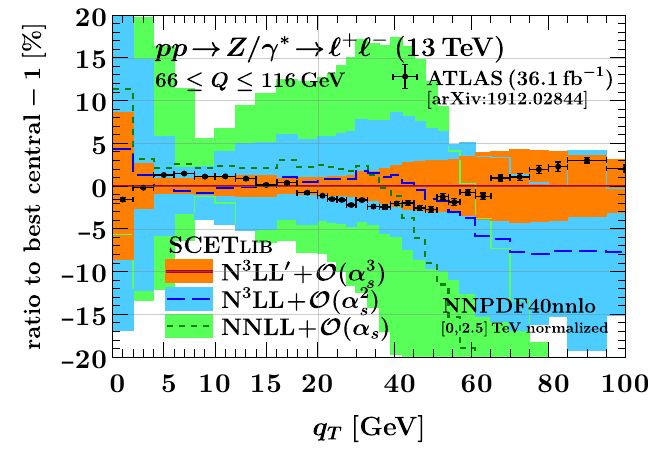}%
\caption{
Same as the center left panel of \fig{Z_qT_fid_predictions},
but normalizing to the whole kinematic range of $0\leq q_T\leq 2.5\TeV$ instead.
For the purposes of this plot, we in addition switched
to the \texttt{NNPDF40nnlo} PDF set as baseline,
cf.\ \fig{Z_qT_as_pdf_uncertainties},
and turned off the nonperturbative model for use as a reference.
Further modifications are described in the text.
The prediction at the highest order still visibly
overshoots the data for $q_T \geq 20 \GeV$
with these settings,
and is not compatible with it within uncertainty.
}
\label{fig:Z_qT_nnlojet_spect_ratio_norm_full_range}
\end{figure*}

\begin{table}
\begin{center}
\begin{tabular}{c | c | c | c}& NNLL$+\ord{\as}$ & N$^3$LL$+\ord{\as^2}$ & N$^3$LL$^\prime\!+\!\ord{\as^3}$
\\
\hline
$\sigma~[\mathrm{pb}]$ & 724.01 & 728.9 & 722.1
\end{tabular}
\caption{Normalization factors for the fiducial $p_T^Z$ spectrum shown in \fig{Z_qT_nnlojet_spect_ratio_norm_full_range}.}
\label{tab:norm_factors_fid_pTZ_full_range}
\end{center}
\end{table}

In this appendix we provide reference results for the fiducial $p_T^Z$ spectrum
normalized on the whole kinematic range, i.e., extending all the way
into the far fixed-order tail.
To do so, we directly use the fixed-order data
from \refcite{Chen:2022cgv} to perform the matching without additional modification,
also adopting the original slicing cut
of $q_T^\mathrm{cut} = 0.447 \GeV$ used in that reference,
and switch to \texttt{NNPDF40nnlo} as the baseline PDF set.
For consistency and ease of comparison with the resummed component of \refcite{Chen:2022cgv}
(and the integrated counterterm,
i.e., the fixed-order singular cross section,
in the N$^3$LO results of \refcite{Chen:2022cgv}),
we furthermore switch off all singlet terms in the hard function
and switch off the nonperturbative model by setting
\begin{align}
\omega_{\nu,q} = \Omega_Z = 0
\,.\end{align}
Our results with these settings are shown in \fig{Z_qT_nnlojet_spect_ratio_norm_full_range},
with normalization factors (i.e., total fiducial Drell-Yan cross sections)
reported in \tab{norm_factors_fid_pTZ_full_range}.
As further consistency checks against the fixed-order results reported
in \refcite{Chen:2022cgv},
we have verified that the LO cross section agrees in all significant digits,
and that the $\ord{\as^3}$ increment to the fixed-order cross section,
reported as $-18.7(1.1) \pb$ in \refcite{Chen:2022cgv},
is in numerical agreement with our result of $-19.0(1.1) \pb$ within numerical uncertainty.%
\footnote{
We note that the bulk of the numerical uncertainty here is due to the common \texttt{NNLOjet}
nonsingular contribution, and thus is fully correlated between the two results.
The relevant uncertainties instead are the integration uncertainties
on the respective fixed-order singular components,
which for the \texttt{RadISH} calculation in \refcite{Chen:2022cgv}
come out to $\pm 0.4 \pb$ (and are negligible on the \texttt{SCETlib} side).
We would like to thank Luca Rottoli and Pier Monni for their help with this comparison.
}

\addcontentsline{toc}{section}{References}
\bibliographystyle{jhep}
\bibliography{refs}

\providecommand{\href}[2]{#2}\begingroup\raggedright\begin{thebibliography}{100}

\bibitem{Aad:2011fp}
{\scshape ATLAS} collaboration, \emph{{Measurement of the Transverse Momentum
  Distribution of $W$ Bosons in $pp$ Collisions at $\sqrt{s}=7$ TeV with the
  ATLAS Detector}},
  \href{https://doi.org/10.1103/PhysRevD.85.012005}{\emph{Phys. Rev. D}
  {\bfseries 85} (2012) 012005}
  [\href{https://arxiv.org/abs/1108.6308}{{\ttfamily 1108.6308}}].

\bibitem{Aad:2014xaa}
{\scshape ATLAS} collaboration, \emph{{Measurement of the $Z/\gamma^*$ boson
  transverse momentum distribution in $pp$ collisions at $\sqrt{s}$ = 7 TeV
  with the ATLAS detector}},
  \href{https://doi.org/10.1007/JHEP09(2014)145}{\emph{JHEP} {\bfseries 09}
  (2014) 145} [\href{https://arxiv.org/abs/1406.3660}{{\ttfamily 1406.3660}}].

\bibitem{Aad:2015auj}
{\scshape ATLAS} collaboration, \emph{{Measurement of the transverse momentum
  and $\phi^*_{\eta }$ distributions of Drell-Yan lepton pairs in proton-proton
  collisions at $\sqrt{s}=8$ TeV with the ATLAS detector}},
  \href{https://doi.org/10.1140/epjc/s10052-016-4070-4}{\emph{Eur. Phys. J.}
  {\bfseries C76} (2016) 291}
  [\href{https://arxiv.org/abs/1512.02192}{{\ttfamily 1512.02192}}].

\bibitem{Aaboud:2017ffb}
{\scshape ATLAS} collaboration, \emph{{Measurement of the Drell-Yan
  triple-differential cross section in $pp$ collisions at $\sqrt{s} = 8$ TeV}},
  \href{https://doi.org/10.1007/JHEP12(2017)059}{\emph{JHEP} {\bfseries 12}
  (2017) 059} [\href{https://arxiv.org/abs/1710.05167}{{\ttfamily
  1710.05167}}].

\bibitem{Aad:2019wmn}
{\scshape ATLAS} collaboration, \emph{{Measurement of the transverse momentum
  distribution of Drell-Yan lepton pairs in proton-proton collisions at
  $\sqrt{s}=13$ TeV with the ATLAS detector}},
  \href{https://doi.org/10.1140/epjc/s10052-020-8001-z}{\emph{Eur. Phys. J. C}
  {\bfseries 80} (2020) 616}
  [\href{https://arxiv.org/abs/1912.02844}{{\ttfamily 1912.02844}}].

\bibitem{ATLAS:2023lsr}
{\scshape ATLAS} collaboration, \emph{{A precise measurement of the Z-boson
  double-differential transverse momentum and rapidity distributions in the
  full phase space of the decay leptons with the ATLAS experiment
  at~$\sqrt{s}=8$~TeV}},
  \href{https://doi.org/10.1140/epjc/s10052-024-12438-w}{\emph{Eur. Phys. J. C}
  {\bfseries 84} (2024) 315}
  [\href{https://arxiv.org/abs/2309.09318}{{\ttfamily 2309.09318}}].

\bibitem{Chatrchyan:2011wt}
{\scshape CMS} collaboration, \emph{{Measurement of the Rapidity and Transverse
  Momentum Distributions of $Z$ Bosons in $pp$ Collisions at $\sqrt{s}=7$
  TeV}}, \href{https://doi.org/10.1103/PhysRevD.85.032002}{\emph{Phys. Rev. D}
  {\bfseries 85} (2012) 032002}
  [\href{https://arxiv.org/abs/1110.4973}{{\ttfamily 1110.4973}}].

\bibitem{Khachatryan:2015oaa}
{\scshape CMS} collaboration, \emph{{Measurement of the Z boson differential
  cross section in transverse momentum and rapidity in proton--proton
  collisions at 8 TeV}},
  \href{https://doi.org/10.1016/j.physletb.2015.07.065}{\emph{Phys. Lett. B}
  {\bfseries 749} (2015) 187}
  [\href{https://arxiv.org/abs/1504.03511}{{\ttfamily 1504.03511}}].

\bibitem{Khachatryan:2016nbe}
{\scshape CMS} collaboration, \emph{{Measurement of the transverse momentum
  spectra of weak vector bosons produced in proton-proton collisions at $
  \sqrt{s}=8 $ TeV}},
  \href{https://doi.org/10.1007/JHEP02(2017)096}{\emph{JHEP} {\bfseries 02}
  (2017) 096} [\href{https://arxiv.org/abs/1606.05864}{{\ttfamily
  1606.05864}}].

\bibitem{Sirunyan:2017igm}
{\scshape CMS} collaboration, \emph{{Measurement of differential cross sections
  in the kinematic angular variable $\phi^*$ for inclusive Z boson production
  in pp collisions at $\sqrt{s}=$ 8 TeV}},
  \href{https://doi.org/10.1007/JHEP03(2018)172}{\emph{JHEP} {\bfseries 03}
  (2018) 172} [\href{https://arxiv.org/abs/1710.07955}{{\ttfamily
  1710.07955}}].

\bibitem{Sirunyan:2019bzr}
{\scshape CMS} collaboration, \emph{{Measurements of differential Z boson
  production cross sections in proton-proton collisions at $\sqrt{s}=$ 13
  TeV}}, \href{https://doi.org/10.1007/JHEP12(2019)061}{\emph{JHEP} {\bfseries
  12} (2019) 061} [\href{https://arxiv.org/abs/1909.04133}{{\ttfamily
  1909.04133}}].

\bibitem{LHCb:2015mad}
{\scshape LHCb} collaboration, R.~Aaij et~al., \emph{{Measurement of forward W
  and Z boson production in $pp$ collisions at $ \sqrt{s}=8 $ TeV}},
  \href{https://doi.org/10.1007/JHEP01(2016)155}{\emph{JHEP} {\bfseries 01}
  (2016) 155} [\href{https://arxiv.org/abs/1511.08039}{{\ttfamily
  1511.08039}}].

\bibitem{LHCb:2016fbk}
{\scshape LHCb} collaboration, R.~Aaij et~al., \emph{{Measurement of the
  forward Z boson production cross-section in pp collisions at $\sqrt{s} = 13$
  TeV}}, \href{https://doi.org/10.1007/JHEP09(2016)136}{\emph{JHEP} {\bfseries
  09} (2016) 136} [\href{https://arxiv.org/abs/1607.06495}{{\ttfamily
  1607.06495}}].

\bibitem{Lindert:2017olm}
J.~M. Lindert et~al., \emph{{Precise predictions for $V+$ jets dark matter
  backgrounds}},
  \href{https://doi.org/10.1140/epjc/s10052-017-5389-1}{\emph{Eur. Phys. J. C}
  {\bfseries 77} (2017) 829}
  [\href{https://arxiv.org/abs/1705.04664}{{\ttfamily 1705.04664}}].

\bibitem{Camarda:2022qdg}
S.~Camarda, G.~Ferrera and M.~Schott, \emph{{Determination of the
  strong-coupling constant from the Z-boson transverse-momentum distribution}},
  \href{https://doi.org/10.1140/epjc/s10052-023-12373-2}{\emph{Eur. Phys. J. C}
  {\bfseries 84} (2024) 39} [\href{https://arxiv.org/abs/2203.05394}{{\ttfamily
  2203.05394}}].

\bibitem{ATLAS:2023lhg}
{\scshape ATLAS} collaboration, \emph{{A precise determination of the
  strong-coupling constant from the recoil of $Z$ bosons with the ATLAS
  experiment at $\sqrt{s} = 8$ TeV}},
  \href{https://arxiv.org/abs/2309.12986}{{\ttfamily 2309.12986}}.

\bibitem{Aaboud:2017svj}
{\scshape ATLAS} collaboration, \emph{{Measurement of the $W$-boson mass in pp
  collisions at $\sqrt{s}=7$ TeV with the ATLAS detector}},
  \href{https://doi.org/10.1140/epjc/s10052-018-6354-3,
  10.1140/epjc/s10052-017-5475-4}{\emph{Eur. Phys. J.} {\bfseries C78} (2018)
  110} [\href{https://arxiv.org/abs/1701.07240}{{\ttfamily 1701.07240}}].

\bibitem{LHCb:2021bjt}
{\scshape LHCb} collaboration, R.~Aaij et~al., \emph{{Measurement of the W
  boson mass}}, \href{https://doi.org/10.1007/JHEP01(2022)036}{\emph{JHEP}
  {\bfseries 01} (2022) 036}
  [\href{https://arxiv.org/abs/2109.01113}{{\ttfamily 2109.01113}}].

\bibitem{CDF:2022hxs}
{\scshape CDF} collaboration, T.~Aaltonen et~al., \emph{{High-precision
  measurement of the $W$ boson mass with the CDF II detector}},
  \href{https://doi.org/10.1126/science.abk1781}{\emph{Science} {\bfseries 376}
  (2022) 170}.

\bibitem{ATLAS:2024erm}
{\scshape ATLAS} collaboration, \emph{{Measurement of the W-boson mass and
  width with the ATLAS detector using proton-proton collisions at $\sqrt{s}$ =
  7 TeV}},  \href{https://arxiv.org/abs/2403.15085}{{\ttfamily 2403.15085}}.

\bibitem{CMS:2024lrd}
{\scshape CMS} collaboration, \emph{{High-precision measurement of the W boson
  mass with the CMS experiment at the LHC}},
  \href{https://arxiv.org/abs/2412.13872}{{\ttfamily 2412.13872}}.

\bibitem{Ebert:2020dfc}
M.~A. Ebert, J.~K.~L. Michel, I.~W. Stewart and F.~J. Tackmann,
  \emph{{Drell-Yan $q_{T}$ resummation of fiducial power corrections at
  N$^{3}$LL}}, \href{https://doi.org/10.1007/JHEP04(2021)102}{\emph{JHEP}
  {\bfseries 04} (2021) 102}
  [\href{https://arxiv.org/abs/2006.11382}{{\ttfamily 2006.11382}}].

\bibitem{Billis:2021ecs}
G.~Billis, B.~Dehnadi, M.~A. Ebert, J.~K.~L. Michel and F.~J. Tackmann,
  \emph{{Higgs $p_T$ Spectrum and Total Cross Section with Fiducial Cuts at
  Third Resummed and Fixed Order in QCD}},
  \href{https://doi.org/10.1103/PhysRevLett.127.072001}{\emph{Phys. Rev. Lett.}
  {\bfseries 127} (2021) 072001}
  [\href{https://arxiv.org/abs/2102.08039}{{\ttfamily 2102.08039}}].

\bibitem{Salam:2021tbm}
G.~P. Salam and E.~Slade, \emph{{Cuts for two-body decays at colliders}},
  \href{https://doi.org/10.1007/JHEP11(2021)220}{\emph{JHEP} {\bfseries 11}
  (2021) 220} [\href{https://arxiv.org/abs/2106.08329}{{\ttfamily
  2106.08329}}].

\bibitem{Amoroso:2022lxw}
S.~Amoroso, L.~A. Bella, M.~Boonekamp, S.~Camarda, A.~Glazov, A.~Guida et~al.,
  \emph{{Drell-Yan cross-sections with fiducial cuts: impact of linear power
  corrections and $q_T$-resummation in PDF determination}},
  \href{https://arxiv.org/abs/2209.13535}{{\ttfamily 2209.13535}}.

\bibitem{Alekhin:2024mrq}
S.~Alekhin et~al., \emph{{Status of QCD precision predictions for Drell-Yan
  processes}},  \href{https://arxiv.org/abs/2405.19714}{{\ttfamily
  2405.19714}}.

\bibitem{McGowan:2022nag}
J.~McGowan, T.~Cridge, L.~A. Harland-Lang and R.~S. Thorne, \emph{{Approximate
  N$^{3}$LO parton distribution functions with theoretical uncertainties:
  MSHT20aN$^3$LO PDFs}},
  \href{https://doi.org/10.1140/epjc/s10052-023-11236-0}{\emph{Eur. Phys. J. C}
  {\bfseries 83} (2023) 185}
  [\href{https://arxiv.org/abs/2207.04739}{{\ttfamily 2207.04739}}].

\bibitem{NNPDF:2024nan}
{\scshape NNPDF} collaboration, R.~D. Ball et~al., \emph{{The path to $\hbox
  {N}^3\hbox {LO}$ parton distributions}},
  \href{https://doi.org/10.1140/epjc/s10052-024-12891-7}{\emph{Eur. Phys. J. C}
  {\bfseries 84} (2024) 659}
  [\href{https://arxiv.org/abs/2402.18635}{{\ttfamily 2402.18635}}].

\bibitem{Cridge:2024exf}
T.~Cridge, L.~A. Harland-Lang and R.~S. Thorne, \emph{{A first determination of
  the strong coupling $\alpha _S$ at approximate $\textrm{N}^3$LO order in a
  global PDF fit}},
  \href{https://doi.org/10.1140/epjc/s10052-024-13320-5}{\emph{Eur. Phys. J. C}
  {\bfseries 84} (2024) 1009}
  [\href{https://arxiv.org/abs/2404.02964}{{\ttfamily 2404.02964}}].

\bibitem{Cooper-Sarkar:2024crx}
A.~Cooper-Sarkar, T.~Cridge, F.~Giuli, L.~A. Harland-Lang, F.~Hekhorn,
  J.~Huston et~al., \emph{{A Benchmarking of QCD Evolution at Approximate
  $N^3LO$}},  \href{https://arxiv.org/abs/2406.16188}{{\ttfamily 2406.16188}}.

\bibitem{MSHT:2024tdn}
{\scshape MSHT, NNPDF} collaboration, T.~Cridge et~al., \emph{{Combination of
  aN$^3$LO PDFs and implications for Higgs production cross-sections at the
  LHC}},  \href{https://arxiv.org/abs/2411.05373}{{\ttfamily 2411.05373}}.

\bibitem{Ridder:2015dxa}
A.~Gehrmann-De~Ridder, T.~Gehrmann, E.~Glover, A.~Huss and T.~Morgan,
  \emph{{Precise QCD predictions for the production of a Z boson in association
  with a hadronic jet}},
  \href{https://doi.org/10.1103/PhysRevLett.117.022001}{\emph{Phys. Rev. Lett.}
  {\bfseries 117} (2016) 022001}
  [\href{https://arxiv.org/abs/1507.02850}{{\ttfamily 1507.02850}}].

\bibitem{Ridder:2016nkl}
A.~Gehrmann-De~Ridder, T.~Gehrmann, E.~Glover, A.~Huss and T.~Morgan,
  \emph{{The NNLO QCD corrections to Z boson production at large transverse
  momentum}}, \href{https://doi.org/10.1007/JHEP07(2016)133}{\emph{JHEP}
  {\bfseries 07} (2016) 133}
  [\href{https://arxiv.org/abs/1605.04295}{{\ttfamily 1605.04295}}].

\bibitem{Boughezal:2015dva}
R.~Boughezal, C.~Focke, X.~Liu and F.~Petriello, \emph{{$W$-boson production in
  association with a jet at next-to-next-to-leading order in perturbative
  QCD}}, \href{https://doi.org/10.1103/PhysRevLett.115.062002}{\emph{Phys. Rev.
  Lett.} {\bfseries 115} (2015) 062002}
  [\href{https://arxiv.org/abs/1504.02131}{{\ttfamily 1504.02131}}].

\bibitem{Boughezal:2015ded}
R.~Boughezal, J.~M. Campbell, R.~Ellis, C.~Focke, W.~T. Giele, X.~Liu et~al.,
  \emph{{Z-boson production in association with a jet at
  next-to-next-to-leading order in perturbative QCD}},
  \href{https://doi.org/10.1103/PhysRevLett.116.152001}{\emph{Phys. Rev. Lett.}
  {\bfseries 116} (2016) 152001}
  [\href{https://arxiv.org/abs/1512.01291}{{\ttfamily 1512.01291}}].

\bibitem{Boughezal:2016isb}
R.~Boughezal, X.~Liu and F.~Petriello, \emph{{Phenomenology of the Z-boson plus
  jet process at NNLO}},
  \href{https://doi.org/10.1103/PhysRevD.94.074015}{\emph{Phys. Rev. D}
  {\bfseries 94} (2016) 074015}
  [\href{https://arxiv.org/abs/1602.08140}{{\ttfamily 1602.08140}}].

\bibitem{Boughezal:2016dtm}
R.~Boughezal, X.~Liu and F.~Petriello, \emph{{W-boson plus jet differential
  distributions at NNLO in QCD}},
  \href{https://doi.org/10.1103/PhysRevD.94.113009}{\emph{Phys. Rev. D}
  {\bfseries 94} (2016) 113009}
  [\href{https://arxiv.org/abs/1602.06965}{{\ttfamily 1602.06965}}].

\bibitem{Gehrmann-DeRidder:2017mvr}
A.~Gehrmann-De~Ridder, T.~Gehrmann, E.~Glover, A.~Huss and D.~Walker,
  \emph{{Next-to-Next-to-Leading-Order QCD Corrections to the Transverse
  Momentum Distribution of Weak Gauge Bosons}},
  \href{https://doi.org/10.1103/PhysRevLett.120.122001}{\emph{Phys. Rev. Lett.}
  {\bfseries 120} (2018) 122001}
  [\href{https://arxiv.org/abs/1712.07543}{{\ttfamily 1712.07543}}].

\bibitem{Campbell:2019gmd}
J.~M. Campbell, R.~K. Ellis and S.~Seth, \emph{{H + 1 jet production
  revisited}}, \href{https://doi.org/10.1007/JHEP10(2019)136}{\emph{JHEP}
  {\bfseries 10} (2019) 136}
  [\href{https://arxiv.org/abs/1906.01020}{{\ttfamily 1906.01020}}].

\bibitem{Neumann:2022lft}
T.~Neumann and J.~Campbell, \emph{{Fiducial Drell-Yan production at the LHC
  improved by transverse-momentum resummation at N$^4$LL$^\prime$+N$^3$LO}},
  \href{https://doi.org/10.1103/PhysRevD.107.L011506}{\emph{Phys. Rev. D}
  {\bfseries 107} (2023) L011506}
  [\href{https://arxiv.org/abs/2207.07056}{{\ttfamily 2207.07056}}].

\bibitem{Ju:2021lah}
W.-L. Ju and M.~Sch\"onherr, \emph{{The q$_{T}$ and
  \ensuremath{\Delta}\ensuremath{\phi} spectra in W and Z production at the LHC
  at N$^{3}$LL'+N$^{2}$LO}},
  \href{https://doi.org/10.1007/JHEP10(2021)088}{\emph{JHEP} {\bfseries 10}
  (2021) 088} [\href{https://arxiv.org/abs/2106.11260}{{\ttfamily
  2106.11260}}].

\bibitem{Re:2021con}
E.~Re, L.~Rottoli and P.~Torrielli, \emph{{Fiducial Higgs and Drell-Yan
  distributions at N$^3$LL$^\prime$+NNLO with RadISH}},
  \href{https://arxiv.org/abs/2104.07509}{{\ttfamily 2104.07509}}.

\bibitem{Chen:2022cgv}
X.~Chen, T.~Gehrmann, E.~W.~N. Glover, A.~Huss, P.~F. Monni, E.~Re et~al.,
  \emph{{Third-Order Fiducial Predictions for Drell-Yan Production at the
  LHC}}, \href{https://doi.org/10.1103/PhysRevLett.128.252001}{\emph{Phys. Rev.
  Lett.} {\bfseries 128} (2022) 252001}
  [\href{https://arxiv.org/abs/2203.01565}{{\ttfamily 2203.01565}}].

\bibitem{Camarda:2023dqn}
S.~Camarda, L.~Cieri and G.~Ferrera, \emph{{Drell\textendash{}Yan lepton-pair
  production: $q_T$ resummation at N$^4$LL accuracy}},
  \href{https://doi.org/10.1016/j.physletb.2023.138125}{\emph{Phys. Lett. B}
  {\bfseries 845} (2023) 138125}
  [\href{https://arxiv.org/abs/2303.12781}{{\ttfamily 2303.12781}}].

\bibitem{Moos:2023yfa}
V.~Moos, I.~Scimemi, A.~Vladimirov and P.~Zurita, \emph{{Extraction of
  unpolarized transverse momentum distributions from the fit of Drell-Yan data
  at N$^{4}$LL}}, \href{https://doi.org/10.1007/JHEP05(2024)036}{\emph{JHEP}
  {\bfseries 05} (2024) 036}
  [\href{https://arxiv.org/abs/2305.07473}{{\ttfamily 2305.07473}}].

\bibitem{Piloneta:2024aac}
S.~Piloneta and A.~Vladimirov, \emph{{Angular distributions of Drell-Yan
  leptons in the TMD factorization approach}},
  \href{https://arxiv.org/abs/2407.06277}{{\ttfamily 2407.06277}}.

\bibitem{Alioli:2015toa}
S.~Alioli, C.~W. Bauer, C.~Berggren, F.~J. Tackmann and J.~R. Walsh,
  \emph{{Drell-Yan production at NNLL$'+$NNLO matched to parton showers}},
  \href{https://doi.org/10.1103/PhysRevD.92.094020}{\emph{Phys. Rev. D}
  {\bfseries 92} (2015) 094020}
  [\href{https://arxiv.org/abs/1508.01475}{{\ttfamily 1508.01475}}].

\bibitem{Alioli:2021qbf}
S.~Alioli, C.~W. Bauer, A.~Broggio, A.~Gavardi, S.~Kallweit, M.~A. Lim et~al.,
  \emph{{Matching NNLO predictions to parton showers using N3LL color-singlet
  transverse momentum resummation in geneva}},
  \href{https://doi.org/10.1103/PhysRevD.104.094020}{\emph{Phys. Rev. D}
  {\bfseries 104} (2021) 094020}
  [\href{https://arxiv.org/abs/2102.08390}{{\ttfamily 2102.08390}}].

\bibitem{Alioli:2023rxx}
S.~Alioli, G.~Bell, G.~Billis, A.~Broggio, B.~Dehnadi, M.~A. Lim et~al.,
  \emph{{N3LL resummation of one-jettiness for Z-boson plus jet production at
  hadron colliders}},
  \href{https://doi.org/10.1103/PhysRevD.109.094009}{\emph{Phys. Rev. D}
  {\bfseries 109} (2024) 094009}
  [\href{https://arxiv.org/abs/2312.06496}{{\ttfamily 2312.06496}}].

\bibitem{Bacchetta:2022awv}
{\scshape MAP (Multi-dimensional Analyses of Partonic distributions)}
  collaboration, A.~Bacchetta, V.~Bertone, C.~Bissolotti, G.~Bozzi, M.~Cerutti,
  F.~Piacenza et~al., \emph{{Unpolarized transverse momentum distributions from
  a global fit of Drell-Yan and semi-inclusive deep-inelastic scattering
  data}}, \href{https://doi.org/10.1007/JHEP10(2022)127}{\emph{JHEP} {\bfseries
  10} (2022) 127} [\href{https://arxiv.org/abs/2206.07598}{{\ttfamily
  2206.07598}}].

\bibitem{Bacchetta:2024qre}
{\scshape MAP} collaboration, A.~Bacchetta, V.~Bertone, C.~Bissolotti,
  G.~Bozzi, M.~Cerutti, F.~Delcarro et~al., \emph{{Flavor dependence of
  unpolarized quark transverse momentum distributions from a global fit}},
  \href{https://doi.org/10.1007/JHEP08(2024)232}{\emph{JHEP} {\bfseries 08}
  (2024) 232} [\href{https://arxiv.org/abs/2405.13833}{{\ttfamily
  2405.13833}}].

\bibitem{Dittmaier:2014qza}
S.~Dittmaier, A.~Huss and C.~Schwinn, \emph{{Mixed QCD-electroweak
  $O(\alpha_s\alpha)$ corrections to Drell-Yan processes in the resonance
  region: pole approximation and non-factorizable corrections}},
  \href{https://doi.org/10.1016/j.nuclphysb.2014.05.027}{\emph{Nucl. Phys.}
  {\bfseries B885} (2014) 318}
  [\href{https://arxiv.org/abs/1403.3216}{{\ttfamily 1403.3216}}].

\bibitem{Dittmaier:2015rxo}
S.~Dittmaier, A.~Huss and C.~Schwinn, \emph{{Dominant mixed QCD-electroweak
  O($\alpha_s\alpha$) corrections to Drell–Yan processes in the resonance
  region}}, \href{https://doi.org/10.1016/j.nuclphysb.2016.01.006}{\emph{Nucl.
  Phys.} {\bfseries B904} (2016) 216}
  [\href{https://arxiv.org/abs/1511.08016}{{\ttfamily 1511.08016}}].

\bibitem{deFlorian:2018wcj}
D.~de~Florian, M.~Der and I.~Fabre, \emph{{QCD$\oplus$QED NNLO corrections to
  Drell Yan production}},
  \href{https://doi.org/10.1103/PhysRevD.98.094008}{\emph{Phys. Rev.}
  {\bfseries D98} (2018) 094008}
  [\href{https://arxiv.org/abs/1805.12214}{{\ttfamily 1805.12214}}].

\bibitem{Delto:2019ewv}
M.~Delto, M.~Jaquier, K.~Melnikov and R.~R{\"o}ntsch, \emph{{Mixed
  QCD$\otimes$QED corrections to on-shell $Z$ boson production at the LHC}},
  \href{https://doi.org/10.1007/JHEP01(2020)043}{\emph{JHEP} {\bfseries 01}
  (2020) 043} [\href{https://arxiv.org/abs/1909.08428}{{\ttfamily
  1909.08428}}].

\bibitem{Bonciani:2019nuy}
R.~Bonciani, F.~Buccioni, N.~Rana, I.~Triscari and A.~Vicini, \emph{{NNLO
  QCD$\times$EW corrections to Z production in the $q\bar{q}$ channel}},
  \href{https://doi.org/10.1103/PhysRevD.101.031301}{\emph{Phys. Rev.}
  {\bfseries D101} (2020) 031301}
  [\href{https://arxiv.org/abs/1911.06200}{{\ttfamily 1911.06200}}].

\bibitem{Cieri:2020ikq}
L.~Cieri, D.~de~Florian, M.~Der and J.~Mazzitelli, \emph{{Mixed
  QCD\ensuremath{\otimes}QED corrections to exclusive Drell Yan production
  using the q$_{T}$ -subtraction method}},
  \href{https://doi.org/10.1007/JHEP09(2020)155}{\emph{JHEP} {\bfseries 09}
  (2020) 155} [\href{https://arxiv.org/abs/2005.01315}{{\ttfamily
  2005.01315}}].

\bibitem{Buccioni:2020cfi}
F.~Buccioni, F.~Caola, M.~Delto, M.~Jaquier, K.~Melnikov and R.~R\"ontsch,
  \emph{{Mixed QCD-electroweak corrections to on-shell Z production at the
  LHC}}, \href{https://doi.org/10.1016/j.physletb.2020.135969}{\emph{Phys.
  Lett. B} {\bfseries 811} (2020) 135969}
  [\href{https://arxiv.org/abs/2005.10221}{{\ttfamily 2005.10221}}].

\bibitem{Cieri:2018sfk}
L.~Cieri, G.~Ferrera and G.~F. Sborlini, \emph{{Combining QED and QCD
  transverse-momentum resummation for Z boson production at hadron colliders}},
  \href{https://doi.org/10.1007/JHEP08(2018)165}{\emph{JHEP} {\bfseries 08}
  (2018) 165} [\href{https://arxiv.org/abs/1805.11948}{{\ttfamily
  1805.11948}}].

\bibitem{Bacchetta:2018dcq}
A.~Bacchetta and M.~G. Echevarria, \emph{{QCD$\times$QED evolution of TMDs}},
  \href{https://doi.org/10.1016/j.physletb.2018.11.019}{\emph{Phys. Lett. B}
  {\bfseries 788} (2019) 280}
  [\href{https://arxiv.org/abs/1810.02297}{{\ttfamily 1810.02297}}].

\bibitem{Billis:2019evv}
G.~Billis, F.~J. Tackmann and J.~Talbert, \emph{{Higher-Order Sudakov
  Resummation in Coupled Gauge Theories}},
  \href{https://doi.org/10.1007/JHEP03(2020)182}{\emph{JHEP} {\bfseries 03}
  (2020) 182} [\href{https://arxiv.org/abs/1907.02971}{{\ttfamily
  1907.02971}}].

\bibitem{Buonocore:2024xmy}
L.~Buonocore, L.~Rottoli and P.~Torrielli, \emph{{Resummation of combined
  QCD-electroweak effects in Drell Yan lepton-pair production}},
  \href{https://doi.org/10.1007/JHEP07(2024)193}{\emph{JHEP} {\bfseries 07}
  (2024) 193} [\href{https://arxiv.org/abs/2404.15112}{{\ttfamily
  2404.15112}}].

\bibitem{scetlib}
M.~A. Ebert, J.~K.~L. Michel, F.~J. Tackmann et~al., \emph{{SCETlib: A C++
  Package for Numerical Calculations in QCD and Soft-Collinear Effective
  Theory}}, {\emph{DESY-17-099} (2018) }.

\bibitem{Collins:1977iv}
J.~C. Collins and D.~E. Soper, \emph{{Angular Distribution of Dileptons in
  High-Energy Hadron Collisions}},
  \href{https://doi.org/10.1103/PhysRevD.16.2219}{\emph{Phys. Rev.} {\bfseries
  D16} (1977) 2219}.

\bibitem{Collins:1981uk}
J.~C. Collins and D.~E. Soper, \emph{{Back-To-Back Jets in QCD}},
  \href{https://doi.org/10.1016/0550-3213(81)90339-4}{\emph{Nucl. Phys.}
  {\bfseries B193} (1981) 381}.

\bibitem{Collins:1981va}
J.~C. Collins and D.~E. Soper, \emph{{Back-To-Back Jets: Fourier Transform from
  B to K-Transverse}},
  \href{https://doi.org/10.1016/0550-3213(82)90453-9}{\emph{Nucl. Phys.}
  {\bfseries B197} (1982) 446}.

\bibitem{Collins:1984kg}
J.~C. Collins, D.~E. Soper and G.~F. Sterman, \emph{{Transverse Momentum
  Distribution in Drell-Yan Pair and W and Z Boson Production}},
  \href{https://doi.org/10.1016/0550-3213(85)90479-1}{\emph{Nucl. Phys.}
  {\bfseries B250} (1985) 199}.

\bibitem{Collins:1350496}
J.~Collins, \emph{{Foundations of perturbative QCD}}, Cambridge monographs on
  particle physics, nuclear physics, and cosmology. Cambridge Univ. Press, New
  York, NY, 2011.

\bibitem{Bauer:2000ew}
C.~W. Bauer, S.~Fleming and M.~E. Luke, \emph{{Summing Sudakov logarithms in $B
  \to X_s\gamma$ in effective field theory}},
  \href{https://doi.org/10.1103/PhysRevD.63.014006}{\emph{Phys. Rev.}
  {\bfseries D63} (2000) 014006}
  [\href{https://arxiv.org/abs/hep-ph/0005275}{{\ttfamily hep-ph/0005275}}].

\bibitem{Bauer:2000yr}
C.~W. Bauer, S.~Fleming, D.~Pirjol and I.~W. Stewart, \emph{{An Effective field
  theory for collinear and soft gluons: Heavy to light decays}},
  \href{https://doi.org/10.1103/PhysRevD.63.114020}{\emph{Phys. Rev.}
  {\bfseries D63} (2001) 114020}
  [\href{https://arxiv.org/abs/hep-ph/0011336}{{\ttfamily hep-ph/0011336}}].

\bibitem{Bauer:2001yt}
C.~W. Bauer, D.~Pirjol and I.~W. Stewart, \emph{{Soft collinear factorization
  in effective field theory}},
  \href{https://doi.org/10.1103/PhysRevD.65.054022}{\emph{Phys. Rev.}
  {\bfseries D65} (2002) 054022}
  [\href{https://arxiv.org/abs/hep-ph/0109045}{{\ttfamily hep-ph/0109045}}].

\bibitem{Bauer:2002nz}
C.~W. Bauer, S.~Fleming, D.~Pirjol, I.~Z. Rothstein and I.~W. Stewart,
  \emph{{Hard scattering factorization from effective field theory}},
  \href{https://doi.org/10.1103/PhysRevD.66.014017}{\emph{Phys. Rev.}
  {\bfseries D66} (2002) 014017}
  [\href{https://arxiv.org/abs/hep-ph/0202088}{{\ttfamily hep-ph/0202088}}].

\bibitem{Becher:2010tm}
T.~Becher and M.~Neubert, \emph{{Drell-Yan Production at Small $q_T$,
  Transverse Parton Distributions and the Collinear Anomaly}},
  \href{https://doi.org/10.1140/epjc/s10052-011-1665-7}{\emph{Eur. Phys. J.}
  {\bfseries C71} (2011) 1665}
  [\href{https://arxiv.org/abs/1007.4005}{{\ttfamily 1007.4005}}].

\bibitem{GarciaEchevarria:2011rb}
M.~G. Echevarria, A.~Idilbi and I.~Scimemi, \emph{{Factorization Theorem For
  Drell-Yan At Low $q_T$ And Transverse Momentum Distributions
  On-The-Light-Cone}},
  \href{https://doi.org/10.1007/JHEP07(2012)002}{\emph{JHEP} {\bfseries 07}
  (2012) 002} [\href{https://arxiv.org/abs/1111.4996}{{\ttfamily 1111.4996}}].

\bibitem{Chiu:2012ir}
J.-Y. Chiu, A.~Jain, D.~Neill and I.~Z. Rothstein, \emph{{A Formalism for the
  Systematic Treatment of Rapidity Logarithms in Quantum Field Theory}},
  \href{https://doi.org/10.1007/JHEP05(2012)084}{\emph{JHEP} {\bfseries 05}
  (2012) 084} [\href{https://arxiv.org/abs/1202.0814}{{\ttfamily 1202.0814}}].

\bibitem{Li:2016axz}
Y.~Li, D.~Neill and H.~X. Zhu, \emph{{An exponential regulator for rapidity
  divergences}},
  \href{https://doi.org/10.1016/j.nuclphysb.2020.115193}{\emph{Nucl. Phys. B}
  {\bfseries 960} (2020) 115193}
  [\href{https://arxiv.org/abs/1604.00392}{{\ttfamily 1604.00392}}].

\bibitem{Bacchetta:2008xw}
A.~Bacchetta, D.~Boer, M.~Diehl and P.~J. Mulders, \emph{{Matches and
  mismatches in the descriptions of semi-inclusive processes at low and high
  transverse momentum}},
  \href{https://doi.org/10.1088/1126-6708/2008/08/023}{\emph{JHEP} {\bfseries
  08} (2008) 023} [\href{https://arxiv.org/abs/0803.0227}{{\ttfamily
  0803.0227}}].

\bibitem{vonKuk:2023jfd}
R.~von Kuk, J.~K.~L. Michel and Z.~Sun, \emph{{Transverse momentum
  distributions of heavy hadrons and polarized heavy quarks}},
  \href{https://doi.org/10.1007/JHEP09(2023)205}{\emph{JHEP} {\bfseries 09}
  (2023) 205} [\href{https://arxiv.org/abs/2305.15461}{{\ttfamily
  2305.15461}}].

\bibitem{Gehrmann:2010ue}
T.~Gehrmann, E.~W.~N. Glover, T.~Huber, N.~Ikizlerli and C.~Studerus,
  \emph{{Calculation of the quark and gluon form factors to three loops in
  QCD}}, \href{https://doi.org/10.1007/JHEP06(2010)094}{\emph{JHEP} {\bfseries
  06} (2010) 094} [\href{https://arxiv.org/abs/1004.3653}{{\ttfamily
  1004.3653}}].

\bibitem{Baikov:2009bg}
P.~A. Baikov, K.~G. Chetyrkin, A.~V. Smirnov, V.~A. Smirnov and M.~Steinhauser,
  \emph{{Quark and gluon form factors to three loops}},
  \href{https://doi.org/10.1103/PhysRevLett.102.212002}{\emph{Phys. Rev. Lett.}
  {\bfseries 102} (2009) 212002}
  [\href{https://arxiv.org/abs/0902.3519}{{\ttfamily 0902.3519}}].

\bibitem{Dicus:1985wx}
D.~A. Dicus and S.~S. Willenbrock, \emph{{Radiative Corrections to the Ratio of
  $Z$ and $W$ Boson Production}},
  \href{https://doi.org/10.1103/PhysRevD.34.148}{\emph{Phys. Rev. D} {\bfseries
  34} (1986) 148}.

\bibitem{Kniehl:1989qu}
B.~A. Kniehl and J.~H. Kuhn, \emph{{QCD Corrections to the Z Decay Rate}},
  \href{https://doi.org/10.1016/0550-3213(90)90070-T}{\emph{Nucl. Phys.}
  {\bfseries B329} (1990) 547}.

\bibitem{Bernreuther:2005rw}
W.~Bernreuther, R.~Bonciani, T.~Gehrmann, R.~Heinesch, T.~Leineweber and
  E.~Remiddi, \emph{{Two-loop QCD corrections to the heavy quark form-factors:
  Anomaly contributions}},
  \href{https://doi.org/10.1016/j.nuclphysb.2005.06.025}{\emph{Nucl. Phys. B}
  {\bfseries 723} (2005) 91}
  [\href{https://arxiv.org/abs/hep-ph/0504190}{{\ttfamily hep-ph/0504190}}].

\bibitem{Gehrmann:2021ahy}
T.~Gehrmann and A.~Primo, \emph{{The three-loop singlet contribution to the
  massless axial-vector quark form factor}},
  \href{https://doi.org/10.1016/j.physletb.2021.136223}{\emph{Phys. Lett. B}
  {\bfseries 816} (2021) 136223}
  [\href{https://arxiv.org/abs/2102.12880}{{\ttfamily 2102.12880}}].

\bibitem{Chen:2021rft}
L.~Chen, M.~Czakon and M.~Niggetiedt, \emph{{The complete singlet contribution
  to the massless quark form factor at three loops in QCD}},
  \href{https://doi.org/10.1007/JHEP12(2021)095}{\emph{JHEP} {\bfseries 12}
  (2021) 095} [\href{https://arxiv.org/abs/2109.01917}{{\ttfamily
  2109.01917}}].

\bibitem{Collins:1992tv}
J.~C. Collins and F.~V. Tkachov, \emph{{Breakdown of dimensional regularization
  in the Sudakov problem}},
  \href{https://doi.org/10.1016/0370-2693(92)91541-G}{\emph{Phys. Lett.}
  {\bfseries B294} (1992) 403}
  [\href{https://arxiv.org/abs/hep-ph/9208209}{{\ttfamily hep-ph/9208209}}].

\bibitem{Collins:2008ht}
J.~Collins, \emph{{Rapidity divergences and valid definitions of parton
  densities}}, \href{https://doi.org/10.22323/1.061.0028}{\emph{PoS} {\bfseries
  LC2008} (2008) 028} [\href{https://arxiv.org/abs/0808.2665}{{\ttfamily
  0808.2665}}].

\bibitem{Chiu:2011qc}
J.-y. Chiu, A.~Jain, D.~Neill and I.~Z. Rothstein, \emph{{The Rapidity
  Renormalization Group}},
  \href{https://doi.org/10.1103/PhysRevLett.108.151601}{\emph{Phys. Rev. Lett.}
  {\bfseries 108} (2012) 151601}
  [\href{https://arxiv.org/abs/1104.0881}{{\ttfamily 1104.0881}}].

\bibitem{Collins:1981uw}
J.~C. Collins and D.~E. Soper, \emph{{Parton Distribution and Decay
  Functions}}, \href{https://doi.org/10.1016/0550-3213(82)90021-9}{\emph{Nucl.
  Phys.} {\bfseries B194} (1982) 445}.

\bibitem{Catani:2011kr}
S.~Catani and M.~Grazzini, \emph{{Higgs Boson Production at Hadron Colliders:
  Hard-Collinear Coefficients at the NNLO}},
  \href{https://doi.org/10.1140/epjc/s10052-012-2013-2,
  10.1140/epjc/s10052-012-2132-9}{\emph{Eur. Phys. J.} {\bfseries C72} (2012)
  2013} [\href{https://arxiv.org/abs/1106.4652}{{\ttfamily 1106.4652}}].

\bibitem{Catani:2012qa}
S.~Catani, L.~Cieri, D.~de~Florian, G.~Ferrera and M.~Grazzini, \emph{{Vector
  boson production at hadron colliders: hard-collinear coefficients at the
  NNLO}}, \href{https://doi.org/10.1140/epjc/s10052-012-2195-7}{\emph{Eur.
  Phys. J.} {\bfseries C72} (2012) 2195}
  [\href{https://arxiv.org/abs/1209.0158}{{\ttfamily 1209.0158}}].

\bibitem{Gehrmann:2014yya}
T.~Gehrmann, T.~L{\"u}bbert and L.~L. Yang, \emph{{Calculation of the
  transverse parton distribution functions at next-to-next-to-leading order}},
  \href{https://doi.org/10.1007/JHEP06(2014)155}{\emph{JHEP} {\bfseries 06}
  (2014) 155} [\href{https://arxiv.org/abs/1403.6451}{{\ttfamily 1403.6451}}].

\bibitem{Luebbert:2016itl}
T.~L{\"u}bbert, J.~Oredsson and M.~Stahlhofen, \emph{{Rapidity renormalized TMD
  soft and beam functions at two loops}},
  \href{https://doi.org/10.1007/JHEP03(2016)168}{\emph{JHEP} {\bfseries 03}
  (2016) 168} [\href{https://arxiv.org/abs/1602.01829}{{\ttfamily
  1602.01829}}].

\bibitem{Echevarria:2015byo}
M.~G. Echevarria, I.~Scimemi and A.~Vladimirov, \emph{{Universal transverse
  momentum dependent soft function at NNLO}},
  \href{https://doi.org/10.1103/PhysRevD.93.054004}{\emph{Phys. Rev.}
  {\bfseries D93} (2016) 054004}
  [\href{https://arxiv.org/abs/1511.05590}{{\ttfamily 1511.05590}}].

\bibitem{Echevarria:2016scs}
M.~G. Echevarria, I.~Scimemi and A.~Vladimirov, \emph{{Unpolarized Transverse
  Momentum Dependent Parton Distribution and Fragmentation Functions at
  next-to-next-to-leading order}},
  \href{https://doi.org/10.1007/JHEP09(2016)004}{\emph{JHEP} {\bfseries 09}
  (2016) 004} [\href{https://arxiv.org/abs/1604.07869}{{\ttfamily
  1604.07869}}].

\bibitem{Li:2016ctv}
Y.~Li and H.~X. Zhu, \emph{{Bootstrapping Rapidity Anomalous Dimensions for
  Transverse-Momentum Resummation}},
  \href{https://doi.org/10.1103/PhysRevLett.118.022004}{\emph{Phys. Rev. Lett.}
  {\bfseries 118} (2017) 022004}
  [\href{https://arxiv.org/abs/1604.01404}{{\ttfamily 1604.01404}}].

\bibitem{Luo:2019hmp}
M.-X. Luo, X.~Wang, X.~Xu, L.~L. Yang, T.-Z. Yang and H.~X. Zhu,
  \emph{{Transverse Parton Distribution and Fragmentation Functions at NNLO:
  the Quark Case}}, \href{https://doi.org/10.1007/JHEP10(2019)083}{\emph{JHEP}
  {\bfseries 10} (2019) 083}
  [\href{https://arxiv.org/abs/1908.03831}{{\ttfamily 1908.03831}}].

\bibitem{Luo:2019szz}
M.-x. Luo, T.-Z. Yang, H.~X. Zhu and Y.~J. Zhu, \emph{{Quark Transverse Parton
  Distribution at the Next-to-Next-to-Next-to-Leading Order}},
  \href{https://doi.org/10.1103/PhysRevLett.124.092001}{\emph{Phys. Rev. Lett.}
  {\bfseries 124} (2020) 092001}
  [\href{https://arxiv.org/abs/1912.05778}{{\ttfamily 1912.05778}}].

\bibitem{Ebert:2020yqt}
M.~A. Ebert, B.~Mistlberger and G.~Vita, \emph{{Transverse momentum dependent
  PDFs at N$^3$LO}}, \href{https://doi.org/10.1007/JHEP09(2020)146}{\emph{JHEP}
  {\bfseries 09} (2020) 146}
  [\href{https://arxiv.org/abs/2006.05329}{{\ttfamily 2006.05329}}].

\bibitem{Lubbert:2016rku}
T.~L\"ubbert, J.~Oredsson and M.~Stahlhofen, \emph{{Rapidity renormalized TMD
  soft and beam functions at two loops}},
  \href{https://doi.org/10.1007/JHEP03(2016)168}{\emph{JHEP} {\bfseries 03}
  (2016) 168} [\href{https://arxiv.org/abs/1602.01829}{{\ttfamily
  1602.01829}}].

\bibitem{Frixione:1998dw}
S.~Frixione, P.~Nason and G.~Ridolfi, \emph{{Problems in the resummation of
  soft gluon effects in the transverse momentum distributions of massive vector
  bosons in hadronic collisions}},
  \href{https://doi.org/10.1016/S0550-3213(98)00853-0}{\emph{Nucl. Phys.}
  {\bfseries B542} (1999) 311}
  [\href{https://arxiv.org/abs/hep-ph/9809367}{{\ttfamily hep-ph/9809367}}].

\bibitem{Monni:2016ktx}
P.~F. Monni, E.~Re and P.~Torrielli, \emph{{Higgs Transverse-Momentum
  Resummation in Direct Space}},
  \href{https://doi.org/10.1103/PhysRevLett.116.242001}{\emph{Phys. Rev. Lett.}
  {\bfseries 116} (2016) 242001}
  [\href{https://arxiv.org/abs/1604.02191}{{\ttfamily 1604.02191}}].

\bibitem{Ebert:2016gcn}
M.~A. Ebert and F.~J. Tackmann, \emph{{Resummation of Transverse Momentum
  Distributions in Distribution Space}},
  \href{https://doi.org/10.1007/JHEP02(2017)110}{\emph{JHEP} {\bfseries 02}
  (2017) 110} [\href{https://arxiv.org/abs/1611.08610}{{\ttfamily
  1611.08610}}].

\bibitem{Bhattacharya:2022dtm}
A.~Bhattacharya, M.~D. Schwartz and X.~Zhang, \emph{{Sudakov shoulder
  resummation for thrust and heavy jet mass}},
  \href{https://doi.org/10.1103/PhysRevD.106.074011}{\emph{Phys. Rev. D}
  {\bfseries 106} (2022) 074011}
  [\href{https://arxiv.org/abs/2205.05702}{{\ttfamily 2205.05702}}].

\bibitem{Bhattacharya:2023qet}
A.~Bhattacharya, J.~K.~L. Michel, M.~D. Schwartz, I.~W. Stewart and X.~Zhang,
  \emph{{NNLL resummation of Sudakov shoulder logarithms in the heavy jet mass
  distribution}}, \href{https://doi.org/10.1007/JHEP11(2023)080}{\emph{JHEP}
  {\bfseries 11} (2023) 080}
  [\href{https://arxiv.org/abs/2306.08033}{{\ttfamily 2306.08033}}].

\bibitem{Korchemsky:1987wg}
G.~P. Korchemsky and A.~V. Radyushkin, \emph{{Renormalization of the Wilson
  Loops Beyond the Leading Order}},
  \href{https://doi.org/10.1016/0550-3213(87)90277-X}{\emph{Nucl. Phys.}
  {\bfseries B283} (1987) 342}.

\bibitem{Moch:2004pa}
S.~Moch, J.~A.~M. Vermaseren and A.~Vogt, \emph{{The Three loop splitting
  functions in QCD: The Nonsinglet case}},
  \href{https://doi.org/10.1016/j.nuclphysb.2004.03.030}{\emph{Nucl. Phys.}
  {\bfseries B688} (2004) 101}
  [\href{https://arxiv.org/abs/hep-ph/0403192}{{\ttfamily hep-ph/0403192}}].

\bibitem{Vogt:2004mw}
A.~Vogt, S.~Moch and J.~A.~M. Vermaseren, \emph{{The Three-loop splitting
  functions in QCD: The Singlet case}},
  \href{https://doi.org/10.1016/j.nuclphysb.2004.04.024}{\emph{Nucl. Phys.}
  {\bfseries B691} (2004) 129}
  [\href{https://arxiv.org/abs/hep-ph/0404111}{{\ttfamily hep-ph/0404111}}].

\bibitem{Lee:2016ixa}
J.~Henn, A.~V. Smirnov, V.~A. Smirnov, M.~Steinhauser and R.~N. Lee,
  \emph{{Four-loop photon quark form factor and cusp anomalous dimension in the
  large-$N_c$ limit of QCD}},
  \href{https://doi.org/10.1007/JHEP03(2017)139}{\emph{JHEP} {\bfseries 03}
  (2017) 139} [\href{https://arxiv.org/abs/1612.04389}{{\ttfamily
  1612.04389}}].

\bibitem{Moch:2017uml}
S.~Moch, B.~Ruijl, T.~Ueda, J.~Vermaseren and A.~Vogt, \emph{{Four-Loop
  Non-Singlet Splitting Functions in the Planar Limit and Beyond}},
  \href{https://doi.org/10.1007/JHEP10(2017)041}{\emph{JHEP} {\bfseries 10}
  (2017) 041} [\href{https://arxiv.org/abs/1707.08315}{{\ttfamily
  1707.08315}}].

\bibitem{Lee:2019zop}
R.~N. Lee, A.~V. Smirnov, V.~A. Smirnov and M.~Steinhauser, \emph{{Four-loop
  quark form factor with quartic fundamental colour factor}},
  \href{https://doi.org/10.1007/JHEP02(2019)172}{\emph{JHEP} {\bfseries 02}
  (2019) 172} [\href{https://arxiv.org/abs/1901.02898}{{\ttfamily
  1901.02898}}].

\bibitem{Henn:2019rmi}
J.~Henn, T.~Peraro, M.~Stahlhofen and P.~Wasser, \emph{{Matter dependence of
  the four-loop cusp anomalous dimension}},
  \href{https://doi.org/10.1103/PhysRevLett.122.201602}{\emph{Phys. Rev. Lett.}
  {\bfseries 122} (2019) 201602}
  [\href{https://arxiv.org/abs/1901.03693}{{\ttfamily 1901.03693}}].

\bibitem{Bruser:2019auj}
R.~Brüser, A.~Grozin, J.~M. Henn and M.~Stahlhofen, \emph{{Matter dependence
  of the four-loop QCD cusp anomalous dimension: from small angles to all
  angles}}, \href{https://doi.org/10.1007/JHEP05(2019)186}{\emph{JHEP}
  {\bfseries 05} (2019) 186}
  [\href{https://arxiv.org/abs/1902.05076}{{\ttfamily 1902.05076}}].

\bibitem{Henn:2019swt}
J.~M. Henn, G.~P. Korchemsky and B.~Mistlberger, \emph{{The full four-loop cusp
  anomalous dimension in $\mathcal{N}=4$ super Yang-Mills and QCD}},
  \href{https://doi.org/10.1007/JHEP04(2020)018}{\emph{JHEP} {\bfseries 04}
  (2020) 018} [\href{https://arxiv.org/abs/1911.10174}{{\ttfamily
  1911.10174}}].

\bibitem{vonManteuffel:2020vjv}
A.~von Manteuffel, E.~Panzer and R.~M. Schabinger, \emph{{Analytic four-loop
  anomalous dimensions in massless QCD from form factors}},
  \href{https://doi.org/10.1103/PhysRevLett.124.162001}{\emph{Phys. Rev. Lett.}
  {\bfseries 124} (2020) 162001}
  [\href{https://arxiv.org/abs/2002.04617}{{\ttfamily 2002.04617}}].

\bibitem{Tarasov:1980au}
O.~V. Tarasov, A.~A. Vladimirov and A.~{\relax Yu}. Zharkov, \emph{{The
  Gell-Mann-Low Function of QCD in the Three Loop Approximation}},
  \href{https://doi.org/10.1016/0370-2693(80)90358-5}{\emph{Phys. Lett.}
  {\bfseries 93B} (1980) 429}.

\bibitem{Larin:1993tp}
S.~A. Larin and J.~A.~M. Vermaseren, \emph{{The Three loop QCD Beta function
  and anomalous dimensions}},
  \href{https://doi.org/10.1016/0370-2693(93)91441-O}{\emph{Phys. Lett.}
  {\bfseries B303} (1993) 334}
  [\href{https://arxiv.org/abs/hep-ph/9302208}{{\ttfamily hep-ph/9302208}}].

\bibitem{vanRitbergen:1997va}
T.~van Ritbergen, J.~A.~M. Vermaseren and S.~A. Larin, \emph{{The Four loop
  beta function in quantum chromodynamics}},
  \href{https://doi.org/10.1016/S0370-2693(97)00370-5}{\emph{Phys. Lett.}
  {\bfseries B400} (1997) 379}
  [\href{https://arxiv.org/abs/hep-ph/9701390}{{\ttfamily hep-ph/9701390}}].

\bibitem{Czakon:2004bu}
M.~Czakon, \emph{{The Four-loop QCD beta-function and anomalous dimensions}},
  \href{https://doi.org/10.1016/j.nuclphysb.2005.01.012}{\emph{Nucl. Phys.}
  {\bfseries B710} (2005) 485}
  [\href{https://arxiv.org/abs/hep-ph/0411261}{{\ttfamily hep-ph/0411261}}].

\bibitem{Herzog:2017ohr}
F.~Herzog, B.~Ruijl, T.~Ueda, J.~A.~M. Vermaseren and A.~Vogt, \emph{{The
  five-loop beta function of Yang-Mills theory with fermions}},
  \href{https://doi.org/10.1007/JHEP02(2017)090}{\emph{JHEP} {\bfseries 02}
  (2017) 090} [\href{https://arxiv.org/abs/1701.01404}{{\ttfamily
  1701.01404}}].

\bibitem{Herzog:2018kwj}
F.~Herzog, S.~Moch, B.~Ruijl, T.~Ueda, J.~A.~M. Vermaseren and A.~Vogt,
  \emph{{Five-loop contributions to low-N non-singlet anomalous dimensions in
  QCD}}, \href{https://doi.org/10.1016/j.physletb.2019.01.060}{\emph{Phys.
  Lett. B} {\bfseries 790} (2019) 436}
  [\href{https://arxiv.org/abs/1812.11818}{{\ttfamily 1812.11818}}].

\bibitem{Davies:2016jie}
J.~Davies, A.~Vogt, B.~Ruijl, T.~Ueda and J.~Vermaseren, \emph{{Large-$n_f$
  contributions to the four-loop splitting functions in QCD}},
  \href{https://doi.org/10.1016/j.nuclphysb.2016.12.012}{\emph{Nucl. Phys. B}
  {\bfseries 915} (2017) 335}
  [\href{https://arxiv.org/abs/1610.07477}{{\ttfamily 1610.07477}}].

\bibitem{Moch:2018wjh}
S.~Moch, B.~Ruijl, T.~Ueda, J.~M. Vermaseren and A.~Vogt, \emph{{On quartic
  colour factors in splitting functions and the gluon cusp anomalous
  dimension}},
  \href{https://doi.org/10.1016/j.physletb.2018.06.017}{\emph{Phys. Lett. B}
  {\bfseries 782} (2018) 627}
  [\href{https://arxiv.org/abs/1805.09638}{{\ttfamily 1805.09638}}].

\bibitem{Das:2019btv}
G.~Das, S.-O. Moch and A.~Vogt, \emph{{Soft corrections to inclusive
  deep-inelastic scattering at four loops and beyond}},
  \href{https://doi.org/10.1007/JHEP03(2020)116}{\emph{JHEP} {\bfseries 03}
  (2020) 116} [\href{https://arxiv.org/abs/1912.12920}{{\ttfamily
  1912.12920}}].

\bibitem{Das:2020adl}
G.~Das, S.~Moch and A.~Vogt, \emph{{Approximate four-loop QCD corrections to
  the Higgs-boson production cross section}},
  \href{https://doi.org/10.1016/j.physletb.2020.135546}{\emph{Phys. Lett. B}
  {\bfseries 807} (2020) 135546}
  [\href{https://arxiv.org/abs/2004.00563}{{\ttfamily 2004.00563}}].

\bibitem{Agarwal:2021zft}
B.~Agarwal, A.~von Manteuffel, E.~Panzer and R.~M. Schabinger, \emph{{Four-loop
  collinear anomalous dimensions in QCD and $\mathcal{N}$=4 super Yang-Mills}},
  \href{https://doi.org/10.1016/j.physletb.2021.136503}{\emph{Phys. Lett. B}
  {\bfseries 820} (2021) 136503}
  [\href{https://arxiv.org/abs/2102.09725}{{\ttfamily 2102.09725}}].

\bibitem{Billis:2019vxg}
G.~Billis, M.~A. Ebert, J.~K.~L. Michel and F.~J. Tackmann, \emph{{A toolbox
  for $q_{T}$ and 0-jettiness subtractions at $\hbox {N}^3\hbox {LO}$}},
  \href{https://doi.org/10.1140/epjp/s13360-021-01155-y}{\emph{Eur. Phys. J.
  Plus} {\bfseries 136} (2021) 214}
  [\href{https://arxiv.org/abs/1909.00811}{{\ttfamily 1909.00811}}].

\bibitem{Bruser:2018rad}
R.~Br{\"u}ser, Z.~L. Liu and M.~Stahlhofen, \emph{{Three-Loop Quark Jet
  Function}}, \href{https://doi.org/10.1103/PhysRevLett.121.072003}{\emph{Phys.
  Rev. Lett.} {\bfseries 121} (2018) 072003}
  [\href{https://arxiv.org/abs/1804.09722}{{\ttfamily 1804.09722}}].

\bibitem{Ebert:2020unb}
M.~A. Ebert, B.~Mistlberger and G.~Vita, \emph{{$N$-jettiness beam functions at
  N$^{3}$LO}}, \href{https://doi.org/10.1007/JHEP09(2020)143}{\emph{JHEP}
  {\bfseries 09} (2020) 143}
  [\href{https://arxiv.org/abs/2006.03056}{{\ttfamily 2006.03056}}].

\bibitem{Vladimirov:2016dll}
A.~A. Vladimirov, \emph{{Correspondence between Soft and Rapidity Anomalous
  Dimensions}},
  \href{https://doi.org/10.1103/PhysRevLett.118.062001}{\emph{Phys. Rev. Lett.}
  {\bfseries 118} (2017) 062001}
  [\href{https://arxiv.org/abs/1610.05791}{{\ttfamily 1610.05791}}].

\bibitem{Duhr:2022yyp}
C.~Duhr, B.~Mistlberger and G.~Vita, \emph{{Four-Loop Rapidity Anomalous
  Dimension and Event Shapes to Fourth Logarithmic Order}},
  \href{https://doi.org/10.1103/PhysRevLett.129.162001}{\emph{Phys. Rev. Lett.}
  {\bfseries 129} (2022) 162001}
  [\href{https://arxiv.org/abs/2205.02242}{{\ttfamily 2205.02242}}].

\bibitem{Moult:2022xzt}
I.~Moult, H.~X. Zhu and Y.~J. Zhu, \emph{{The four loop QCD rapidity anomalous
  dimension}}, \href{https://doi.org/10.1007/JHEP08(2022)280}{\emph{JHEP}
  {\bfseries 08} (2022) 280}
  [\href{https://arxiv.org/abs/2205.02249}{{\ttfamily 2205.02249}}].

\bibitem{Duhr:2022cob}
C.~Duhr, B.~Mistlberger and G.~Vita, \emph{{Soft integrals and soft anomalous
  dimensions at N$^{3}$LO and beyond}},
  \href{https://doi.org/10.1007/JHEP09(2022)155}{\emph{JHEP} {\bfseries 09}
  (2022) 155} [\href{https://arxiv.org/abs/2205.04493}{{\ttfamily
  2205.04493}}].

\bibitem{Ebert:2017uel}
M.~A. Ebert, J.~K.~L. Michel and F.~J. Tackmann, \emph{{Resummation Improved
  Rapidity Spectrum for Gluon Fusion Higgs Production}},
  \href{https://doi.org/10.1007/JHEP05(2017)088}{\emph{JHEP} {\bfseries 05}
  (2017) 088} [\href{https://arxiv.org/abs/1702.00794}{{\ttfamily
  1702.00794}}].

\bibitem{Luo:2020epw}
M.-x. Luo, T.-Z. Yang, H.~X. Zhu and Y.~J. Zhu, \emph{{Unpolarized quark and
  gluon TMD PDFs and FFs at N$^{3}$LO}},
  \href{https://doi.org/10.1007/JHEP06(2021)115}{\emph{JHEP} {\bfseries 06}
  (2021) 115} [\href{https://arxiv.org/abs/2012.03256}{{\ttfamily
  2012.03256}}].

\bibitem{Lee:2010cga}
R.~N. Lee, A.~V. Smirnov and V.~A. Smirnov, \emph{{Analytic Results for
  Massless Three-Loop Form Factors}},
  \href{https://doi.org/10.1007/JHEP04(2010)020}{\emph{JHEP} {\bfseries 04}
  (2010) 020} [\href{https://arxiv.org/abs/1001.2887}{{\ttfamily 1001.2887}}].

\bibitem{Bertone:2022sso}
V.~Bertone, G.~Bozzi and F.~Hautmann, \emph{{Perturbative hysteresis and
  emergent resummation scales}},
  \href{https://doi.org/10.1103/PhysRevD.105.096003}{\emph{Phys. Rev. D}
  {\bfseries 105} (2022) 096003}
  [\href{https://arxiv.org/abs/2202.03380}{{\ttfamily 2202.03380}}].

\bibitem{Bertone:2024snr}
V.~Bertone, G.~Bozzi and F.~Hautmann, \emph{{Perturbative RGE systematics in
  precision observables}},  \href{https://arxiv.org/abs/2407.20842}{{\ttfamily
  2407.20842}}.

\bibitem{Ebert:2021aoo}
M.~A. Ebert, \emph{{Analytic results for Sudakov form factors in QCD}},
  \href{https://doi.org/10.1007/JHEP02(2022)136}{\emph{JHEP} {\bfseries 02}
  (2022) 136} [\href{https://arxiv.org/abs/2110.11360}{{\ttfamily
  2110.11360}}].

\bibitem{Catani:2007vq}
S.~Catani and M.~Grazzini, \emph{{An NNLO subtraction formalism in hadron
  collisions and its application to Higgs boson production at the LHC}},
  \href{https://doi.org/10.1103/PhysRevLett.98.222002}{\emph{Phys. Rev. Lett.}
  {\bfseries 98} (2007) 222002}
  [\href{https://arxiv.org/abs/hep-ph/0703012}{{\ttfamily hep-ph/0703012}}].

\bibitem{Gaunt:2015pea}
J.~Gaunt, M.~Stahlhofen, F.~J. Tackmann and J.~R. Walsh, \emph{{N-jettiness
  Subtractions for NNLO QCD Calculations}},
  \href{https://doi.org/10.1007/JHEP09(2015)058}{\emph{JHEP} {\bfseries 09}
  (2015) 058} [\href{https://arxiv.org/abs/1505.04794}{{\ttfamily
  1505.04794}}].

\bibitem{Cleymans:1978ip}
J.~Cleymans and M.~Kuroda, \emph{{Angular Distribution of Dileptons in Hadronic
  Collisions}}, \href{https://doi.org/10.1016/0550-3213(79)90282-7}{\emph{Nucl.
  Phys. B} {\bfseries 155} (1979) 480}.

\bibitem{Chaichian:1981va}
M.~Chaichian, M.~Hayashi and K.~Yamagishi, \emph{{Angular Distributions of High
  Mass Dileptons With Finite Transverse Momentum in High-energy Hadronic
  Collisions}}, \href{https://doi.org/10.1103/PhysRevD.25.130}{\emph{Phys. Rev.
  D} {\bfseries 25} (1982) 130}.

\bibitem{Mirkes:1992hu}
E.~Mirkes, \emph{{Angular decay distribution of leptons from W bosons at NLO in
  hadronic collisions}},
  \href{https://doi.org/10.1016/0550-3213(92)90046-E}{\emph{Nucl. Phys.}
  {\bfseries B387} (1992) 3}.

\bibitem{Campbell:2015qma}
J.~M. Campbell, R.~K. Ellis and W.~T. Giele, \emph{{A Multi-Threaded Version of
  MCFM}}, \href{https://doi.org/10.1140/epjc/s10052-015-3461-2}{\emph{Eur.
  Phys. J.} {\bfseries C75} (2015) 246}
  [\href{https://arxiv.org/abs/1503.06182}{{\ttfamily 1503.06182}}].

\bibitem{Boughezal:2016wmq}
R.~Boughezal, J.~M. Campbell, R.~K. Ellis, C.~Focke, W.~Giele, X.~Liu et~al.,
  \emph{{Color singlet production at NNLO in MCFM}},
  \href{https://doi.org/10.1140/epjc/s10052-016-4558-y}{\emph{Eur. Phys. J.}
  {\bfseries C77} (2017) 7} [\href{https://arxiv.org/abs/1605.08011}{{\ttfamily
  1605.08011}}].

\bibitem{Bailey:2020ooq}
S.~Bailey, T.~Cridge, L.~A. Harland-Lang, A.~D. Martin and R.~S. Thorne,
  \emph{{Parton distributions from LHC, HERA, Tevatron and fixed target data:
  MSHT20 PDFs}},
  \href{https://doi.org/10.1140/epjc/s10052-021-09057-0}{\emph{Eur. Phys. J. C}
  {\bfseries 81} (2021) 341}
  [\href{https://arxiv.org/abs/2012.04684}{{\ttfamily 2012.04684}}].

\bibitem{NNPDF:2021njg}
{\scshape NNPDF} collaboration, R.~D. Ball et~al., \emph{{The path to proton
  structure at 1\% accuracy}},
  \href{https://doi.org/10.1140/epjc/s10052-022-10328-7}{\emph{Eur. Phys. J. C}
  {\bfseries 82} (2022) 428}
  [\href{https://arxiv.org/abs/2109.02653}{{\ttfamily 2109.02653}}].

\bibitem{Moult:2016fqy}
I.~Moult, L.~Rothen, I.~W. Stewart, F.~J. Tackmann and H.~X. Zhu,
  \emph{{Subleading Power Corrections for N-Jettiness Subtractions}},
  \href{https://doi.org/10.1103/PhysRevD.95.074023}{\emph{Phys. Rev. D}
  {\bfseries 95} (2017) 074023}
  [\href{https://arxiv.org/abs/1612.00450}{{\ttfamily 1612.00450}}].

\bibitem{Moult:2017jsg}
I.~Moult, L.~Rothen, I.~W. Stewart, F.~J. Tackmann and H.~X. Zhu,
  \emph{{N-jettiness subtractions for $gg\to H$ at subleading power}},
  \href{https://doi.org/10.1103/PhysRevD.97.014013}{\emph{Phys. Rev. D}
  {\bfseries 97} (2018) 014013}
  [\href{https://arxiv.org/abs/1710.03227}{{\ttfamily 1710.03227}}].

\bibitem{Bizon:2019zgf}
W.~Bizon, A.~Gehrmann-De~Ridder, T.~Gehrmann, N.~Glover, A.~Huss, P.~F. Monni
  et~al., \emph{{The transverse momentum spectrum of weak gauge bosons at
  N$^3$LL + NNLO}},
  \href{https://doi.org/10.1140/epjc/s10052-019-7324-0}{\emph{Eur. Phys. J. C}
  {\bfseries 79} (2019) 868}
  [\href{https://arxiv.org/abs/1905.05171}{{\ttfamily 1905.05171}}].

\bibitem{Ebert:2022cku}
M.~A. Ebert, J.~K.~L. Michel, I.~W. Stewart and Z.~Sun, \emph{{Disentangling
  long and short distances in momentum-space TMDs}},
  \href{https://doi.org/10.1007/JHEP07(2022)129}{\emph{JHEP} {\bfseries 07}
  (2022) 129} [\href{https://arxiv.org/abs/2201.07237}{{\ttfamily
  2201.07237}}].

\bibitem{Lustermans:2019plv}
G.~Lustermans, J.~K.~L. Michel, F.~J. Tackmann and W.~J. Waalewijn,
  \emph{{Joint two-dimensional resummation in $q_{T}$ and $0$-jettiness at
  NNLL}}, \href{https://doi.org/10.1007/JHEP03(2019)124}{\emph{JHEP} {\bfseries
  03} (2019) 124} [\href{https://arxiv.org/abs/1901.03331}{{\ttfamily
  1901.03331}}].

\bibitem{Bertolini:2017eui}
D.~Bertolini, M.~P. Solon and J.~R. Walsh, \emph{{Integrated and Differential
  Accuracy in Resummed Cross Sections}},
  \href{https://doi.org/10.1103/PhysRevD.95.054024}{\emph{Phys. Rev. D}
  {\bfseries 95} (2017) 054024}
  [\href{https://arxiv.org/abs/1701.07919}{{\ttfamily 1701.07919}}].

\bibitem{Bacchetta:2017gcc}
A.~Bacchetta, F.~Delcarro, C.~Pisano, M.~Radici and A.~Signori,
  \emph{{Extraction of partonic transverse momentum distributions from
  semi-inclusive deep-inelastic scattering, Drell-Yan and Z-boson production}},
  \href{https://doi.org/10.1007/JHEP06(2017)081}{\emph{JHEP} {\bfseries 06}
  (2017) 081} [\href{https://arxiv.org/abs/1703.10157}{{\ttfamily
  1703.10157}}].

\bibitem{NNPDF:2017mvq}
{\scshape NNPDF} collaboration, R.~D. Ball et~al., \emph{{Parton distributions
  from high-precision collider data}},
  \href{https://doi.org/10.1140/epjc/s10052-017-5199-5}{\emph{Eur. Phys. J. C}
  {\bfseries 77} (2017) 663}
  [\href{https://arxiv.org/abs/1706.00428}{{\ttfamily 1706.00428}}].

\bibitem{Hou:2019efy}
T.-J. Hou et~al., \emph{{New CTEQ global analysis of quantum chromodynamics
  with high-precision data from the LHC}},
  \href{https://doi.org/10.1103/PhysRevD.103.014013}{\emph{Phys. Rev. D}
  {\bfseries 103} (2021) 014013}
  [\href{https://arxiv.org/abs/1912.10053}{{\ttfamily 1912.10053}}].

\bibitem{Boussarie:2023izj}
R.~Boussarie et~al., \emph{{TMD Handbook}},
  \href{https://arxiv.org/abs/2304.03302}{{\ttfamily 2304.03302}}.

\bibitem{Scimemi:2018xaf}
I.~Scimemi and A.~Vladimirov, \emph{{Systematic analysis of double-scale
  evolution}}, \href{https://doi.org/10.1007/JHEP08(2018)003}{\emph{JHEP}
  {\bfseries 08} (2018) 003}
  [\href{https://arxiv.org/abs/1803.11089}{{\ttfamily 1803.11089}}].

\bibitem{Vladimirov:2020umg}
A.~A. Vladimirov, \emph{{Self-contained definition of the Collins-Soper
  kernel}}, \href{https://doi.org/10.1103/PhysRevLett.125.192002}{\emph{Phys.
  Rev. Lett.} {\bfseries 125} (2020) 192002}
  [\href{https://arxiv.org/abs/2003.02288}{{\ttfamily 2003.02288}}].

\bibitem{Scimemi:2016ffw}
I.~Scimemi and A.~Vladimirov, \emph{{Power corrections and renormalons in
  Transverse Momentum Distributions}},
  \href{https://doi.org/10.1007/JHEP03(2017)002}{\emph{JHEP} {\bfseries 03}
  (2017) 002} [\href{https://arxiv.org/abs/1609.06047}{{\ttfamily
  1609.06047}}].

\bibitem{Collins:2014jpa}
J.~Collins and T.~Rogers, \emph{{Understanding the large-distance behavior of
  transverse-momentum-dependent parton densities and the Collins-Soper
  evolution kernel}},
  \href{https://doi.org/10.1103/PhysRevD.91.074020}{\emph{Phys. Rev. D}
  {\bfseries 91} (2015) 074020}
  [\href{https://arxiv.org/abs/1412.3820}{{\ttfamily 1412.3820}}].

\bibitem{Schlemmer:2021aij}
M.~Schlemmer, A.~Vladimirov, C.~Zimmermann, M.~Engelhardt and A.~Sch\"afer,
  \emph{{Determination of the Collins-Soper Kernel from Lattice QCD}},
  \href{https://doi.org/10.1007/JHEP08(2021)004}{\emph{JHEP} {\bfseries 08}
  (2021) 004} [\href{https://arxiv.org/abs/2103.16991}{{\ttfamily
  2103.16991}}].

\bibitem{LatticePartonLPC:2022eev}
{\scshape Lattice Parton (LPC)} collaboration, M.-H. Chu et~al.,
  \emph{{Nonperturbative determination of the Collins-Soper kernel from
  quasitransverse-momentum-dependent wave functions}},
  \href{https://doi.org/10.1103/PhysRevD.106.034509}{\emph{Phys. Rev. D}
  {\bfseries 106} (2022) 034509}
  [\href{https://arxiv.org/abs/2204.00200}{{\ttfamily 2204.00200}}].

\bibitem{Shu:2023cot}
H.-T. Shu, M.~Schlemmer, T.~Sizmann, A.~Vladimirov, L.~Walter, M.~Engelhardt
  et~al., \emph{{Universality of the Collins-Soper kernel in lattice
  calculations}},
  \href{https://doi.org/10.1103/PhysRevD.108.074519}{\emph{Phys. Rev. D}
  {\bfseries 108} (2023) 074519}
  [\href{https://arxiv.org/abs/2302.06502}{{\ttfamily 2302.06502}}].

\bibitem{LatticePartonLPC:2023pdv}
{\scshape Lattice Parton (LPC)} collaboration, M.-H. Chu et~al., \emph{{Lattice
  calculation of the intrinsic soft function and the Collins-Soper kernel}},
  \href{https://doi.org/10.1007/JHEP08(2023)172}{\emph{JHEP} {\bfseries 08}
  (2023) 172} [\href{https://arxiv.org/abs/2306.06488}{{\ttfamily
  2306.06488}}].

\bibitem{Avkhadiev:2024mgd}
A.~Avkhadiev, P.~E. Shanahan, M.~L. Wagman and Y.~Zhao, \emph{{Determination of
  the Collins-Soper Kernel from Lattice QCD}},
  \href{https://doi.org/10.1103/PhysRevLett.132.231901}{\emph{Phys. Rev. Lett.}
  {\bfseries 132} (2024) 231901}
  [\href{https://arxiv.org/abs/2402.06725}{{\ttfamily 2402.06725}}].

\bibitem{Bury:2022czx}
M.~Bury, F.~Hautmann, S.~Leal-Gomez, I.~Scimemi, A.~Vladimirov and P.~Zurita,
  \emph{{PDF bias and flavor dependence in TMD distributions}},
  \href{https://doi.org/10.1007/JHEP10(2022)118}{\emph{JHEP} {\bfseries 10}
  (2022) 118} [\href{https://arxiv.org/abs/2201.07114}{{\ttfamily
  2201.07114}}].

\bibitem{Scimemi:2018mmi}
I.~Scimemi and A.~Vladimirov, \emph{{Matching of transverse momentum dependent
  distributions at twist-3}},
  \href{https://doi.org/10.1140/epjc/s10052-018-6263-5}{\emph{Eur. Phys. J.}
  {\bfseries C78} (2018) 802}
  [\href{https://arxiv.org/abs/1804.08148}{{\ttfamily 1804.08148}}].

\bibitem{Boer:1999mm}
D.~Boer, \emph{{Investigating the origins of transverse spin asymmetries at
  RHIC}}, \href{https://doi.org/10.1103/PhysRevD.60.014012}{\emph{Phys. Rev.}
  {\bfseries D60} (1999) 014012}
  [\href{https://arxiv.org/abs/hep-ph/9902255}{{\ttfamily hep-ph/9902255}}].

\bibitem{Gao:2024xxx}
A.~Gao, J.~K.~L. Michel and I.~W. Stewart, \emph{to appear},  2024.

\bibitem{Gao:2022bzi}
A.~Gao, J.~K.~L. Michel, I.~W. Stewart and Z.~Sun, \emph{{Better angle on
  hadron transverse momentum distributions at the Electron-Ion Collider}},
  \href{https://doi.org/10.1103/PhysRevD.107.L091504}{\emph{Phys. Rev. D}
  {\bfseries 107} (2023) L091504}
  [\href{https://arxiv.org/abs/2209.11211}{{\ttfamily 2209.11211}}].

\bibitem{Tackmann:2024xxx}
F.~J. Tackmann, \emph{Theory uncertainties and correlations from theory
  nuisance parameters},  in \emph{SCET 2024: XXIst annual workshop on
  Soft-Collinear Effective Theory}, (2024).

\bibitem{Tackmann:2024kci}
F.~J. Tackmann, \emph{{Beyond Scale Variations: Perturbative Theory
  Uncertainties from Nuisance Parameters}},
  \href{https://arxiv.org/abs/2411.18606}{{\ttfamily 2411.18606}}.

\bibitem{Stewart:2013faa}
I.~W. Stewart, F.~J. Tackmann, J.~R. Walsh and S.~Zuberi, \emph{{Jet $p_T$
  resummation in Higgs production at NNLL$'+$NNLO}},
  \href{https://doi.org/10.1103/PhysRevD.89.054001}{\emph{Phys. Rev.}
  {\bfseries D89} (2014) 054001}
  [\href{https://arxiv.org/abs/1307.1808}{{\ttfamily 1307.1808}}].

\bibitem{Cal:2023mib}
P.~Cal, R.~von Kuk, M.~A. Lim and F.~J. Tackmann, \emph{{$q_T$ spectrum for
  Higgs boson production via heavy quark annihilation at N3LL$'$+aN3LO}},
  \href{https://doi.org/10.1103/PhysRevD.110.076005}{\emph{Phys. Rev. D}
  {\bfseries 110} (2024) 076005}
  [\href{https://arxiv.org/abs/2306.16458}{{\ttfamily 2306.16458}}].

\bibitem{Lam:1978pu}
C.~S. Lam and W.-K. Tung, \emph{{A Systematic Approach to Inclusive Lepton Pair
  Production in Hadronic Collisions}},
  \href{https://doi.org/10.1103/PhysRevD.18.2447}{\emph{Phys. Rev.} {\bfseries
  D18} (1978) 2447}.

\bibitem{Shanahan:2021tst}
P.~Shanahan, M.~Wagman and Y.~Zhao, \emph{{Lattice QCD calculation of the
  Collins-Soper kernel from quasi-TMDPDFs}},
  \href{https://doi.org/10.1103/PhysRevD.104.114502}{\emph{Phys. Rev. D}
  {\bfseries 104} (2021) 114502}
  [\href{https://arxiv.org/abs/2107.11930}{{\ttfamily 2107.11930}}].

\bibitem{Pietrulewicz:2017gxc}
P.~Pietrulewicz, D.~Samitz, A.~Spiering and F.~J. Tackmann,
  \emph{{Factorization and Resummation for Massive Quark Effects in Exclusive
  Drell-Yan}}, \href{https://doi.org/10.1007/JHEP08(2017)114}{\emph{JHEP}
  {\bfseries 08} (2017) 114}
  [\href{https://arxiv.org/abs/1703.09702}{{\ttfamily 1703.09702}}].

\bibitem{Butterworth:2015oua}
J.~Butterworth et~al., \emph{{PDF4LHC recommendations for LHC Run II}},
  \href{https://doi.org/10.1088/0954-3899/43/2/023001}{\emph{J. Phys. G}
  {\bfseries 43} (2016) 023001}
  [\href{https://arxiv.org/abs/1510.03865}{{\ttfamily 1510.03865}}].

\bibitem{ParticleDataGroup:2024cfk}
{\scshape Particle Data Group} collaboration, S.~Navas et~al., \emph{{Review of
  particle physics}},
  \href{https://doi.org/10.1103/PhysRevD.110.030001}{\emph{Phys. Rev. D}
  {\bfseries 110} (2024) 030001}.

\bibitem{Abbate:2010xh}
R.~Abbate, M.~Fickinger, A.~H. Hoang, V.~Mateu and I.~W. Stewart, \emph{{Thrust
  at N$^3$LL with Power Corrections and a Precision Global Fit for
  $\alpha_s(m_Z)$}},
  \href{https://doi.org/10.1103/PhysRevD.83.074021}{\emph{Phys. Rev. D}
  {\bfseries 83} (2011) 074021}
  [\href{https://arxiv.org/abs/1006.3080}{{\ttfamily 1006.3080}}].

\bibitem{Abbate:2012jh}
R.~Abbate, M.~Fickinger, A.~H. Hoang, V.~Mateu and I.~W. Stewart,
  \emph{{Precision Thrust Cumulant Moments at $N^3$LL}},
  \href{https://doi.org/10.1103/PhysRevD.86.094002}{\emph{Phys. Rev. D}
  {\bfseries 86} (2012) 094002}
  [\href{https://arxiv.org/abs/1204.5746}{{\ttfamily 1204.5746}}].

\bibitem{Falcioni:2023luc}
G.~Falcioni, F.~Herzog, S.~Moch and A.~Vogt, \emph{{Four-loop splitting
  functions in QCD \textendash{} The quark-quark case}},
  \href{https://doi.org/10.1016/j.physletb.2023.137944}{\emph{Phys. Lett. B}
  {\bfseries 842} (2023) 137944}
  [\href{https://arxiv.org/abs/2302.07593}{{\ttfamily 2302.07593}}].

\bibitem{Falcioni:2023tzp}
G.~Falcioni, F.~Herzog, S.~Moch, J.~Vermaseren and A.~Vogt, \emph{{The double
  fermionic contribution to the four-loop quark-to-gluon splitting function}},
  \href{https://doi.org/10.1016/j.physletb.2023.138351}{\emph{Phys. Lett. B}
  {\bfseries 848} (2024) 138351}
  [\href{https://arxiv.org/abs/2310.01245}{{\ttfamily 2310.01245}}].

\bibitem{Moch:2023tdj}
S.~Moch, B.~Ruijl, T.~Ueda, J.~Vermaseren and A.~Vogt, \emph{{Additional
  moments and x-space approximations of four-loop splitting functions in QCD}},
  \href{https://doi.org/10.1016/j.physletb.2024.138468}{\emph{Phys. Lett. B}
  {\bfseries 849} (2024) 138468}
  [\href{https://arxiv.org/abs/2310.05744}{{\ttfamily 2310.05744}}].

\bibitem{Falcioni:2024xyt}
G.~Falcioni, F.~Herzog, S.~Moch, A.~Pelloni and A.~Vogt, \emph{{Four-loop
  splitting functions in QCD \textendash{} The quark-to-gluon case}},
  \href{https://doi.org/10.1016/j.physletb.2024.138906}{\emph{Phys. Lett. B}
  {\bfseries 856} (2024) 138906}
  [\href{https://arxiv.org/abs/2404.09701}{{\ttfamily 2404.09701}}].

\bibitem{Falcioni:2024qpd}
G.~Falcioni, F.~Herzog, S.~Moch, A.~Pelloni and A.~Vogt, \emph{{Four-loop
  splitting functions in QCD -- The gluon-gluon case --}},
  \href{https://arxiv.org/abs/2410.08089}{{\ttfamily 2410.08089}}.

\bibitem{Boughezal:2017nla}
R.~Boughezal, A.~Guffanti, F.~Petriello and M.~Ubiali, \emph{{The impact of the
  LHC Z-boson transverse momentum data on PDF determinations}},
  \href{https://doi.org/10.1007/JHEP07(2017)130}{\emph{JHEP} {\bfseries 07}
  (2017) 130} [\href{https://arxiv.org/abs/1705.00343}{{\ttfamily
  1705.00343}}].

\bibitem{Duhr:2020seh}
C.~Duhr, F.~Dulat and B.~Mistlberger, \emph{{Drell-Yan Cross Section to Third
  Order in the Strong Coupling Constant}},
  \href{https://doi.org/10.1103/PhysRevLett.125.172001}{\emph{Phys. Rev. Lett.}
  {\bfseries 125} (2020) 172001}
  [\href{https://arxiv.org/abs/2001.07717}{{\ttfamily 2001.07717}}].

\end{thebibliography}\endgroup

\end{document}